\documentclass[a4paper,11pt]{article}
\usepackage{bbold}
\usepackage[x11names]{xcolor}
\usepackage{pos}
\usepackage{tcolorbox}
\usepackage{enumitem}
\usepackage{graphicx}
\usepackage{cleveref}

\usepackage{tikz}
\usetikzlibrary{patterns,decorations.pathmorphing}

\tikzset{
  hatchblob/.style={
    circle, draw=black,
    minimum size=5mm, inner sep=0pt,
    preaction={fill=white},
    pattern=north east lines,
    pattern color=black
  },
  cpblob/.style={circle, fill=black, draw=black, minimum size=4mm, inner sep=0pt},
  pion/.style={dashed, thick},
  nucleon/.style={thick},
  photon/.style={decorate, decoration={snake, amplitude=1.2pt, segment length=4pt}, thick},
}

\usepackage{bibentry}
\nobibliography{biblio} 

\definecolor{pos}{RGB}{102, 156, 22}

\title{Lectures on Light Particles and Compact Objects}

\newcommand{\nobel}{\raisebox{-0.2ex}{\includegraphics[height=2ex]{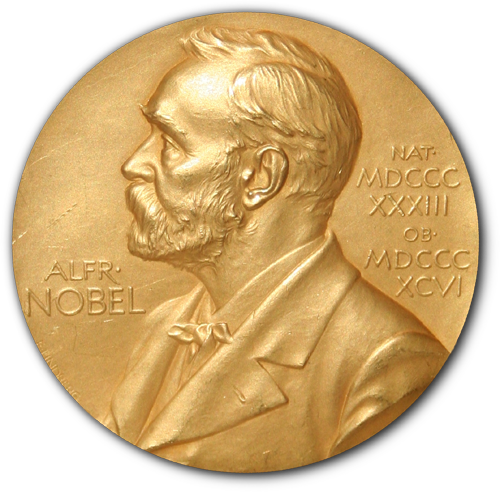}}}

\author[b,c]{Alessandro Lella}

\author[a]{Jamie McDonald}

\affiliation[b]{Dipartimento Interateneo di Fisica “Michelangelo Merlin”, Via Amendola 173, 70126 Bari, Italy}

\affiliation[c]{Istituto Nazionale di Fisica Nucleare - Sezione di Bari, Via Orabona 4, 70126 Bari, Italy}

\affiliation[a]{Department of Physics and Astronomy, University of Manchester, Oxford Road, Manchester M13 9PL, United Kingdom}

\affiliation[aux]{\vspace{-1.5em}}  

\emailAdd{alessandro.lella@ba.infn.it}

\emailAdd{jamie.mcdonald@manchester.ac.uk}

\abstract{This document is based on lectures delivered at a recent COSMIC WISPers COST Action training school in Annecy in September 2025. They examine detection of weakly interacting slim particles (WISPs), specifically axions and high-frequency gravitational waves, with compact objects. These slightly expanded notes focus on searches for axion dark matter and axion-like particles with neutron stars, superradiance, white dwarfs and astrophysical searches for high-frequency gravitational waves. They are accompanied by a set of practical exercises. Comments on these notes are gratefully received. 
}

\FullConference{3rd Training School of the COST Action COSMIC WISPers (CA21106)\\
16-19 September 2025\\
Annecy (France)\\}

\begin{document}
\maketitle

\newpage
\oldTableofcontents

 \newpage

\section{Introduction}

Compact objects describe a class of astrophysical bodies consisting of neutron stars, white dwarfs and black holes, whose density greatly exceeds that of other stars. For many decades, compact objects have been a leading source of progress in the development of fundamental particle physics and cosmology. Neutron stars arguably provided early indirect evidence for quantum mechanics via neutron degeneracy pressure. Without black holes and neutron stars, LIGO and VIRGO's detectors would remain silent. Pulsars have provided indirect evidence for nHz gravitational waves, as well as giving us numerous precision test of general relativity, including the first indirect evidence for gravitational radiation in the form of the Hulse-Taylor binary. A new window on neutrino physics was provided by a core collapse supernova in 1987. This same event also allowed us to place constraints on axions. We owe the discovery of dark energy to type Ia supernovae which act as standard candles. Spinning black holes allow us to place constraints on light axion-like particles. The list goes on. 

Below, we review below some of the key developments in the history of compact objects and their role in understanding the fundamental interactions of nature.

\vspace{1cm}
\noindent \textbf{Some Highlights from Compact Objects and Fundamental Physics}
\vspace{2pt}
\begin{enumerate}[label={}, leftmargin=0pt, itemsep=0pt, parsep=1.5pt, topsep=1pt]
   
 \item \textbf{1006 } First recorded SN\footnote{Earlier records date to 185 CE but these are not confirmed.} with confirmed present-day remnant \cite{1965AJ.....70..754G}.
    
    \item \textbf{1784 } Mitchell \cite{Michell} and Laplace (1796) hypothesise dense objects from which light cannot escape.
    
    \item  \textbf{1916 } Schwarzschild publishes first black hole solution to Einstein Equation 
    \cite{Schwarzschild:1916uq} 
    
    \item \textbf{1934 } Baade \& Zwicky \cite{Baade:1934wuu} propose neutron stars as the endpoint of supernovae.
    
    \item \textbf{1963 } Kerr derives solution for spinning black holes \cite{Kerr:1963ud}.
    
    \item \textbf{1967 } Pulsars discovered by Hewish \textit{et al.}~\cite{Hewish:1968bj}  at the Mullard Radio Astronomy Observatory (\nobel 1974).
    
    \item \textbf{1968 } Neutron stars proposed  \cite{Gold:1968zf}  as origin of radio pulses, building on earlier work \cite{Woltjer1964,Pacini:1967epn}
    
    \item \textbf{1974 } Hulse-Taylor binary discovered \cite{Hulse:1974eb}.   Indirect probe \cite{HulseTaylorGWs} of gravitational waves (\nobel 1993).

     \item \textbf{1987} Neutrino detectors \cite{Hirata:1987hu}, \cite{Bionta:1987qt} receive burst from SN 1987A.
    
    \item \textbf{2015} Direct detection of gravitational waves from black hole mergers by LIGO \cite{LIGOScientific:2016aoc} (\nobel 2017)

    \item \textbf{2023} Possible indirect observation of gravitational waves by pulsar timing \cite{NANOGrav:2020bcs,Goncharov:2021oub,EPTA:2021crs,Tarafdar:2022toa}

\end{enumerate}
\vspace{1cm}

The purpose of these lectures notes is to outline some contemporary topics in the search for light particles with  compact objects. These notes can only briefly capture a snapshot of the huge breadth of work in this area. Many excellent reviews touching on the topics can be found below: 

\newpage 

\noindent \textbf{Further Reading and Reviews}

\begin{quote}
 \bibentry{Bekenstein:1998nt}~\cite{Bekenstein:1998nt}
 \\\\
 \bibentry{Brito:2015oca}~\cite{Brito:2015oca}
\\\\
\bibentry{Raffelt:1996wa}~\cite{Raffelt:1996wa}
\\\\
\bibentry{Caputo:2024oqc}~\cite{Caputo:2024oqc}
\end{quote}

The remainder of these lecture notes are organized as follows. In sec.~\ref{sec:WISPtheory} we review the underlying theory of axions describing briefly both QCD axions and the string axiverse. Then, in sec.~\ref{sec:superradiance} we outline black hole superradiance and the constraints which can be place on light-particles using spinning black holes. We also outline recent developments in our understanding of neutron stars as sources of superradiance. In sec.~\ref{sec: NeutronStars} we describe the role of neutron stars in constraining WISP properties. Section \ref{sec:WDs} reviews the role of white dwarfs in FIP searches. Finally, in sec.~\ref{sec:HFGWs}, we review recent development in the search for high-frequency gravitational waves in astrophysical and cosmological environments. We conclude with a set of exercises delivered at the training school.

\section{WISP Theory}\label{sec:WISPtheory}

Since these lecture note mainly focus on axion searches, in this section we describe briefly the main theoretical motivations for the existence of axions, discussing both the QCD axion and axion-like particles (ALPs). Our starting point is the path integral for QCD, which reads
\begin{equation}
Z[J]
=\int \mathcal{D}A\,\mathcal{D}q\,\mathcal{D}\bar q\,
\exp\!\left\{i\int d^4x\,
\Big[\mathcal{L}_{\rm QCD}(A,q,\bar q) \Big]\right\}\, , 
\end{equation}
where the $QCD$ Lagrangian is given by 
\begin{equation}
\mathcal L_{\rm QCD}
=
-\frac{1}{4}\, G^a_{\mu\nu} G^{a\,\mu\nu}
+\sum_{f=1}^{N_f}
\bar q_f\,(i\gamma^\mu D_\mu - m_f)\,q_f
+\theta\,\frac{g_s^2}{32\pi^2}\,
G^a_{\mu\nu}\,\tilde G^{a\,\mu\nu}.
\end{equation}
Here $A^a_\mu$ is the gluon gauge field, $q_f$ are the quark fields of flavour $f=1,\dots,N_f$, and
\begin{equation}
D_\mu = \partial_\mu - i g_s\, T^a A^a_\mu
\end{equation}
is the $SU(3)_c$ covariant derivative acting on quarks (with $T^a=\lambda^a/2$ in the fundamental representation). The non-abelian field-strength tensor is
\begin{equation}
G_{\mu\nu}
=\partial_\mu A_\nu
-\partial_\nu A_\mu
- i g_s [A_\mu,A_\nu].
\end{equation}
and its dual is
\begin{equation}
\tilde G^{a\,\mu\nu}=\frac{1}{2}\,\epsilon^{\mu\nu\rho\sigma}G^a_{\rho\sigma}\,,
\end{equation}
with $\epsilon^{0123}=+1$. The final term proportional to $\theta$ is the topological $G\tilde G$ term, which violates $P$ and $CP$. Under chiral transformations the left and right-handed fermions transform as \begin{equation}
\bar q_i \;\to\; \bar q_i\, e^{\,i \alpha_i \gamma_5},
\qquad
\bar q_{iL} \;\to\; \bar q_{iL}\, e^{+i\alpha_i},
\qquad
\bar q_{iR} \;\to\; \bar q_{iR}\, e^{-i\alpha_i}.
\end{equation}
For massless quarks, this corresponds to the classical chiral symmetry conservation, with associated Noether current $ 
J_5^\mu \equiv \bar q \gamma^\mu \gamma_5 q.$ which is classically conserved. However, fermion masses and quantum effects, via the chiral anomaly, violate this conservation, leading to the relation
\begin{equation}
\partial_\mu J_5^\mu
=
2\,\bar q\, M\, i\gamma_5\, q
+
2N_f\,\frac{g_s^2}{32\pi^2}\,
G^a_{\mu\nu}\,\tilde G^{a\,\mu\nu},
\end{equation}
Under chiral transformations, the measure of the path integral, and the fermion mass matrix, transform as, 
\begin{equation}
\mathcal{D}q\,\mathcal{D}\bar q
\;\rightarrow\;
\mathcal{D}q\,\mathcal{D}\bar q\;
\exp\!\left[
\,i\,2\!\left(\sum_{i=1}^{N_f}\alpha_i\right)
\frac{g_s^2}{32\pi^2}
\int d^4x\;
G^a_{\mu\nu}\tilde G^{a\,\mu\nu}
\right]
\qquad\qquad
M \;\rightarrow\; e^{-2i\alpha}\, M .
\end{equation}
We see that the quantity
\begin{equation}
    \bar{\theta} = \theta + \arg \det M, \, 
\end{equation}
is invariant under field redefinitions from chiral rotations and contains information about the coupling of $G \tilde{G}$. Together, this suggests that $\bar{\theta}$ is an invariant which parameterize CP-violation. We will now show how $\bar{\theta}$ acts as a parameter controlling the presence of CP violation in physical observables. To demonstrate this, it is therefore necessary to show the appearance of $\theta$ in CP-violating observables. An excellent overview of CP violation and calculations within QCD and chiral perturbation theory is given in the recent lecture notes by Sannino \cite{Sannino:2026wgx}. 

We will begin by computing the neutron electric dipole moment. A classic estimate using chiral perturbation theory (ChiPT), which describes the behaviour of hadrons (i.e. QCD at low energy) was given in Ref.~\cite{Crewther:1979pi}. 

\begin{figure}
\centering

\begin{tikzpicture}

\begin{scope}[xshift=0cm]
  \coordinate (n_in)  at (-1.6,-1.7);
  \coordinate (p_out) at (-1.6, 2.3);

\def\s{1.2} 

\coordinate (L1) at (-0.9,-0.3);
\coordinate (L2) at (-0.9,{-0.3+\s});
\coordinate (R)  at ({-0.9+0.8660254*\s},{-0.3+0.5*\s});

  \draw[nucleon,->] (n_in) -- (L1) -- (L2) -- (p_out);

  \draw[pion] (L1) to[out=20,in=200] (R);
  \draw[pion] (R)  to[out=160,in=340] (L2);

  \draw[photon] (R) -- (2.0,0.35) node[right] {$\gamma$};

  \node[hatchblob] at (L1) {};
  \node[cpblob]    at (L2) {};
  \node[hatchblob] at (R)  {};

  \node[left]  at (-1.65, 0.55) {$p$};
  \node[left]  at (-1.65,-1.45) {$n$};
  \node[below] at (0.15,-0.65) {$\pi^-$};
\end{scope}

\begin{scope}[xshift=6.6cm]

\def\s{1.2} 

  \coordinate (n_in)  at (-1.6,-1.7);
  \coordinate (p_out) at (-1.6, 2.3);
  \coordinate (L1) at (-0.9,-0.3);
  \coordinate (L2) at (-0.9,{-0.3+\s});
  \coordinate (R)  at ({-0.9+0.8660254*\s},{-0.3+0.5*\s});

  \draw[nucleon,->] (n_in) -- (L1) -- (L2) -- (p_out);

  \draw[pion] (L1) to[out=20,in=200] (R);
  \draw[pion] (R)  to[out=160,in=340] (L2);

  \draw[photon] (R) -- (2.0,0.35);

  \node[cpblob]    at (L1) {};
  \node[hatchblob] at (L2) {};
  \node[hatchblob] at (R)  {};
\end{scope}

\end{tikzpicture}

\caption{Diagrams contributing to the neutron electric dipole moment dark and hatched circles denote the CP odd and CP even interactions in eq.~\eqref{eq:ChPTCPOdd}, respectively.  }
\label{fig:nEDM}
\end{figure}
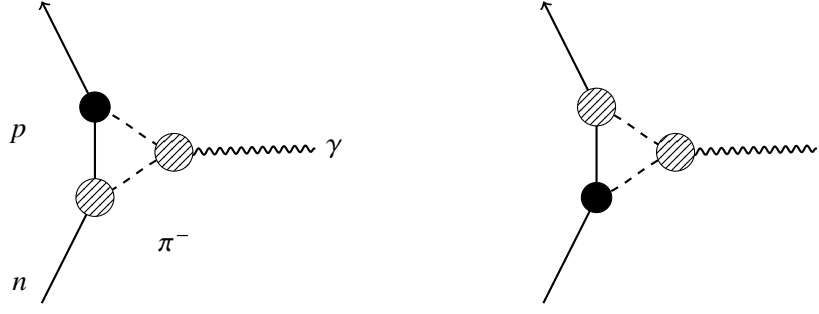

The neutron electric dipole moment corresponds to compute the vertex correction shown in Fig.~\ref{fig:nEDM}. Within the validity of chiral perturbation theory, the leading order corrections can be captured by the following vertices\footnote{It is important to note that the authors of Ref.~\cite{Ai:2020ptm} point out a general form of the chiral Lagrangian relevant for computing the neutron EDM given by $|\lambda|\, e^{-i\xi}\, f_\pi^4 \det U \;+\; |\lambda|\, e^{i\xi}\, f_\pi^4 \det U^\dagger$. Here $\xi$  can take two values $\xi = \theta$, i.e.\ in general misaligned with the mass terms such that there is CP violation, 
while $\xi = -\bar{\alpha}$, i.e.\ aligned with the mass terms such that there is no CP violation. The latter is argued to be consistent with the treatment of instantaneous in the infinite volume limit, implying no strong-CP problem. The absence of a strong-CP problem has also been argued independently of this approach using canonical quantization in Ref.~\cite{Ai:2024vfa}. The latter approach does not depend on the so-called ``ordering of limits". It is worth remarking that if the $\theta$-term is unphysical, this would naturally neutralise the strong-CP problem, no new physics is required, and to date, the authors have been able to issue rebuttals to critiques of their work. This constitutes an Occam’s razor explanation of why no neutron EDM is observed. }:
\begin{equation}\label{eq:ChPTCPOdd}
\mathcal{L}_{\pi NN}
= \bar{N}
\left(
i\gamma_5\, g_{\pi NN}
+ \bar{g}_{\pi NN}
\right)
\tau^a N \,\pi^a \, .
\end{equation}

At leading order in chiral perturbation theory, the CP-violating
coupling $\bar g_{\pi NN}$ is proportional to $\bar\theta$,
\begin{equation}
\bar g_{\pi NN}
\;\sim\;
\bar\theta\,\frac{m_*}{f_\pi},
\end{equation}
where $m_* = \frac{m_u m_d}{m_u + m_d}$ and $f_\pi \simeq 93\,\text{MeV}$.
The dominant contribution to the neutron electric dipole moment then
arises from the one-loop pion diagram shown in Fig.~\ref{fig:nEDM}.
Evaluating this diagram yields the classic estimate \cite{Crewther:1979pi},
\begin{equation}
d_n
\;\simeq\;
\frac{e\, g_{\pi NN}\,\bar g_{\pi NN}}{4\pi^2 m_N}
\ln\!\left(\frac{m_N}{m_\pi}\right)
\;\sim\;
\mathcal{O}(10^{-16})\,\bar\theta\; e\,\text{cm}.
\end{equation}
Comparing with the current experimental bound
$|d_n| < 1.8 \times 10^{-26}\, e\,\text{cm}$,
one obtains the stringent constraint
\begin{equation}
|\bar\theta| \lesssim 10^{-10},
\end{equation}
which constitutes the strong CP problem.

The solution to this problem was proposed by Peccei and Quinn \cite{Peccei1977}, who introduced an additional global $U(1)$ symmetry that is spontaneously broken, giving rise to a pseudo-Goldstone boson known as the axion. The axion dynamically relaxes the effective $\theta$ parameter to zero, thereby solving the strong CP problem. Explicit realizations of QCD axion models were subsequently developed, most notably the KSVZ \cite{Kim:1979if,Shifman:1979if} and DFSZ \cite{Dine:1981rt,Zhitnitsky:1980tq} models. More generally, axion-like particles (ALPs) can arise in a variety of extensions of the Standard Model, for example in the context of string theory \cite{Arvanitaki:2009fg}.

\section{Black Hole and Stellar Superradiance} \label{sec:superradiance}

Superradiance is a classical wave amplification process whereby bosonic
fields scattering off a rotating body can extract rotational energy.
The phenomenon was first understood in the context of rotating conductors
by Zeldovich, \cite{Zeldovich:1971,Zeldovich:1972}, who showed that a dissipative,
rotating cylinder amplifies incident waves satisfying
\begin{equation}
    \omega < m \Omega ,
\end{equation}
where $\omega$ is the wave frequency, $m$ its azimuthal quantum number,
and $\Omega$ the angular velocity of the body. 

The gravitational analogue arises for Kerr black holes.
Superradiant solutions for massive scalar fields in the Kerr geometry
were first obtained in Ref.~\cite{Detweiler:1980uk}, with early instability
studies in Ref.~\cite{Zouros:1979iw}. A comprehensive modern review can be
found in Ref.~\cite{Brito:2015oca}.  A particularly transparent effective
field theory description, interpreting superradiance in terms of
stimulated emission and absorption processes, was developed in
\cite{Endlich:2016jgc}. Further calculations of the superradiant growth rate were carried out in Ref.~\cite{Leaver:1985ax} using a continued fraction method, which is needed when the Compton wavelength of the field approaches the radius of the black hole. 

Crucially, when fields are endowed with a finite but small mass, they exhibit boundstates, whose amplitude  grows exponentially in time, extracting rotational energy from the central black hole. By requiring the amount of spin extraction to be limited by observations, one can exclude the existence of certain light bosonic fields. Black hole superradiance offers a powerful probe of the so-called string axiverse \cite{Arvanitaki:2010sy}. One important physical effect which is still a subject of active research is the role of self-interactions in influencing the evolution of field profiles, and hence the constraints which can be derived from them - see \cite{Baryakhtar:2020gao} as well as  \cite{Witte:2024drg}.

Two kinds of constraints can be placed: the first comes from
astrophysical (stellar-mass) black holes, leading to constraints in
the mass range $\mu \sim 10^{-13} - 10^{-11}\ {\rm eV}$  whilst the second comes from supermassive black holes, leading to
constraints in the range $\mu \sim 10^{-19} - 10^{-16}\ {\rm eV}$. 

The question of whether neutron stars can support superradiance remains an active area of debate. Unlike the black hole case — where the long-wavelength dynamics of a light bosonic field are fully determined by general relativity, i.e. a minimally coupled scalar propagating in a Kerr background — superradiance in stars necessarily requires dissipation arising from microscopic interactions between the bosonic field and stellar matter. In this respect, the situation is closer to Zeldovich’s rotating cylinder than to a black hole: amplification requires an explicit coupling to internal degrees of freedom.
A phenomenological description of dissipation was introduced in \cite{Cardoso:2015zqa}, where the scalar equation of motion was modified to
\begin{equation}\label{eq:stellarSupEq}
\Box \Phi + \alpha \frac{\partial \Phi}{\partial t} -  \mu^2 \Phi = 0, 
\end{equation}
with $\alpha$ representing an effective damping coefficient in the co-rotating frame of the star and $\mu$ is the mass of the field. As can be shown in exercise 1, Eq.~\eqref{eq:stellarSupEq} leads to superradiant amplification.  While this parametrization captures the essential features needed for superradiant amplification, no first-principles derivation of the equation above was provided at that time. In particular, it was unclear how a damping term emerges from an underlying interaction Lagrangian coupling $\Phi$ to ordinary matter, or how the microphysics of dense stellar interiors determines the magnitude of $\alpha$.
Progress was made in Ref.~\cite{Cardoso:2017kgn}, where an effective damping rate for vector fields was estimated. However, even there the connection to a microscopic derivation from first principles remained incomplete.
A systematic first-principles treatment was developed in \cite{Chadha-Day:2022inf}, where techniques from non-equilibrium quantum field theory were used to derive effective equations of motion including dissipative terms. This approach clarified both the structure of the damping term and its diagrammatic interpretation in terms of scattering processes in dense nuclear matter. In this framework, the damping rate is directly related to the mean free path of the bosonic excitation inside the star.
More recently, however, it has been argued \cite{Bai:2025nzd} that non-perturbative many-body effects — in particular multiple nucleon scatterings — dramatically suppress the effective superradiant growth rate relative to naive perturbative estimates in Ref.~\cite{Chadha-Day:2022inf}. This suppression significantly weakens previously anticipated astrophysical constraints derived from stellar superradiance.
It will therefore be important to further explore macroscopic collective effects in neutron stars. Large-scale perturbations, such as the phonon-based analysis of \cite{Kaplan:2019ako}, may provide complementary avenues for understanding dissipation and amplification in dense stellar environments.

\section{WISP searches with Neutron Stars}
\label{sec: NeutronStars}

Neutron stars (NSs) are among the most compact objects in the Universe (see, e.g., Refs.~\cite{VitaliiGinzburg_1971,Lattimer:2004pg,Potekhin:2010aii,Potekhin:2015qsa} for reviews). They are believed to be the remnants of core-collapse SN explosion events. This marks the terminal phase of massive stars ($M>8\,M_\odot$) which have exhausted their nuclear ``fuel'', giving rise to a degenerate iron core. The subsequent gravitational collapse compresses nuclear matter to densities close to the critical nuclear saturation density $\rho_0\simeq2.8\times10^{14}\,$g cm$^{-3}$. At those densities, nuclear matter becomes incompressible and the core rebounces, giving rise to a shock wave able to strip off the outer envelope of the giant star. The hot and compact remnant rapidly cools down through neutrino emission, releasing an enormous amount of energy $\sim10^{53}\,$erg over time scales of $\sim10\,$s, leading to a strong neutronization of the nuclear matter in the core. The outcome of this catastrophic event is a compact star composed mainly of neutrons, with a typical mass of $1-2\,M_\odot$ compressed within a radius of $R_{\rm NS}\sim10\,$km. 

The newborn NS is a gravitationally stable system, where the gravitational pressure is counterbalanced by the degeneracy pressure of hadrons in the core. Two main qualitatively different regions can be identified: the \emph{core} and the \emph{crust}~(see, e.g., Ref.~\cite{Haensel:2007yy}). The core itself is expected to show two regions characterized by different properties. The \emph{outer core} of a neutron star is several kilometres thick with typical densities around $0.5\rho_0<\rho<2\rho_0$, accounting for the largest fraction of the stellar mass, has well-known qualitative characteristics. It is composed by strongly-degenerate matter, in which neutrons constitute the main component, correlated with an admixture of protons, muons and electrons. Due to the large densities, it is believed that, as the NS cools down, both neutrons and protons may undergo phase transitions to a superfluid and superconductive state, respectively. The \emph{inner core} occupies the central region of the neutron star, which may reach values of the densities as high as $\rho \gtrsim 2 \rho_0$. The composition and the properties of the matter in the inner core are still not well understood. Among the possible hypotheses are the presence of hyperons, kaon and pion condensates and deconfined quarks. Finally, the \emph{stellar crust} is typically $\sim1-2\,$km thick. Here, the density drops to values $\rho<0.5\rho_0$, allowing for the presence of atomic nuclei in addition to degenerate electrons and superfluid neutrons.

Approximately 20 seconds after its birth, the stellar core becomes transparent to neutrinos, which are the main cooling channel during the early evolutionary stages. Shortly thereafter, the high thermal conductivity of the core establishes thermal equilibrium, which is maintained throughout the entire star lifetime. The neutrino cooling stage lasts $\sim10^5$ years, during which the main emission channels are direct and modified Urca processes, as well as neutrino bremsstrahlung radiation~\cite{Gamow:1941gis, Yakovlev:2000jp, Yakovlev:2004iq, Page:2005fq}. At $t\gtrsim10^5\,$years, neutrino emission, which is strongly temperature-dependent, becomes strongly suppressed by the lower core temperature. In this final stage, the star cools down via blackbody radiation through photon emission from the stellar surface~\cite{Yakovlev:2000jp,Page:2006ud}.

Furthermore, NSs are expected to host powerful magnetic fields, with typical values of $B\sim10^{8}-10^{15}\,$G, dependent on their type: isolated active neutron star, millisecond pulsar or magnetar. These fields exhibit a dominant dipole poloidal structure, correlated with small-scale poloidal components and both large- or small- scale toroidal fields in the crust~\cite{Igoshev:2021ewx}. The formation of such strong magnetic fields can originate from the conservation of the magnetic flux of fossil fields present in the SN progenitor. The typical radius of supergiant stars originating the SN explosion event is in the order of $R_*\gtrsim10\,R_\odot$. Then, if the magnetic flux is conserved, the fossil stellar field might be amplified by a factor $\sim(R_*/R_{\rm NS})^2\approx5\times10^{11}$, generating the field magnitudes expected in NS magnetospheres. Dynamo amplification at the proto-NS level may also be at the origin the strong NS magnetization. In particular, the strong convective activity in the proto-NS may produce both poloidal and toroidal components, while a rapid rotation of the newborn NS is required to form a strong dipolar component. 

The extreme interior conditions make NSs powerful laboratories for probing sub-keV WISP candidates~\cite{Caputo:2024oqc,Carenza:2024ehj,Arza:2026rsl}. Owing to their large densities, WISPs can be efficiently produced in NS interiors despite their feeble couplings to Standard Model particles~ Once produced, their weak interactions allow them to escape freely the stellar volume potentially altering the thermal evolution of the stellar cooling process~\cite{Leinson:2014ioa,Sedrakian:2015krq,Sedrakian:2018kdm,Hamaguchi:2018oqw,Leinson:2021ety,Buschmann:2021juv,Fiorillo:2025zzx}.
Notably, WISP candidates, like axions and ALPs can convert into photons in the strong magnetic fields of the NS magnetosphere, possibly producing an unexpected excess of photons~\cite{Morris:1984iz,Fortin:2018ehg,Buschmann:2019pfp}. As we illustrate in the following, depending on the origin of the WISPs converting in the NS magnetic fields, different types of radio- and X-ray signatures can be severely constrained using NS observations~\cite{Pshirkov:2007st,Safdi:2018oeu,Hook:2018iia,Leroy:2019ghm,Battye:2019aco,Foster:2020pgt,Witte:2021arp}.

\subsection{WISP impact on Neutron Star cooling }

Despite their extremely feeble couplings with the standard model, WISP production through several possible mechanisms could be substantially enhanced by the extreme conditions expected in the NS interior. Once produced, WISPs can escape unimpeded from the stellar interior, providing an additional cooling channel for the NS core. Due to a milder temperature dependence compared to neutrino emission rates, WISP emission is typically subdominant during the early neutrino-cooling phase, when the star is hotter. Nonetheless, as the stellar core cools down, WISP emission can become significant, potentially dominating the cooling at intermediate ages ($\sim 10^5$ years).

\begin{figure}
\centering
    \includegraphics[width=0.7\linewidth]{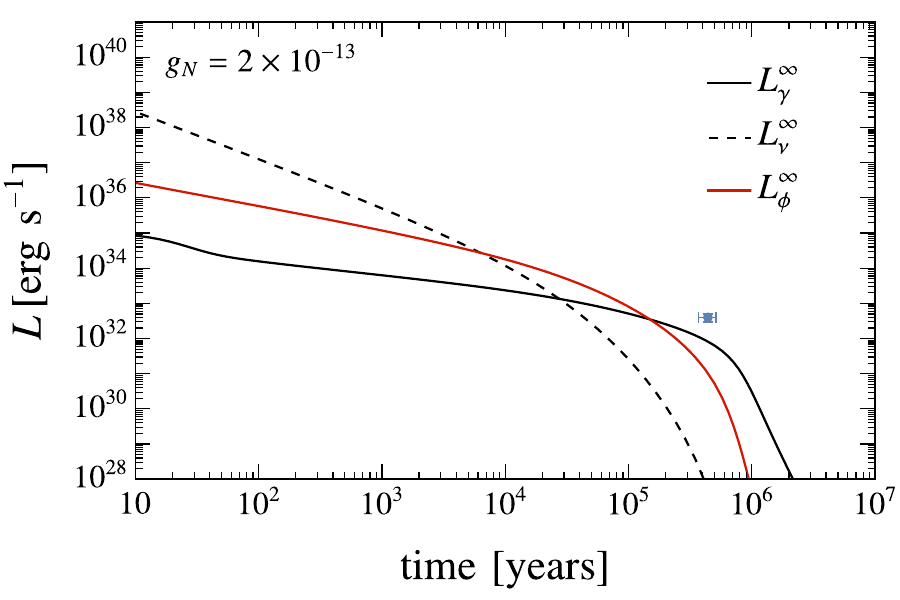}
    \caption{Photon, neutrino and scalar cooling light-curves computed in the best fit model for the isolated NS J1605 derived in Ref.~\cite{Fiorillo:2025zzx}. The scalar emissivity is obtained by assuming a scalar-nucleon coupling $g_N=2\times10^{-13}$.}\label{fig:NSCooling}
\end{figure}

This makes middle-aged, thermally emitting neutron stars, such as the “Magnificent Seven”, particularly suitable for probing anomalous cooling. Fig.~\ref{fig:NSCooling} clearly shows how the impact of an exotic energy-loss channel could significantly modify the standard cooling lightcurve able to fit data of the isolated NS J1605~\cite{Tetzlaff:2012rz, Pires:2019qsk}, accelerating the expected cooling process. Therefore, observations of middle-age NS surface luminosities, combined with theoretical cooling models, have been used to place stringent constraints on WISP couplings.

This possibility has been widely investigated in the context of the QCD axion and ALPs sporting nuclear couplings~\cite{Iwamoto:1984ir,Brinkmann:1988vi,Iwamoto:1992jp}. The dominant ALP production channel in non-superfluid regions is nucleon bremsstrahlung via processes such as $nn \to nn a$, $np \to np a$, and $pp \to pp a$. These processes typically give rise to a modified thermal distribution~\cite{Iwamoto:1984ir}
\begin{equation}
    \frac{dF}{dE} \propto z^3 \frac{z^2 + 4\pi^2}{e^z - 1}\,,
\end{equation}
where $z = E/T$, $E$ is the emitted particle energy, and $T$ is the local core temperature. For typical neutron stars at ages around $\sim10^{5}$ years, this spectrum peaks at energies $\mathcal{O}(\mathrm{keV})$. If the core is in a superfluid state, nucleon bremsstrahlung becomes ineffective as most of the nucleons are bound into Cooper pairs. However, the formation and breaking of these pairs opens an efficient alternative production channel. In this case, axions are emitted as the binding energy of Cooper pairs is released. This mechanism is particularly important when the temperature is close to the critical temperature for superfluidity and can dominate the emission. The corresponding spectrum is generally harder than that from bremsstrahlung, often peaking at $\mathcal{O}(10\ \mathrm{keV})$. Additional production channels may arise from leptonic processes or more exotic phases of dense matter, such as hyperon superfluids or meson condensates. Ref.~\cite{Buschmann:2021juv} investigated the impact of axion emission over five isolated NSs with kinematic ages$\sim10^5$ years, for which thermal
luminosity data are available~\cite{Potekhin:2020ttj,Suzuki:2021ium}, This analysis excludes regions of parameter space corresponding to QCD axion masses at the level of tens of meV, depending on model and the assumptions, setting a limit at the same level as the constraint derived by applying the standard SN cooling based on observations of the SN 1987A neutrino burst~\cite{Raffelt:2006cw,Fischer:2016cyd,Chang:2018rso,Carenza:2019pxu,Carenza:2020cis,Lella:2022uwi,Lella:2023bfb,Caputo:2024oqc,Carenza:2024ehj,Springmann:2024ret,Fiorillo:2025gnd}. 

Neutron stars also provide exceptionally strong probes of light scalar particles coupled to nucleons. A key distinction arises in nucleon bremsstrahlung: while axion and neutrino emission require a nucleon spin flip—leading to rates suppressed by the emitted particle momentum—scalar emission does not. Instead, it proceeds through the quadrupole moment of the nucleon system, introducing a dependence on the nucleon momentum as $(p_N/m_N)^4$. This difference becomes particularly important in cold neutron stars, where nucleons are highly degenerate and their momenta are set by the Fermi momentum $p_F \sim 200$ MeV, rather than by thermal scales. Compared to SN environments, where $p_N \sim \sqrt{m_N T}$ with $T \sim 30$ MeV, this leads to a dramatic enhancement of scalar emission. The ratio of scalar to neutrino luminosities in neutron stars can be estimated as
\begin{equation}
    \left(\frac{L_\phi}{L_\nu}\right)_{\rm NS}
    \sim \left(\frac{L_\phi}{L_\nu}\right)_{\rm SN}
    \frac{p_F^4}{m_N^2 T_{\rm NS}^2}
    \sim 10^7 \left(\frac{L_\phi}{L_\nu}\right)_{\rm SN}\,,
\end{equation}
underlying the enhanced sensitivity of neutron stars to scalar emission. Following the approach introduced for axions \cite{Buschmann:2021juv}, Ref.~\cite{Fiorillo:2025zzx} ruled out value of the scalar-nucleon coupling $g_{\phi N} \gtrsim 5 \times 10^{-14}$ for scalar masses below $m_S \lesssim 1$ MeV. These limits are currently the strongest astrophysical constraints on light scalars in the range ${\rm eV} \lesssim m_S \lesssim {\rm MeV}$, and also imply stringent bounds on axion CP-violating couplings~\cite{Raffelt:2012sp,OHare:2020wah,Fiorillo:2025zzx}. These results also extend to Higgs-portal models, setting the bound on the scalar-Higgs mixing angle at $\sin\theta\lesssim 6\times 10^{-11}$, which improves Red giant/Horizontal branch and WD limits from scalar electron couplings~\cite{Hardy:2016kme,Bottaro:2023gep}. Finally, this argument probes effective Yukawa strengths as small as $|\alpha| \sim 10^{-15}$ relative to gravity, setting also the most stringent limit on scalar-mediated fifth forces acting over length scales $\lambda = m_S^{-1}$ in the micron to picometer regime.

\subsection{Axion Detection in Neutron Star Magnetospheres} \label{sec:NSs}
One of the most promising targets for the detection of axion dark matter in astrophysical environments is conversion in NS magnetospheres. This is especially true with the near future arrival of the Square Kilometre Array (SKA) which will greatly increase sensitivity to axion dark matter. In this scenario, axions from the Galactic halo fall into the plasma surrounding neutron stars, where they are converted into photons. This mechanism was proposed in early works, notably the classic reference~\cite{Raffelt:1987im} and later in Ref.~\cite{Pshirkov:2007st}. It was recently popularised by Refs.~\cite{Hook:2018iia,Huang:2018lxq}, which carried out more detailed estimates of the expected signal strength. 

The mechanism is promising because it not only exploits the extremely strong magnetic fields of neutron stars, which may reach up to $10^{15}\,\mathrm{G}$, but also harnesses resonant enhancement. This occurs when the axion mass, $m_a$, becomes approximately degenerate\footnote{Strictly speaking, the precise statement is that the axion and photon momenta become kinematically matched. As explained in Ref.~\cite{McDonald:2023shx}, in a strongly magnetised plasma, the photon dispersion relation becomes anisotropic taking the form $\omega^{2} = \frac{1}{2}\left[|\mathbf{k}|^{2} + \omega_{p}^{2}\;\pm\;\sqrt{|\mathbf{k}|^{4}+ \omega_{p}^{4}
+ 2|\mathbf{k}|^{2}\omega_{p}^{2}
\left(1 - 2\cos^{2}\theta_{B}\right)
}
\right]$, where $\omega$ is the photon frequency, $\textbf{k}$ is its momentum, $\omega_p$ is the plasma mass, and $\theta_B$ is the angle between the background magnetic field and $\textbf{k}$. This means that the condition to match the axion and photon momenta describes a continuous foliation of surfaces parametrised by $\theta_B$, rather than a single surface on which $\omega_{\rm pl} = \omega_{\rm p}$.
However, for non-relativistic axions, these surfaces are nearly degenerate with $\omega_p \simeq m_a$ remaining an adequate description..} with the effective plasma mass of the photon, $\omega_{\rm p}$. 

This can lead to strong production of radio photons from the magnetosphere around neutron stars, which has prompted a number of observations with MeerKAT \cite{Battye:2023oac}, Greenbank and Effelsberg \cite{foster2020, FosterSETI2022}, and the Very Large Array (VLA) \cite{darling2020apj,Darling:2020uyo, Battye2022}. Exercise 3 focuses on computing the sensitivity of radio telescopes to these signals. Much effort has also been made on the theory side, to model the transport of the photons out of the magnetosphere using ray-tracing techniques \cite{Witte:2021arp,Battye:2021xvt, McDonald:2023shx} which include both refraction from a fully anisotropic magnetised plasma as well as curved spacetime effects. The modelling of the conversion process itself also posed a significant challenge for some time, in that formally, the traditional 1D mixing equations \cite{Raffelt:1987im} need not apply when the axion and photon move on distinct curvilinear trajectories. This even led to some \cite{Witte:2021arp} speculation that the signal may experience a suppression due to axions and photons moving on different trajectories (a phenomenon dubbed de-phasing). However, this problem has recently been solved, using both kinetic theory \cite{McDonald:2023ohd}, which describes the phase-space densities of axions and photons and the production of one into another, as well as mixing via classical wave equations \cite{McDonald:2024uuh}. Both lead to the following compact formula, for axion photon conversion in an arbitrary plasma: 
\begin{equation}\label{eq:ProbabilityPhysical}
	P_{a \gamma  } =   \frac{  \pi g_{a \gamma \gamma}^2 \big|\hat{\boldsymbol{\varepsilon}}\cdot \textbf{B}_{\rm ext}\big|^2	}{  \left| \textbf{v}_p \cdot \nabla_\textbf{x} E_\gamma(\textbf{k},\textbf{x})\right|}  \frac{U_E}{U},
\end{equation}
where $\hat{\boldsymbol{\varepsilon}}$ is the photon polarisation vector, $\textbf{B}_{\rm ext}$ is the magnetic field, $\textbf{v}_p$ is the phase velocity of the incoming axion and $E_\gamma$ is the energy of the photon, which is determined by the 3-momentum and the plasma mass. Finally, $U_E$ and $U$ give the electric energy and total electromagnetic energy in the plasma mode. For the Langmuir O modes corresponding to photons produced by axions, we have $U_E/U = 1/2$. Crucially, any suppression from photon refraction is already captured by the energy gradient in the denominator, which describes the forces on, and hence the refraction of the photon. The validity of this expression has been independently confirmed through numerical finite element method simulations of full axion electrodynamics \cite{Gines:2024ekm}. Its derivation is the subject of exercise 2.

Another method for probing axions via neutron stars relies on their production in the polar caps of neutron stars \cite{Prabhu:2020yif, Noordhuis:2022ljw}. In this setup axions are sourced via $\textbf{E}\cdot\textbf{B}$ according to the axion equation of motion: 
\begin{equation}
(\Box + m_a^2)\, a(x) \;=\; -\, g_{a\gamma\gamma}\, (\textbf{E}\cdot\textbf{B})(x)\, .
\end{equation}

Large parts of the magnetospheres of neutron stars, are in a so-called force-free state, meaning that the Lorentz force on particles vanishes, i.e, $\textbf{E} + \textbf{v}\times \textbf{B} =0$, which implies $\textbf{E}\cdot \textbf{B} = 0$. In order to maintain this so-called force-free condition, sufficient charge must be supplied, which defines a characteristic Goldreich-Julian \cite{Goldreich:1969sb} charge density $\rho_{\rm GJ} \propto \Omega B/e$, where $\Omega$ is the rotational angular velocity of the neutron star and $B$ the local magnetic field value. However, in some regions, insufficient charge can be supplied, leading to the formation of unscreened $\textbf{E}\cdot\textbf{B}$ contributions. These regions can occur in the polar caps of neutron stars, defined as those regions where the magnetic dipole axis intersects the neutron star surface. Furthermore, these regions undergo charge-discharge cycles, leading to cyclical growth and decay in the value of $\textbf{E}\cdot\textbf{B}$. This process is driven by the production of pair cascades which rapidly fill the cap regions with charges, screening the parallel component of the electric field. These charges then advect out of the plasma, allowing an unscreened electric field to grow once more. This process repeats, leading to radio emission from polar caps, thought to account for the observed radio spectrum from pulsars. 

These plasma fluctuations source copious amounts of axions via the equation above. These axions then convert into radio photos in the star magnetic field, leading to low frequency radio production. By using the observed radio flux from pulsars, one can place strong constraints on axions \cite{Noordhuis:2022ljw},  independently of whether they are dark matter, at frequencies below typical Haloscope experiments.

\section{WISP searches with White Dwarfs}
\label{sec:WDs}

White Dwarfs~(WDs) constitute the final evolutionary state of the vast majority of stars with masses smaller than $\mathcal{M}\lesssim8-9\,M_\odot$. Close to the end of their lives, such stars ascend the asymptotic giant branch~(AGB), where helium is burnt in the central core. The progressive depletion of helium leads to the development of an electron-degenerate core made up of carbon and oxygen. Because stars in this mass range never attain the temperatures required for carbon–oxygen~(CO) ignition, the CO core remains inert while the surrounding hydrogen- and helium-burning shells gradually cease their  activity. As mass loss intensifies during the late AGB phase, the stellar envelope is expelled thereby unveiling the electron-degenerate core. The remnant left behind is a WD, which represents the final evolutionary outcome of the progenitor star. 

The chemical composition of WDs is primarily determined by the core mass, which defines the stage at which nuclear fusion ceases. In particular, WDs with $M \lesssim 0.4$~M$_\odot$ are made of helium (He-WDs), while those with intermediate mass $0.4\,{\rm M}_\odot \lesssim M \lesssim 1.05$~M$_\odot$ are composed by carbon and oxygen (C/O-WDs). At higher masses $M \gtrsim 1.05$~M$_\odot$, temperatures may have become high enough to produce significant amounts of neon, giving rise to O/Ne-WDs. In the final configuration, electron degeneracy pressure counterbalances gravity, preventing further collapse of the stellar remnant. This mechanism sets an upper limit on the WD mass to the Chandrasekhar limit — approximately $1.44\,M_\odot$— beyond which electron degeneracy pressure cannot support it. The resulting compact object exhibits a small radius, $R\sim10^4\,$km and average densities $\rho\sim10^{5}-10^{7}\,$g cm$^{-3}$.

The remaining evolution of WDs is essentially a gravo-thermal process~\cite{Althaus:2010pi}. In particular, the WD cooling process can be divided into different phases. At early times $t\lesssim10^{4}\,$yrs, remnant activity of hydrogen burning via carbon-nitrogen-oxygen (CNO) cycle is still present, constituting the dominant contribution to WD luminosity. After this stage, nuclear reactions are exhausted and neutrino emission via plasmon decay becomes dominant~\cite{Winget:2003xf}. Due to the steep dependence on the core temperature, neutrino emission becomes strongly suppressed around $10^7-10^8$~yrs and surface cooling via photon losses subsequently dominates the WD evolution. When the temperature drops sufficiently, the weakly-interacting Coulomb plasma~\cite{1994ApJ...434..641S,2010CoPP...50...82P,2021ApJ...913...72J} in the inner core may undergo crystallization~\cite{Isern:1997na}.  The specific heat in WD interior then follows the Debye's law leading to a rapid cooling of the system. Thus, in this phase the thermal evolution is regulated energy content of outer layers, which prevents the sudden disappearance of the WD~\cite{1989ApJ...347..934D}.

During their evolution, WDs may experience departures from hydrostatical equilibrium, causing episodes of variability with a pulsation frequency spectrum which sensitively depend on the internal structure of the star~(see Ref.~\cite{2010A&ARv..18..471A,2019A&ARv..27....7C,2020FrASS...7...47C,2021RvMP...93a5001A} for some reviews about WD asteroseismology). Pulsation modes can be radial, if they preserve spherical symmetry, or non-radial, the latter further classified into spheroidal and toroidal modes. Spheroidal modes can be distinguished on the basis of the restoring force: for $g$- and $f$-modes, the restoring force is gravity, while $p$-modes are driven by pressure gradient. Typically, $g$-modes are characterized by low oscillation frequencies and horizontal displacements, while p-modes have higher frequencies and radial displacements. At least six different classes of pulsating WDs can be identified on the basis of the main element in their atmosphere, which determines their spectral type.  In the following we focus on DAVs, DBVs and DOVs, characterized by hydrogen-dominated~(DA type), helium-dominated~(DB type), and helium-carbon-oxygen mixture atmospheres~(DO type), respectively~\cite{2019A&ARv..27....7C}. These stars exhibit a multifrequency character and period of pulsation in the range $10^2-10^3$~s, classifying them as $g$-mode pulsators with buoyancy as main restoring force.

White Dwarfs are also expected to host powerful magnetic fields, with surface fields ranging within $10^{3}-10^{9}\,$G~\cite{Ferrario:2015oda}, with more massive stars exhibiting the strongest fields. Highly-magnetic WDs may exhibit a complex non-dipolar field structure, with some objects showing the presence of higher-order multipoles. Typically, WDs with strong magnetic fields are characterized by slow rotation periods ($\sim100$yrs) compared to their non-magnetic counterparts ($\sim$hours), suggesting that strong magnetic fields enhance the braking of the stellar core due to efficient losses in angular momentum~\cite{1989MNRAS.237...39H,10.1111/j.1365-8711.1998.t01-1-01913.x}. Several mechanisms have been proposed to account for the origin of magnetic fields in white dwarfs, including fossil fields inherited from the progenitor’s convective core, crystallization-driven dynamos, and dynamos operating during common-envelope phases in binary evolution~\cite{10.1111/j.1365-2966.2004.08603.x,Bagnulo_2022}. Nevertheless, the physical origin of the powerful magnetic fields observed in MWDs remains an open question.

The study of the cooling of WDs provides a valuable tool to constrain WISP properties. In particular, if WISPs are copiously produced in the interior of WDs, they may open an efficient energy-loss channel, thereby modifying the thermal evolution of the star. In this regard, it has been shown that WISP emission may dramatically alter the secular drift of the period of pulsation of variable WDs as well as the shape of the luminosity function (see, e.g., Refs.~\cite{Isern:2019nrg,Isern:2022vdx,Carenza:2024ehj} for some reviews).

The magnetosphere of MWDs may also serve as a powerful laboratory to study WISP phenomenology. Conversions between WISPs and photons taking place in the powerful magnetic domains of these systems may imprint observable signatures in the thermal photon fluxes they emit, possibly leading to X-ray excesses or modifications in the polarization state of the observed light.

In the following, we briefly review the main lines of WISP searches involving WDs, highlighting how these such systems offer many complementary approaches to probe the phenomenology of axions and other WISPs~\cite{Caputo:2024oqc,Carenza:2024ehj,Arza:2026rsl}.

\subsection{WISP signatures on the secular drift of the pulsation period}

Axions and other WISPs can be efficiently produced in compact, electron-degenerate stars such as WDs. The additional energy-loss channel related to their emission may alter several properties of these stars predicted on the basis of standard cooling theory. 
In particular, changes in the thermal and mechanical structure related to the cooling process may affect the star's oscillation period. Although accurate modelling has shown that the WD thermal profile is not significantly affected, the secular drift in the pulsation period $\dot{P}$ is highly sensitive to the introduction of exotic energy sinks due to WISP emission. The temporal variation of the pulsation period can be obtained as~\cite{1983Natur.303..781W}
\begin{equation}
\frac{\dot P}{P}= -a\frac{\dot T}{T}+ b\frac{\dot R}{R}\,,
\label{pdot}
\end{equation}
where $a$ and $b$ are positive constants of the order of unity. In this expression, the first term of the r.h.s. encodes the decrease of the Brunt-V\"ais\"al\"a frequency with the temperature, while the second one is the increase of the frequency induced by the residual gravitational contraction. 

Therefore, if WD standard modelling can be considered reliable enough, observations of the secular drift can be used to probe any physical mechanism leading to changes in the pulsation period of these stars. In this context, the introduction of extra-cooling mechanisms may alter the observed values of $\dot{P}$ as~\cite{Isern:1992gia}
\begin{equation}
\frac{{{L_{0}} + {L_x}}}{{{L_{0}}}} \approx \frac{{{{\dot P}_{\rm obs}}}}{{{{\dot P}_{0}}}}
\label{eqise92}
\end{equation}
where $\dot P_{\rm obs}$ is the observed period drift,  $L_0$ and $\dot P_0$ are the luminosity and period drift obtained from standard models, respectively, and $L_x$ is the extra luminosity contribution.

\begin{figure}[t]
\center
  \includegraphics[width=0.7\linewidth,clip=true,trim=4.5cm 9.3cm 2cm 10cm]{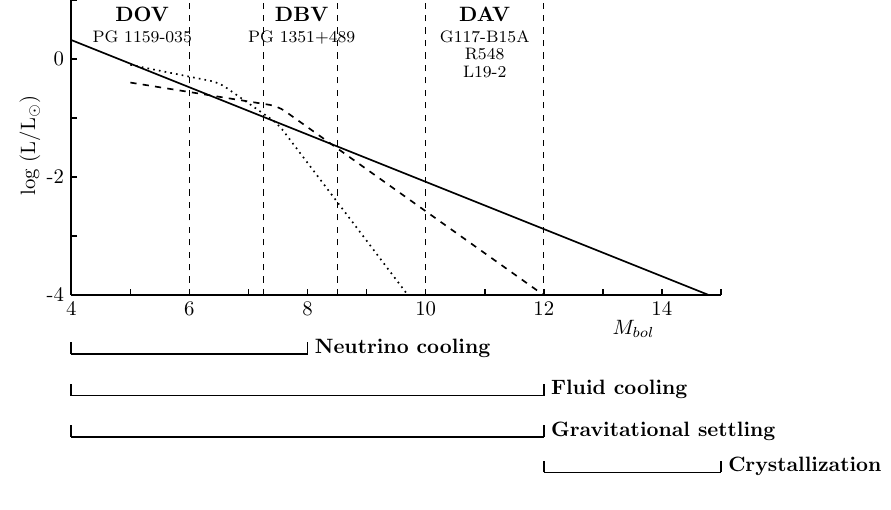}
\caption{Schematic representation of the WD cooling process including photon~(solid), neutrino~(dotted) and axion~(dashed) luminosities as well. Vertical dashed lines identify regions dominated by different kinds of energy sink. The figure also shows the position of the variable white dwarfs. Figure taken from Ref.~\cite{Carenza:2024ehj} with permission.}
\label{Fig: wdsketch}
\end{figure}

Fig.~\ref{Fig: wdsketch} shows that different classes of WD variables are characterized by different dominant cooling mechanisms. While in DOVs and DBVs neutrino emission is competitive with photon emission, the DAVs cooling process is largely dominated by photon emission since the neutrino bremsstrahlung emission is rapidly suppressed with the temperature.

DAVs pulsating WDs have been the first class to be discovered~\cite{1968ApJ...153..151L} and constitutes the most numerous group with more than 400 observed samples~\cite{2022MNRAS.511.1574R}. On the other hand, due to the smallness of the secular drift for this class ($\dot P\sim \mathcal{O}(10^{-15})$~s$^{-1}$), it has been measured only for three DAV stars (G117-B15A, R548, and L19-2). For this sample of stars, the observed values of $\dot P$~\cite{1991ApJ...378L..45K} are actually larger than the one predicted by standard evolutionary models~\cite{1991ASIC..336..153F}, providing indications that some extra-cooling channels might be present.

Axions were first proposed as the main candidate able to solve this puzzle~\cite{Isern:1992gia}. At typical temperatures and densities of DAVs --$T\sim1\,$keV and $\rho\sim10^6-10^7\,$g cm$^{-3}$~\cite{Corsico:2019nmr}-- light axions may be emitted by the degenerate electron plasma by means of electron bremsstrahlung $e + Ze \to e + Ze + a$~\cite{Nakagawa:1987pga,Nakagawa:1988rhp,Raffelt:1996wa,Carenza:2021osu}. Since the axion production rate scales as $\epsilon_a \propto T^4$, axion emissivity remains comparatively efficient in these late evolutionary stages, whereas neutrino emission --scaling as $\epsilon_\nu \propto T^7$~\cite{1983ApJ...275..858I}-- is already strongly suppressed. Therefore, under typical DAV conditions, axions constitute a potentially significant additional energy sink. 
By using Eq.~(\ref{eqise92}) and a simplistic WD model, Ref.~\cite{Isern:1992gia} suggested that the emission DFSZ axions with a mass $m_a\cos^2 \beta \approx 8.5 $~meV ($ g_{ae} \approx 2.36\times 10^{-13}$) could explain the drift in the period observed for G117-B15A. Nevertheless, both measured values~\cite{1991ApJ...378L..45K,2000ApJ...534L.185K,2005ApJ...634.1311K,2012ASPC..462..322K,2021ApJ...906....7K} and the ones predicted by theoretical models~\cite{2001NewA....6..197C,2008ApJ...675.1512B,2012MNRAS.424.2792C,2012MNRAS.420.1462R} have evolved with time. The most up-to-date measurements led to ${\dot P=(5.12 \pm 0.82)\times 10^{-15}}$~s$^{-1}$~\cite{2021ApJ...906....7K} while the most recent evolutionary models predict ${(1.25 \pm 0.09)\times 10^{-15}}$~s$^{-1}$ \cite{2021ApJ...906....7K}. This discrepancy points towards a best fit value for the axion-electron coupling $g_{ae}=(5.66 \pm 0.57)\times 10^{-13}$ ($m_a \cos^2 \beta =20 \pm 2 $~meV if DFSZ type)~\cite{2021ApJ...906....7K}. Similar results~\cite{2008ApJ...675.1512B,2018phos.confE..28B,2012JCAP...12..010C,2016JCAP...07..036C} have been derived from observations of the secular drift of the drift of R548~\cite{2013ApJ...771...17M} and L19-2~\cite{2015ASPC..493..199S}. A similar analysis was performed by employing observations of the pulsation modes of the DBV star PG 1351 + 489, in which discrepancies between measured and theoretical values of the drift are compatible with axion-electron couplings $g_{ae}\lesssim 7 \times 10^{-13}$ ($m_a \cos^2 \beta =20 \pm 2 $~meV for DFSZ axions)~\cite{2016JCAP...08..062B} in agreement with the values obtained with the DAVs.

\subsection{WISP signatures on the luminosity function}

The copious emission of WISPs from the interior of WD may also seriously affect the WD luminosity function~(WDLF), i.e. the number of WDs of a given absolute magnitude (or luminosity) per unit magnitude (luminosity) interval. Under the assumption that WDs are not destroyed and that the ensemble is closed, the number $N$ of WDs with absolute magnitude within the interval $M_{\rm abs} \pm 0.5\,\Delta M_{\rm abs}$ reads~\cite{Isern:2022vdx,Carenza:2024ehj}

\begin{equation}
N\left( M_{\rm abs}\right) = \int\limits_{{M_l}}^{{M_u}} {\Phi \left( M \right)\Psi \left( {{T_G} - {t_{cool}} - {t_{\rm PS}}} \right){\tau _{\rm cool}}\,dM}, 
\label{eq:lf}
\end{equation}
where $t_{\rm cool}$ is the cooling time necessary to reach the magnitude $M_{abs}$, $\tau_{\rm cool} = dt/dM_{\rm abs}$ is the characteristic cooling time of the WD at this magnitude, $t_{\rm PS}$ is the lifetime of the WD progenitor star and $T_G$ is the age of the Galaxy or the population under study. The integral runs over the possible masses of the WD progenitors $M$, with integration boundaries $M_l$ and $M_u$ set by the maximum and minimum mass of Main Sequence~(MS) stars able to produce a WD, respectively. Therefore, $M_l$ fulfils the condition $T_G=t_{\rm cool}(M_{\rm abs},M_l)+t_{\rm PS}(M_l)$. Moreover, this expression accounts for the initial mass function~(IMF) $\Phi(M)$ and the star formation rate~(SFR) $\Psi(t)$ of the population considered. Finally, in Eq.~(\ref{eq:lf}) there is no initial-final-mass-relation (IFMR) function which maps the progenitor's stellar properties onto those of the resulting WD. Since the total density of WD is not yet well known, to compare theoretical predictions to observations it is customary to normalize the computed luminosity function to a bin with a small error bar, usually $\log L/L_\odot \simeq 3$ or the corresponding magnitude. We highlight that Eq.~(\ref{eq:lf}) contains three sets of terms: the observational ones, $N(M_{abs})$, the stellar ones, $t_{\rm cool}, \tau_{\rm cool}, t_{\rm PS}, M_{u}, M_{l}$ plus the IMF, and the galactic terms $\Phi$ and $\Psi$.

\begin{figure}[h]
\center
  \includegraphics[width=1.\linewidth]{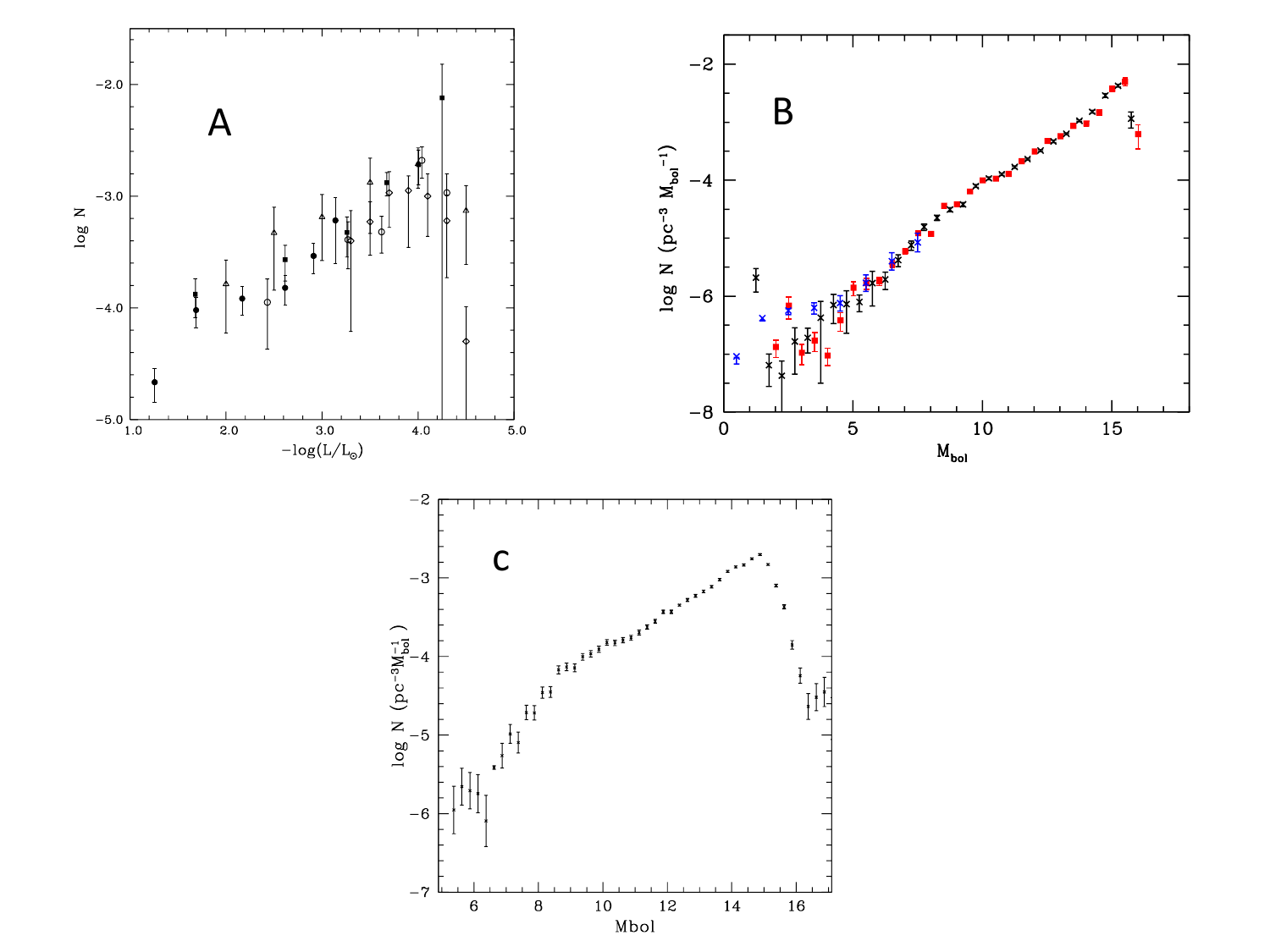}
\caption{{\it Panel A:} Early luminosity functions based on data from Ref.~\cite{1988ApJ...332..891L} (full circles), Ref.~\cite{1992MNRAS.255..521E} (full squares), Ref.~\cite{1996Natur.382..692O} (open triangles), Ref.~\cite{1998ApJ...497..294L} (open diamonds) and Ref.~\cite{1999MNRAS.306..736K} (open circles). {\it Panel B:} WDLFs obtained with the SCSS catalogue~\cite{2011MNRAS.417...93R} (black crosses), the SDSS catalogue~\cite{2006AJ....131..571H} (red squares), and from UV-excesses~\cite{2009A&A...508..339K} (blue crosses) normalized at $M_{\rm bol}\approx12$. {\it Panel C:}  Luminosity function of WDs located within a 100 pc horizon based on the Gaia Early Data Release 3~\cite{2021A&A...649A...6G}. Figure taken from Ref.~\cite{Carenza:2024ehj} with permission.}
\label{Fig: wdlf}
\end{figure}

The first attempt to determine the WDLF dates back to Ref.~\cite{Weidemann:1968uw}, before the advent of large cosmological surveys, with a sample of observed WDs comprised by few hundreds of stars~\cite{1988ApJ...332..891L,1992MNRAS.255..521E,1996Natur.382..692O,1998ApJ...497..294L,1999MNRAS.306..736K}. These early luminosity functions already showed a monotonic growth in the number of stars with their associated bolometric magnitude, followed by a sharp cut-off attributed to the finite age of the Galaxy~\cite{1987ApJ...315L..77W}. Nonetheless, the substantial uncertainties in the measurements and the large dispersion of bins prevented the determination of the slope of the monotonic rise. Large cosmological surveys, such as the Sloan Digital Sky Survey (SDSS)~\cite{2006AJ....131..571H} and the Super COSMOS Sky Survey (SCSS)~\cite{2011MNRAS.417...93R}, enlarged to sample to several thousand WDs, allowing the reconstruction of the WDLF with precision high enough to discriminate a change in the slope due to the transition from the neutrino-dominated to the photon dominated cooling epoch.
Recently, the latest data release of Gaia mission permitted the determination of a statistically complete WDLF with a sample of WDs within 100 pc~\cite{Isern:2022vdx}~(see Fig.~\ref{Fig: wdlf}).

The behaviour of the WDLF is sensibly dependent on the characteristic cooling time. Thus, it may be employed to test the presence of exotic source of energy sink, which may speed up to WD cooling process. Since the bright branch of the luminosity function ($M_{\rm bol}\lesssim13$ or, equivalently $\log{L/L_\odot}\gtrsim-3.5$) is dominated by WDs produced by low-mass MS stars, Eq.~(\ref{eq:lf}) reads

\begin{equation}
{N \propto \left\langle {{\tau _{\rm cool}}} \right\rangle \int\limits_{{M_i}}^{{M_s}} {\Phi \left( m \right)\Psi \left( T_G-t_{\rm PS}-t_{\rm cool}\right)} } \,dm\, .
\label{eq:wdlf2}
\end{equation}

Since bright WDs have had no time to cool down, i.e. $t_{\rm cool}$ is small, and the MS star lifetimes is strongly dependent on their mass, the minimum mass $M_i$ able to produce WD contributing to this region of the WDLF is almost constant and independent on the luminosities under consideration. Therefore, the slope is essentially determined by the averaged characteristic cooling time $\langle t_{\rm cool}\rangle$ and only weakly on the star formation history. Notice that this is not true if recent star formation burst ($t_{\rm burst}\lesssim2\,{\rm Gyr}$) is included in the SFR, since low-mass MS stars have not had sufficient time to become WDs and thus $M_i$ becomes dependent on the luminosity bin considered.

The slope of the bright branch of the WDLF can be employed point out the presence of exotic sources of energy loss in the WD cooling process. This technique has been employed to probe axions for the first time in Ref.~\cite{2006AJ....131..571H} using the preliminary form of the luminosity function displayed in panel B of Fig.~\ref{Fig: wdlf}, suggesting that DFSZ axions with $g_{ae}=(1.4)^{+0.9}_{-0.8}\times10^{-13}$ could improve the agreement between. These results have been re-examined in Ref.~\cite{2014JCAP...10..069M} by employing a self-consistent treatment for neutrino cooling, confirming the hint for axion existence of Ref.~\cite{2006AJ....131..571H} and concluding that value $g_{ae}\gtrsim2.8\times10^{-13}$ could be excluded.

Nevertheless, the degeneracy between stellar and galactic terms in Eq.~\eqref{eq:wdlf2} suggests that the shape in the WDLF may equivalently arise from either axion emission or changes in the SFR like those introduced by recent star-formation bursts. This degeneracy can be broken by observing the WDLF of different stellar populations. Indeed, if axions contribute to the WD cooling process, their impact would be observable in all the independent luminosity functions at roughly the same luminosity bin. Following this argument, Ref.~\cite{Isern:2018uce} showed axions with $g_{ae}\approx2.24\times10^{-13}$~($g_{ae}\approx4.48\times10^{-13}$) may help to improve the discrepancies between theoretical modelling and observations, by assuming a thin (thick) disk halo model~\cite{2011MNRAS.417...93R}. Moreover, Ref.~\cite{2017AJ....153...10M} and Ref.~\cite{2017ApJ...837..162K} improved the determination of the disk luminosity function, obtaining that the value $g_{ae}\approx2.24\times10^{-13}$ is favoured when assuming both constant and variable heights above the Galactic plane, respectively.

Another strategy to break the degeneracy between stellar and galactic terms is suggested by observations of massive WDs. For this class, the progenitor lifetime is short compared to the characteristic cooling time. Thus, their luminosity function closely follows the temporal variation of the SFR, allowing one to directly obtain the SFR $\psi$~\cite{Isern:2019nrg}. Recently, the luminosity function for WDs within distances $d\sim100\,$pc and masses in the range $0.9-1.1$~M$_\odot$ obtained by the Gaia data release 2~\cite{2019Natur.565..202T} has been employed to rule out axions with $g_{ae}\gtrsim4.5\gtrsim10^{-13}$, while values around $g_{ae}\sim2.24\times10^{-13}$~\cite{Isern:2019nrg} are compatible with observations if a time dependent height scale is assumed~\cite{Cukanovaite_2023}.

We underline that axion hints from both the secular drift of the pulsation period and the WDLFs are currently in tension with constraints from the red giant branch tip~\cite{Capozzi:2020cbu,Straniero:2020iyi}. Furthermore, previous constraints from WD cooling have been recently improved in Ref.~\cite{Fleury:2025ahw}, which excluded $g_{ae}\gtrsim0.81\times10^{-13}$ by comparing outcomes from up-to-date WD simulations including axion emission to observations of the WD population in the globular cluster 47 Tucanae, ruling out values of the axion-electron coupling favoured by axion hints.

The same argument concerning the WDLF has been also employed to set limits on the coupling between exotic scalars $\phi$ particles and SM fermions $\psi$, described by the Lagrangian term $\mathcal{L} \supset g_{\psi} \phi \psi \bar{\psi}$. Also in this case, the dominant emission process is constituted by electron bremsstrahlung over protons $e + p \to e + p + \phi$. Nonetheless, since the related energy emission rate shows a milder dependence on the temperature $\epsilon_\phi \propto T^2$, scalars are able to affect the cooling process of colder, and thus older, WDs. In this case, constraints are derived by considering the WDLF only in the range of magnitudes dominated by surface photon emission. Following this rationale, Ref.~\cite{Bottaro:2023gep} excluded $g_{\phi e}\gtrsim 3.9\times 10^{-16}$ and $g_{\phi p}\gtrsim 6.5\times 10^{-13}$ for leptophilic and baryophilic scalars, respectively. Similar results were obtained by \cite{Yamamoto:2023zlu} by assuming a simplified WD one-zone model and requiring the expected scalar luminosity to be smaller than the stellar one.

\subsection{WISP signatures in photon spectra}
The strong magnetic fields hosted by MWDs are powerful testbeds to probe WISPs that couple to photons. In this context, light axions constitutes good candidate to be investigated in such astrophysical environments.
The thin atmosphere of WDs is typically opaque to radiation and X-ray photons cannot escape the core of WDs, which can reach temperatures as high as $T\sim1$ keV. Therefore, the only expected emission comes from optical photons radiated by the stellar surface, where temperatures are much lower $T\sim1\,$eV.
On the other hand, the stellar medium is essentially transparent to weakly coupled axions produced in the core via thermal bremsstrahlung of electrons off ions. As a result, axions with typical energies of $E\sim5\,$keV can freely stream out of the stellar interior. Upon reaching the MWD magnetosphere, where magnetic field strengths may be as large as $B\sim10^{9}\,$G, axion–photon conversion can take place. For typical conditions expected in the MWD magnetosphere, the conversion process is efficient for axion masses $m_a\lesssim10^{-6}\,$eV, potentially leading to the production of X-ray photons. Consequently, the possible observation of an anomalous X-ray flux may serve as a powerful probe such light axions. 

This scenario has been studied for the first time in Ref.~\cite{Dessert:2019sgw}, which analysed observations of the magnetic WD \mbox{RE J0317-853}~\cite{1995MNRAS.277..971B}. This star is expected to be a perfect laboratory to conduct axion searches, as it is relatively nearby ($d=29.38\pm0.02$~\cite{Gaia:2021gsq}), is expected to host a strong magnetic field ($B_{\rm pole} \sim 500$ MG~\cite{Burleigh:1998pqa}) and is particularly hot ($T_{\rm eff} = 4.93_{-0.12}^{+0.22} \times10^4$~K~\cite{1995MNRAS.277..971B, 2010A&A...524A..36K}, leading to a core temperature $T\sim1.5\,$~keV). Observations of this WD with the \textit{Suzaku} telescope did not reveal any astrophysical X-ray emission~\cite{doi:10.1093/pasj/65.4.73} in the energy range $2-10\,$keV, thus excluding values of the product between the axion-photon and axion-electron couplings $g_{a\gamma}g_{ae}\gtrsim1.7\times10^{-24}\,$GeV${-1}$ for axion masses $m_{a}\lesssim10^{-5}\,$eV~\cite{Dessert:2019sgw}. These limits were further improved by using {\it Chandra} observations of the same MWD. The better sensitivity of the instrument employed allowed Ref.~\cite{Dessert:2021bkv} to strengthen the constraints down to $g_{a\gamma}g_{ae}\gtrsim1.3\times10^{-25}\,$GeV$^{-1}$ for axion masses $m_{a}\lesssim10^{-5}\,$eV.

The axion-photon coupling can also be severely constrained through measurements of the linear polarization degree of the thermal radiation emitted by MWD. This idea, originally proposed in Refs.~\cite{Lai:2006af,Gill:2011yp}, relies on the standard picture in which MWDs are not intrinsic sources of linearly polarized optical photon, which are thermally emitted from the stellar surface at typical energies $E\sim$eV. Nevertheless, photons polarized along the direction of the magnetic field may undergo conversion into axions within the strong magnetized environment surrounding the star. Therefore, depending on the efficiency of axion-photon conversions --directly governed by the axion-photon coupling strength-- initially unpolarized stellar radiation may develop a net linear polarization oriented along the direction perpendicular to the magnetic field. 
For this purpose, Ref.~\cite{Gill:2011yp} analysed observations of two MWDs, PG 1031+234 and SDSS J234 605+385337, which were reported to host magnetic fields as high as $B\sim10^{9}\,$~G~\cite{Jordan2009}. Since both objects are expected to exhibit low linear polarization at the level of $\sim1\%$~\cite{1992A&A...259..143P,Vanlandingham_2005}, values of the axion-photon coupling $g_{a\gamma}\gtrsim(5-9)\times10^{-13}\,$GeV$^{-1}$ were ruled out for axion masses  $m_a\lesssim$ few $\times10^{-7}\,$eV. These results were reassessed in Ref.~\cite{Dessert:2022yqq}, where it was shown that upper limits from these MWDs can be relaxed by approximately one order of magnitude once astrophysical uncertainties on the strength of the magnetic fields and their geometry are properly accounted for. In the same work, two other promising MWD candidates, SDSS J135141.13+541947.4 and Grw {70}$^{\circ}$8247 were identified. The absence of linear polarization in measurements of polarized light from these objects yield stringent constraints down to $g_{a\gamma}\lesssim5.4\times10^{-12}\,$GeV$^{-1}$ at 95\% confidence for axion masses $m_a\lesssim3\times10^{-7}\,$eV. Recently, limits on the axion-photon coupling have been further improved using dedicated spectropolarimetric data collected by the Lick and Keck observatories, targeting a sample of five different MWDs candidates. The dominant constraints were derived from Lick observations of SDSS J033320+000720, which excluded $g_{a\gamma}\gtrsim1.7\times10^{-12}\,$GeV$^{-1}$ at the 95\% confidence level for masses $m_a\lesssim2\times10^{-7}\,$eV. This result currently provides the leading bound on the axion-photon coupling in the range masses $10^{-10}\,$eV$\lesssim m_a\lesssim2\times10^{-7}\,$eV.

\section{HFGWs and Compact Objects}

High frequency gravitational waves are a rapidly evolving area of interest - see Ref.~\cite{Aggarwal:2025noe} for the latest community report. This focus is driven by two main factors. The first is that in the last decade, gravitational wave astronomy has shifted from being an aspiration to a practical reality: the measurement of gravitational waves by LIGO \cite{LIGOScientific:2016aoc} in 2015 heralded the era of direct detection. Similarly, pulsar timing \cite{NANOGrav:2020bcs,Goncharov:2021oub,EPTA:2021crs,Tarafdar:2022toa} has recently provided indirect evidence for stochastic backgrounds of gravitational waves in the nHz band. These discoveries therefore motivate the full exploration of the gravitational waves spectrum. Second, much of the emerging technologies used in tabletop detectors for, e.g., axion searches are readily adaptable to searches for gravitational waves above 10kHz. These have come to be known as high-frequency gravitational waves (HFGWs), which remain comparatively underexplored. Transferable experimental approaches include microwave and radio cavities \cite{Berlin:2021txa,Fischer:2024msc,Berlin:2023grv}, lumped element detectors \cite{Domcke:2022rgu,Pappas:2025zld} and magnets acting as Weber bars \cite{Domcke:2024mfu}, as well as a broader landscape of emerging quantum technologies \cite{Aggarwal:2025noe}. 

In this section we briefly discuss \textit{indirect} constraints on high-frequency gravitational waves (HFGWs) arising from astrophysical environments. Our discussion follows that of Ref.~\cite{Aggarwal:2025noe} very closely. Many astrophysical and cosmological probes rely on the inverse Gertsenshtein effect, in which gravitational waves propagating through magnetic fields convert into photons \cite{Raffelt:1987im,Domcke:2025qlw}. 

As we have seen in previous sections, compact objects provide powerful probes of axions. The idea of using neutron stars as detectors of HFGWs, however, has only recently begun to attract attention. In particular, Ref.~\cite{Ito:2023fcr} derived tentative limits on stochastic gravitational waves using radio observations in the frequency range $0.1$--$1\,\mathrm{GHz}$, as well as much higher frequencies between $10^{13}$ and $10^{27}\,\mathrm{Hz}$ spanning the infrared to X-ray bands. Their analysis suggests strain sensitivities of approximately $h_c \lesssim 10^{-14}$--$10^{-18}$ in the radio band and $h_c \lesssim 10^{-16}$--$10^{-26}$ at higher frequencies, assuming non-resonant graviton--photon conversion. More recently, Ref.~\cite{McDonald:2024nxj} examined how resonant conversion could strengthen these constraints and constraints on gravitational waves have also been explored using populations of neutron stars rather than individual objects \citep{Dandoy:2024oqg}. 

Graviton--photon conversion in cosmological magnetic fields has also been investigated as a potential probe of high-frequency gravitational waves \citep{Pshirkov:2009sf, Chen:1994ch, Dolgov:2012be, Cillis:1996qy, Domcke:2020yzq}. Although cosmological magnetic fields are comparatively weak, they can remain coherent over scales of kiloparsecs to megaparsecs, effectively providing a very large detection volume. A particularly interesting frequency window lies between $100\,\mathrm{MHz}$ and $30\,\mathrm{GHz}$, corresponding to the Rayleigh--Jeans tail of the cosmic microwave background and targeted by several existing and upcoming radio experiments. Observations from ARCADE~2 \citep{Fixsen_2011} and EDGES \citep{Bowman:2018yin}, for example, can be reinterpreted as limits of approximately $h_{c,\rm sto} < 10^{-24}\,(10^{-14})$ in the range
$3\,\mathrm{GHz} \lesssim f \lesssim 30\,\mathrm{GHz}$ (ARCADE~2) and
$h_{c,\rm sto}(f \approx 78\,\mathrm{MHz}) < 10^{-12}\,(10^{-21})$ (EDGES),
depending on whether the strongest or weakest cosmological magnetic fields consistent with current data are assumed \citep{Domcke:2020yzq}. The dominant uncertainty in these limits arises from the poorly constrained power spectrum of cosmological magnetic fields in the early Universe, leading to an uncertainty of roughly ten orders of magnitude in the resulting constraints on $h_c$. Improved modelling of primordial magnetic fields will therefore be essential to refine these bounds.
\\\\
Finally, galactic and planetary magnetic fields have also been proposed as environments in which stochastic gravitational-wave backgrounds might be probed \citep{Ito:2023nkq, Lella:2024dus, Liu:2023mll}. Comparable sensitivities in the $100\,\mathrm{TeV}$--$\mathrm{PeV}$ frequency range have also been investigated using LHAASO observations searching for gravitational-wave conversion in the Milky Way \citep{Ramazanov:2023nxz}. Prospects for future radio telescopes and CMB spectral-distortion experiments are discussed in \cite{He:2023xoh}, although these projections typically rely on optimistic assumptions regarding magnetic field strengths and instrumental sensitivities.

\label{sec:HFGWs}

\section{Hands-on tutorials}\label{sec:tutorials}

\subsection*{Exercise 1: Superradiant Scattering off a Rotating Cylinder}
\label{sec:tutorial_ex1}

\begin{tcolorbox}[colback=pos!5!white, colframe=pos!75!black, title=Exercise 1]

Consider the problem discussed in the lectures, which describes a rotating conducting cylinder, for which the governing equation can be written as 

    \begin{equation}
        \square \phi - \sigma (\Omega \partial_\varphi - \partial_t)\phi = 0,
    \end{equation}
    
    where $\sigma $ is constant for cylindrical radial distances $r < R$ and zero for $r > R$.\\\\
    (a) Assume an infinitely tall cylinder, such that you can choose a $z$-independent separable solution of the form $\phi = \phi(t,r,\varphi)  =\psi(r) e^{i m \varphi} e^{i \omega t} $, where $m$ is an integer. Obtain a radial equation for $\psi(r)$.
    \newline
    
    (b) Solve this radial equation to obtain an expression for $\psi(r)$ in terms of known functions. You will need to impose sensible boundary conditions, e.g., regularity at the origin, continuity across $r=R$, etc. 
    \newline
    
    (c) By using the asymptotic forms of these known functions as $r \rightarrow \infty$, express the exterior solutions at $r \rightarrow \infty$ as a sum of outgoing and ingoing waves, with amplitudes $A_{\rm out}$ and $A_{\rm in}$ respectively.
    [\textit{hint: it may be useful to express the asymptotic solutions in terms of Hankel functions}]
    \newline
    
    (d) Finally, derive an expression for $Z = 1 - |A_{\rm out}|^2/|A_{\rm in}|^2$ in the small $\omega R \ll 1$ limit to linear order in $\sigma$.

\end{tcolorbox}

\smallskip

\paragraph{Solution.} 

To solve the proposed exercise, let's consider the EOM for the scalar field $\phi$
\begin{equation}
\Box \phi - \sigma \left( \Omega \partial_\varphi - \partial_t^2 \right)\phi = 0 ,
\end{equation}
where
\begin{equation}
\sigma =
\begin{cases}
\text{const} & r < R, \\
0 & r > R.
\end{cases}
\end{equation}
\vspace{0.2cm}
{\bf\textit{a) Derivation of the radial equation.}}\\
Since the cylinder is infinitely tall, we can assume the solutions to be $z$-independent.
Assuming the separability of variables, let us look for solutions in the following form:
\begin{equation}
\phi(t,r,\varphi) = \psi(r) \, e^{i m \varphi} e^{i \omega t}.
\end{equation}
Let us plug the expression of $\phi$ into the EOM. Then we get
\begin{equation}
\nabla^2 \left[ \psi(r) e^{i m \varphi} e^{i \omega t} \right]
+ \omega^2\,\psi(r) e^{i m \varphi} e^{i \omega t}
+ \sigma \left( \Omega^2 - \omega^2 \right)\,
\psi(r) e^{i m \varphi} e^{i \omega t}= 0.
\label{Eq:EOM}
\end{equation}
Neglecting the dependence on $z$, the Laplacian in cylindrical coordinates reads
\begin{equation}
\begin{split}
\nabla^2&=\frac{1}{r} \frac{\partial}{\partial r}
\left(r \frac{\partial}{\partial r}
\right)+\frac{1}{r^2}\frac{\partial^2}{\partial \varphi^2}\\
&=\frac{\partial^2}{\partial r^2}+\frac{1}{r}\frac{\partial}{\partial r}+\frac{1}{r^2}\frac{\partial^2}{\partial \varphi^2}.
\end{split}
\end{equation}
Therefore, by substituting in Eq.~\eqref{Eq:EOM}, we get
\begin{equation}
    \psi''+\frac{1}{r}\,\psi'+\left[\omega^2-\frac{m^2}{r^2}-i\sigma\,(m\Omega-\omega) \right]\psi=0\,,
\label{Eq: RadialEq}
\end{equation}
and most general linearly-independent solutions for the radial equation are Bessel functions of the first and the second kind. \vspace{.7cm}\\
{\bf\textit{b--c) Solutions inside and outside the cylinder}}

\paragraph{$r<R.$}
Inside the cylinder the radial equation reads
\begin{equation}
\psi'' + \frac{1}{r}\psi' 
+ \left( k^2 - \frac{m^2}{r^2} \right)\psi = 0\,\qquad r<R\,\,,
\label{Eq: RadialEqIn}
\end{equation}
where we have defined
\begin{equation}
k^2 = \omega^2 - i\sigma (m\Omega - \omega).
\end{equation}
The most general solution of Eq.~\eqref{Eq: RadialEqIn} is a linear combination of Bessel functions of first and second kind, which we denote here as $J_m$ and $Y_m$, respectively:
\begin{equation}
\psi(r) = A J_m(kr) + B Y_m(kr).
\end{equation}
Imposing regularity at the origin $r=0$, we note that the asymptotic behaviour of the Bessel functions in the small-argument limit is given by
\begin{equation}
J_m(z) \xrightarrow[z\to 0]{} 0,
\qquad
Y_m(z) \xrightarrow[z\to 0]{} \infty\,.
\end{equation}
Therefore, only Bessel functions of the first kind can be taken as well-behaved solutions in the region $r<R$. 
\begin{equation}
\psi(r) = A J_m(kr), \qquad r < R.
\end{equation}
\smallskip

\paragraph{$r > R$.}

Outside of the cylinder ($\sigma=0$), we need to solve the equation
\begin{equation}
\psi'' + \frac{1}{r}\psi' 
+ \left( \omega^2 - \frac{m^2}{r^2} \right)\psi = 0 .
\end{equation}
Since this region does not contain the origin, both Bessel functions are admissible:
\begin{equation}
\psi(r) = c_1 J_m(\omega r) + c_2 Y_m(\omega r).
\end{equation}
We note that for $r\to+\infty$,
\begin{equation}
J_m(x) \sim \cos x,
\qquad
Y_m(x) \sim \sin x.
\end{equation}
Imposing plane-wave behaviour at large distances from the cylinder $\psi(r)\sim e^{\pm i\omega r}$, we adopt the Hankel functions of the first and second kind as a basis of linearly independent solutions
\begin{equation}
    \begin{split}
        H_m^{(1)}(\omega r) &= J_m(\omega r) + i Y_m(\omega r) \\
        H_m^{(2)}(\omega r) &= J_m(\omega r) - i Y_m(\omega r)
    \end{split}
\end{equation}
Thus, in the exterior of the cylinder the solution of Eq.~\eqref{Eq: RadialEq} reads
\begin{equation}
\psi(r) = A_{\text{in}}\,H_m^{(1)}(\omega r)
+ A_{\text{out}}\,H_m^{(2)}(\omega r),
\qquad r > R\,,
\end{equation}
where $A_{\text{in}}$ and $A_{\text{out}}$ encode the amplitudes of the incoming and outgoing plane waves, respectively.

\paragraph{Matching at $r=R$.}

The regularity of the solution across the surface of the rotating cylinder at $r=R$ requires

\begin{equation}
    \begin{cases}
    CJ_m(kR)=A_{\rm in}H^{(1)}_m(\omega R)+A_{\rm out}H^{(2)}_m(\omega R)\\
    Ck\,J_m'(kR)=\omega\,\left[A_{\rm in}{H^{(1)}_m}'(\omega R)+A_{\rm out}{H^{(2)}_m}'(\omega R)\right]\,,
    \end{cases}
\end{equation}
which gives
\begin{equation}
    \begin{split}
        &C=\frac{A_{\rm in}H^{(1)}_m(\omega R)+A_{\rm out}H^{(2)}_m(\omega R)}{J_m(kR)}\,,\\
        &\frac{A_{\rm out}}{A_{\rm in}}=-\frac{k\frac{J'_m(kR)}{J_m(kR)}H^{(2)}_m(\omega R)-\omega{H^{(2)}_m}'(\omega R)}{k\frac{J'_m(kR)}{J_m(kR)}H^{(1)}_m(\omega R)-\omega{H^{(1)}_m}'(\omega R)}\,.
    \end{split}
\end{equation} \vspace{.3cm}\\
{\bf\textit{d) Amplification factor in the $\omega R\ll1$ limit.}}\\
Let us expand the previous expressions in the limit $\omega R\ll1$. In the small $\sigma$ limit, this also implies that $kR \ll1$. Then, at lowest order in $k R$, Bessel functions of the first kind can be expanded as
\begin{equation}
    J_m(kR)=(kR)^m\left[ \frac{2^{-m}}{\Gamma(1+m)}-\frac{2^{-2-m}}{(1+m)\Gamma(1+m)}+\ldots\right]\,,    
\end{equation}
where $\Gamma$ is the Euler gamma function. Therefore, the ratio $k{J'_m(kR)}/{J_m(kR)}$ in the small $kR$ expansion reads
\begin{equation}
    \begin{split}
    k\frac{J'_m(kR)}{J_m(kR)}&=k\left(\frac{m}{kR}-\frac{kR}{2(1+m)}\right)\\
    &=\frac{m}{R}+\Delta F\,,
    \end{split}
\end{equation}
where $\Delta F=-{Rk^2}/{2 (m+1)}$ is at the first non vanishing term in the $kR$ expansion. It is now convenient to define
\begin{equation}
    \begin{cases}
    D_1 = \left( \frac{m}{R} + \Delta F \right) H_1- \omega H_1'\,, \\
    D_2 = \left( \frac{m}{R} + \Delta F \right) H_2- \omega H_2'\,,
    \end{cases}
\end{equation}
where we used the shortcut notation
\begin{equation}
H_{1,2} = H_m^{(1,2)}(\omega R),
\end{equation}
and we have defined
\begin{equation}
D_{1,2}^{(0)} = \frac{m}{R} H_{1,2} - \omega H_{1,2}'\,.
\end{equation}
With these definitions
\begin{equation}
\frac{A_{\text{out}}}{A_{\text{in}}}
=
-\frac{D_2^{(0)} + \Delta F H_2'}
{D_1^{(0)} + \Delta F H_1'}.
\end{equation}
Expanding to first order in $\Delta F$,
\begin{equation}
\frac{A_{\text{out}}}{A_{\text{in}}}
=
\frac{D_2^{(0)}}{D_1^{(0)}}
\left[
1 + \Delta F
\left(
\frac{H_2'}{D_2^{(0)}}
-
\frac{H_1'}{D_1^{(0)}}
\right)
\right].
\end{equation}
Moreover, at lowest order in $\omega R$,
\begin{equation}
\frac{D_2^{(0)}}{D_1^{(0)}} = -1\,,
\end{equation}
and the ratio between the amplitude of the outgoing and incoming waves reads
\begin{equation}
\frac{A_{\text{out}}}{A_{\text{in}}}
=
1 + \Delta F
\left(
\frac{H_2'}{D_2^{(0)}}
-
\frac{H_1'}{D_1^{(0)}}
\right).
\end{equation}
Ultimately, the amplification factor is given by
\begin{equation}
Z = 1 - \left| \frac{A_{\text{out}}}{A_{\text{in}}} \right|^2\,.
\end{equation}
Thus, at first order in $\Delta F$ 
\begin{equation}
Z\approx-2\,\mathrm{Re}\left[\Delta F\left(\frac{H_2'}{D_2^{(0)}}-\frac{H_1'}{D_1^{(0)}}\right)\right]\,.
\end{equation}
Expanding the ratio in the parenthesis at lowest order in $\omega R$, we have
\begin{equation}
\frac{H_2(\omega R)}{D_2^{(0)}(\omega R)}-
\frac{H_1(\omega R)}{D_1^{(0)}(\omega R)}
\sim-\frac{i \pi (\omega R)^{2m}}{\omega^{2m+1} \Gamma^2(1+m)}\,.
\end{equation}
The final expression for the amplification factor is then provided by
\begin{equation}
    \begin{split}
        Z&=-2i\,\Im\left[\Delta F\right]\left(\frac{H_2(\omega R)}{D_2^{(0)}(\omega R)}-\frac{H_1(\omega R)}{D_1^{(0)}(\omega R)}\right)\\
        &=-\frac{m\Omega-\omega}{\omega}R\sigma\,(\omega R)^{2m+1}\,\frac{m+1}{4^m\,\Gamma(m+2)^2}
    \end{split}
\end{equation}
\clearpage

\subsection*{Exercise 2: Resonant Axion-Photon Conversion in 3D Astrophysical Plasmas}
\label{sec:tutorial_ex2}

\begin{tcolorbox}[colback=pos!5!white, colframe=pos!75!black, title=Exercise 2]

Consider the Boltzmann-like equation: 
\begin{equation}\label{eq:BoltzmannExplicit}
   \partial_k \mathcal{H} \cdot \partial_x f_\gamma	- \partial_x \mathcal{H}  \cdot \partial_k f_\gamma  =   g_{a \gamma \gamma}^2 \big| k \cdot F_{\rm ext} \cdot \varepsilon \big|^2 2 \pi \delta \left(E_\gamma ( \textbf{k},x)^2 - E_\phi(\textbf{k})^2 \right) f_\phi \, .
\end{equation}
Here $k=k_\mu$ and $x = x^\mu$ are the 4-momentum and spacetime position of photons respectively, which have a phase-space density $f = f_\gamma(k,x)$. Note $\varepsilon$ is the 4-polarisation vector of the photon and $f_\phi$ gives the phase-space density of axions. \\\\
(a) Solve this equation by a method of characteristics by using characteristic curves $x = x(\lambda)$ and $k = k(\lambda)$ which give the characteristic curves of photons corresponding to the operator appearing on the left-hand side of Eq.~\eqref{eq:BoltzmannExplicit}. What is the interpretation of the characteristic equations for $x'(\lambda)$ and $k'(\lambda)$, have you seen such equations before? You should find that you end up with an equation for the form
\begin{equation}
    \frac{d f (k(\lambda), x(\lambda))}{d \lambda }  = \cdots 
\end{equation}
\newline
(b) By integrating this equation and assuming a stationary background, derive the following expression for the conversion probability $P_{a \gamma} = f_\gamma(k_c,x_c)/f_\phi(k_c,x_c)$ where $(k_c,x_c)$ are the points where the resonance occurs, i.e. where the argument of the delta-function above vanishes. 
\begin{equation}\label{eq:ProbabilityCovariant}
	P_{a \gamma  } =   \frac{  \pi g_{a \gamma \gamma}^2 \big| k \cdot F_{\rm ext} \cdot \varepsilon \big|^2	}{ E_\gamma \partial_{k_0} \mathcal{H} \left| \textbf{v}_p \cdot \nabla_\textbf{x} E_\gamma(\textbf{k},\textbf{x})\right|} ,
\end{equation}
where $\textbf{v}_p = \textbf{k}/k_0$ is the phase-velocity. You may find the chain-rule result $\partial_\textbf{x} \mathcal{H}/\partial_{k_0} \mathcal{H} = \nabla_\textbf{x} E_\gamma$ helpful. 
\end{tcolorbox}

\paragraph{Solution.} 

Let us consider the Boltzmann-like equation
\begin{equation}
\partial_k \mathcal{H} \cdot \partial_x f_\gamma	- \partial_x \mathcal{H}  \cdot \partial_k f_\gamma  =g_{a\gamma\gamma}^2
\left|
\mathbf{k}\cdot \mathbf{F}_{\rm ext}\cdot \boldsymbol{\varepsilon}
\right|^2
\,2\pi\,
\delta\!\left(E_\gamma^2(k,x)-E_\phi^2(x)\right)
f_\phi ,
\end{equation}
where parametric evolution of the photon worldline in the photon phase-space is specified through $\left(k^\mu(\lambda),x^\mu(\lambda)\right)$.\vspace{.7cm}\\
{\bf\textit{a) Solution of the equation by the method of characteristic}}\\
Let us first consider the left–hand side of the Boltzmann-like equation.
Along the photon worldline, classical Hamilton's equations apply:
\begin{equation}
\frac{\partial \mathcal{H}}{\partial k^\mu}
=
\frac{\partial x^\mu}{\partial \lambda},
\qquad
\frac{\partial \mathcal{H}}{\partial x^\mu}
=
-
\frac{\partial k^\mu}{\partial \lambda}.
\end{equation}
Using these relations, the left–hand side becomes
\begin{equation}
\frac{\partial x^\mu}{\partial \lambda}\frac{\partial f_\gamma}{\partial x^\mu}
+
\frac{\partial k^\mu}{\partial \lambda}\frac{\partial f_\gamma}{\partial k^\mu}
=
\frac{d}{d\lambda}
f_\gamma(k^\mu(\lambda),x^\mu(\lambda)).
\end{equation}
Therefore we obtain
\begin{equation}
\frac{d f_\gamma(k^\mu(\lambda),x^\mu(\lambda))}{d\lambda}
=
g_{a\gamma\gamma}^2
\left|
\mathbf{k}\cdot \mathbf{F}_{\rm ext}\cdot \boldsymbol{\varepsilon}
\right|^2
\,2\pi\,
\delta\!\left(E_\gamma^2(k,x)-E_\phi^2(x)\right)
f_\phi .
\end{equation} \vspace{0.3cm}\\
{\bf\textit{b) Resonant conversion probability}}\\
We denote by $(k_c,x_c)$ the point in phase space where the
resonance occurs, namely
\begin{equation}
E_\gamma^2(k_c,x_c)-E_\phi^2(x_c)=0\,,
\end{equation}
in which we have defined $k_c = k(\lambda_c)$ and $x_c = x(\lambda_c)$. Let us recall the property of the Dirac delta function
\begin{equation}
\delta(f(x)) =
\sum_i
\frac{\delta(x-x_i)}{|f'(x_i)|},
\end{equation}
where $x_i$ are the zeros of the function $f(x)$. In particular, at the resonance point we have
\begin{equation}
E_\gamma^2(k(\lambda_c),x(\lambda_c))
-
E_\phi^2(x(\lambda_c)) = 0 .
\end{equation}
and 
\begin{equation}
\left.
\frac{d}{d\lambda}
\left(E_\gamma^2-E_\phi^2\right)
\right|_{\lambda=\lambda_c}
=
2E_\gamma\left(E_\gamma'-E_\phi'\right)|_{\lambda=\lambda_c}\,.
\end{equation}
Therefore, the integration of the Boltzmann equation around the resonance point gives
\begin{equation}
f_\gamma(k_c,x_c)
=
g_{a\gamma\gamma}^2
\pi
\left|
\mathbf{k}\cdot \mathbf{F}_{\rm ext}\cdot \boldsymbol{\varepsilon}
\right|^2
\frac{1}{E_\gamma|E_\gamma'-E_\phi'|}
\,f_\phi(k_c,x_c).
\end{equation}
Under the assumption of a stationary background, the photon energy must be conserved along its worldline, implying $E_\gamma' = 0$. The axion energy is given by $E_\phi^2 = \mathbf{k}(\lambda)\cdot\mathbf{k}(\lambda) + \mu_\phi^2$, leading to
\begin{equation}
E_\phi'
=
\frac{\mathbf{k}\cdot\mathbf{k}'}{E_\phi}\,.
\end{equation}
Using Hamilton's equation $\mathbf{k}' = -\nabla_x \mathcal{H}$, we obtain
\begin{equation}
E_\phi'
=
-\mathbf{v}_p\cdot\nabla_x \mathcal{H},
\end{equation}
where $\mathbf{v}_p$ is the axion-photon beam phase velocity.
Finally, we derive the expression for the conversion probability around the resonance point
\begin{equation}
P_{a\gamma}
=
\frac{f_\gamma(k_c,x_c)}{f_\phi(k_c,x_c)}
=
g_{a\gamma\gamma}^2
\pi
\frac{
\left|
\mathbf{k}\cdot \mathbf{F}_{\rm ext}\cdot \boldsymbol{\varepsilon}
\right|^2
}{
E_\gamma
\left|
\mathbf{v}_p\cdot\nabla_x E_\gamma
\right|
}.
\end{equation}

\subsection*{Exercise 3: Radio Telescope Sensitivity to Axion Dark Matter}\label{sec:tutorial_ex3}

\begin{tcolorbox}[colback=pos!5!white, colframe=pos!75!black, title=Exercise 3]
In general, the Hamiltonian $\mathcal{H}$ for a photon in a magnetised plasma can be quite complicated. Instead we'll do something simpler and assume a weakly magnetised plasma where
\begin{equation}
    \mathcal{H} = k_\mu k^\mu  + \omega_p^2, 
\end{equation}
where $\omega_p$ is the plasma mass. We will model the plasma around a neutron star as spherical toy model:  
\begin{equation}
    \omega_p^2 = \frac{4 \pi \alpha n_e}{m_e} , \qquad n_e = \frac{2 \Omega B}{e}, \qquad  B = B_s \left(\frac{R}{r}\right)^3\, , 
\end{equation}
where $R$ is the neutron star radius, $\Omega = 2\pi/P$ is the angular frequency with which the neutron star is rotating, and $P$ is its corresponding period.  $B_s$ is the surface magnetic field. Let's assume a stationary setup, in which case, by integrating  the energy conservation equation
\begin{equation}\label{eq:EnergyConservation}
	\frac{{\rm d}}{{\rm d}t} \int {\rm d}\mathcal{V} \int {\rm d}^3\textbf{k} \, \omega f_\gamma + \int {\rm d}^3 \textbf{k} \int {\rm d}\textbf{A} \cdot \textbf{v}_g \, \omega f_\gamma +  \int {\rm d}^3 \textbf{k} \int {\rm d} \mathcal{V} \, \partial_t E_\gamma f_\gamma = \int {\rm d}\mathcal{V} \, Q
\end{equation}
over phase space, we arrive at
\begin{equation}\label{eq:EnergyConservationStatoinary}
	 \int {\rm d}^3 \textbf{k} \int {\rm d}\textbf{A} \cdot \textbf{v}_g \, \omega f_\gamma =\int {\rm d}^3 \textbf{k} \int {\rm d}  \boldsymbol{\Sigma}_\textbf{k}\cdot \textbf{v}_p \omega P_{a \gamma} f_\phi  \equiv \mathcal{P}
\end{equation} 
where $\mathcal{P}$ is the power (i.e. energy per unit time) produced by axions converting into photons. Hence by deriving an expression for the right-hand side of \eqref{eq:EnergyConservationStatoinary}, you will be able to derive an expression for the total power $\mathcal{P}$ emitted by resonantly produced photons. You may estimate the flux density (that is, power, per unit area, per unit frequency) arriving on earth as
\begin{equation}
    \mathcal{S} = \frac{\mathcal{P}}{ 4 \pi d^2} \frac{1}{ \Delta f},
\end{equation}
where $d$ is the distance to source, and $\Delta f$ is the bandwidth of the signal, which for an axion line signal you can take to be $\Delta f = v_0^2 \, m_a$.\\\\ 
(a) Argue that for the toy model above, the critical surface is spherical, with radius $r_c$, and drive an expression of $r_c$ in terms of $m_a$ and other quantities. 
\newline

(b) By evaluating the expression in the right-hand side of Eq.~\eqref{eq:EnergyConservationStatoinary} using the model described above, obtain, and evaluate an integral for the total power $\mathcal{P}$ in this model. You may take the axion density to be $f_a(\mathbf{x}, \mathbf{k}) = v_a \rho_{\text{DM}}^{r_c}/m_a \, \frac{\delta(|\textbf{k}| - \omega_c)}{4 \pi k^2}$ where $\omega_c = \sqrt{m_a^2 + k_c^2}
$ and $k_c = m_a v_a$. Where $\rho_{\text{DM}}^{r_c} = \rho_{\text{DM}}^{\infty} \, \frac{2}{\sqrt{\pi}} \, \frac{1}{v_0} \, \sqrt{\frac{2 G M_{\text{NS}}}{r_c}}$.

\end{tcolorbox}

\begin{tcolorbox}[colback=pos!5!white, colframe=pos!75!black]
(c) Finally, let's consider a pulsar PSR J2144-3933 which is 
$d = 180 \, \text{parsec}$ from Earth. 
You can assume a stellar radius of $R = 10 \, \text{km}$, 
$v_a \simeq \sqrt{\tfrac{GM}{r_c}}$, 
$B_s = 2 \times 10^{12} \, \text{Gauss}$, 
and $\rho_{\text{DM}}^{\infty} \simeq 0.45 \times   \text{GeV cm}^{-3}$, $P=8.5 {\rm s}$ and $v_0\sim 200$ km/sec. Using these numbers and the results above, derive an expression for $\mathcal{S}$ for a generic axion mass.
\newline
\newline
(d) Finally, the minimal detectable signal is 
\[
S_{\min} = \text{SNR}_{\min} \, \frac{\text{SEFD}}{\sqrt{n_{\text{pol}} \Delta f  t_{\text{obs}}}}. 
\]
By taking a typical system equivalent flux density of a telescope to be $\text{SEFD} =2$ Jy, an observing time of 100 hours and an SNR of $3$, derive the value of $g_{a \gamma \gamma}$ to which you would be sensitive for an axion mass of $m_a = \mu eV$. You may also wish to find a generic expression for fiducial values of different parameters of the form $g_{a \gamma \gamma} \sim (\mu eV/m_a)^{\rm index}  (B_s/10^{14})^{\rm index}  \cdot \cdot $ etc to get some feel for the scaling.  
\end{tcolorbox}

\paragraph{Solution.} Let us assume a spherical toy model for the magnetized plasma surrounding the neutron star. The plasma photon mass, the electron number density and the radial profile of the magnetic field are then defined as
\begin{equation}
\omega_p^2 = \frac{4\pi \alpha n_e}{m_e},
\qquad
n_e = \frac{2\Omega B}{e},
\qquad
B(r) = \left(\frac{R}{r}\right)^3 B_s ,
\end{equation}
where $\Omega$ is the angular velocity,$\Omega = {2\pi}/{P}$,
with $P$ being the rotation period. By assuming a stationary setup, the total power radiated in photons due to resonant conversions through the resonant surface  $\overline{\Sigma}_{\mathbf{k}}$ is given by
\begin{equation}
\mathcal{P} =
\int d^3\mathbf{k}
\int_{\Sigma_{\rm res}}
d\boldsymbol{\Sigma}_{\mathbf{k}}
\cdot
\mathbf{v}_p \,
\omega \,
P_{a\gamma}\,
f_\phi ,
\end{equation}
where $\mathbf{v}_p = {\mathbf{k}}/{E_\gamma}$ is the phase velocity. The photon flux density arriving at Earth is given by
\begin{equation}
S = \frac{Q}{4\pi d^2}\frac{1}{\Delta f}.
\end{equation}\vspace{0.3cm}\\
{\bf\textit{a) Sphericity of the critical surface}}\\
In the toy model considered, the neutron-star magnetosphere shows
no dependence on angular variables. Since all the quantities entering
the determination of the photon energy are spherically symmetric,
the resonant surface, defined by
\begin{equation}
E_\gamma(k_e,r_e) = E_\phi(k_e,r_e)\,,
\end{equation}
must also be characterized by spherical symmetry. Therefore, the resonant surface unit vector is radial, $\overline{\Sigma}_{\mathbf{k}} = {\Sigma}_{\mathbf{k}}\,\hat{r}$. Let us derive an expression for the resonant conversion radius.
\begin{equation}
\omega_p^2
=
\frac{4\pi\alpha n_e}{m_e}
=
\frac{8\pi\alpha\Omega}{m_e e}
B_s \left(\frac{R}{r}\right)^3 .
\end{equation}
Assuming a weakly magnetized plasma, at the resonant surface $r=r_c$ the axion and photon energies must coincide
\begin{equation}
E_\gamma = E_a
\qquad\Longleftrightarrow\qquad
\omega_p^2 = m_a^2 .
\end{equation}
Thus
\begin{equation}
m_a^2
=
\frac{4\pi\alpha\Omega B_s}{m_e}
\left(\frac{R}{r_c}\right)^3 ,
\end{equation}
which implies
\begin{equation}
r_c =
\left(
\frac{4\pi\alpha\Omega B_s}{m_em_a^2}
\right)^{1/3} R .
\end{equation}\vspace{0.3cm}\\
{\bf\textit{b) Total radiated power.}}\\
Let us evaluate explicitly the integral determining the total
radiated power
\begin{equation}
\begin{split}
    \mathcal{P} &= \int d^3\mathbf{k} \int_{\Sigma_{\rm res}} d\boldsymbol{\Sigma}_{\mathbf{k}}\cdot\mathbf{v}_p\,\omega\,P_{a\gamma}\,f_\phi\\
    &=8\pi^2\int\sin\theta_B d\theta_B\int dk\,k^2\,r_e^2\cos\theta_B|\mathbf{v}_p|\,\omega\,\frac{\pi}{2}g_{a\gamma\gamma}^2\frac{|\mathbf{B}_{\rm ext}\cdot\hat{\varepsilon}|^2}{|\mathbf{v}_p\cdot\nabla_x E_\gamma|}\,f_\phi\,,
\end{split}
\end{equation}
where $\theta_B$ is the angle between the radial direction and the
phase velocity. The photon dispersion relation is given by
\begin{equation}
E_\gamma^2 = |\mathbf{k}|^2 + \omega_p^2\,.
\end{equation}
Therefore,
\begin{equation}
\nabla_x E_\gamma =
\frac{\omega_p}{E_\gamma}\nabla_x \omega_p\,.
\end{equation}
and
\begin{equation}
\mathbf{v}_p\cdot\nabla_x E_\gamma=\frac{\omega_p}{E_\gamma}
\cos\theta_B|\mathbf{v}_p||\nabla_x\omega_p| .
\end{equation}
Moreover, since photons are transverse in weakly magnetized plasmas, we have
\begin{equation}
|\mathbf{B}_{\rm ext}\cdot\hat{\varepsilon}|^2
=
\sin^2\theta_B\,B^2\,,
\end{equation}
Performing the angular integration
\begin{equation}
\int_0^\pi \sin^3\theta_B d\theta_B = \frac{4}{3}\,,
\end{equation}
we obtain
\begin{equation}
\mathcal{P} =
\frac{4\pi^2}{3}
g_{a\gamma\gamma}^2
r_c^2
\frac{\omega_c^2}{\omega_p}
\frac{B^2}{|\nabla_x\omega_p|}
v_a
\frac{\rho_{\rm DM}}{m_a}.
\end{equation}
Where $v_a$ is the phase velocity of the axion dark matter halo. Using spherical symmetry,
\begin{equation}
|\nabla\omega_p|
=
\left|\frac{\partial\omega_p}{\partial r}\right|
=
\frac{3}{2}\frac{m_a}{r_c}.
\end{equation}
Finally
\begin{equation}
\omega_c^2 = m_a^2 + k^2
\approx m_a^2
\qquad (v_a^2\ll1).
\end{equation}
Thus the total radiated power is
\begin{equation}
    \begin{split}
        \mathcal{P}&=\frac{8\pi^2}{9}g_{a\gamma\gamma}^2r_c^3
        \frac{v_a}{m_a}B^2(r_c)\rho_{\rm DM}\\
        &=\frac{2^{11/6}\pi^{5/6} m_e^{4/3} G}{9\,\alpha^{2/3}}\,g_{a\gamma\gamma}^2\,\frac{\rho_{\rm DM}^{\infty}}{v_0}\,\frac{M_{\rm NS} R^{2} B_s^{2/3}}{\Omega^{4/3}}\,m_a^{5/3}.
    \end{split}
\end{equation}\vspace{0.3cm}\\
{\bf\textit{c) Flux density from PSR J2144$-$3933.}}\\
Now let us consider the case of PSR J2144$-$3933, at a distance
$d = 180 \ \mathrm{pc}$, and rotating with angular velocity $\Omega = \frac{2\pi}{P}$, with rotation period $P = 8.5 \ \mathrm{s}$. The total flux density received at Earth is
\begin{equation}
S = \frac{Q}{4\pi d^2}\frac{1}{\Delta f}
    = \frac{Q}{4\pi d^2}\frac{1}{v_0^2 \mu_a}.
\end{equation}
Evaluating numerically we obtain
\begin{align}
S =
1.2\times10^{-4} \, {\rm Jy}
\left(\frac{g_{a\gamma\gamma}}{10^{-12}\,{\rm GeV}^{-1}}\right)^2
\left(\frac{\mu_a}{1\,\mu{\rm eV}}\right)^{2/3}
\left(\frac{B_s}{10^{14}\,{\rm G}}\right)^{2/3}
\left(\frac{R}{10\,{\rm km}}\right)^2  \nonumber\\
\times
\left(\frac{d}{180\,{\rm pc}}\right)^{-2}
\left(\frac{\rho_{\rm NS}}{0.4\,{\rm GeV/cm^3}}\right)
\left(\frac{\Omega}{1\,{\rm s^{-1}}}\right)^{-4/3}.
\end{align}
Hence, the flux density can be estimated as
\begin{equation}
S =
7.8\times10^{-5}\,{\rm Jy}
\left(\frac{g_{a\gamma\gamma}}{10^{-12}\,{\rm GeV}^{-1}}\right)^2
\left(\frac{\mu_a}{1\,\mu{\rm eV}}\right)^{2/3}.
\end{equation}\vspace{0.3cm}\\
{\bf\textit{d) Sensitivity to the axion-photon coupling.}}\\
The minimum detectable signal for an observation time $t_{\rm obs}=100\,{\rm h}$ is given by
\begin{equation}
S_{\rm min} =
{\rm SNR}_{\rm req}
\frac{S_{\rm EFD}}
{\sqrt{\Delta\nu\,t_{\rm obs}}}.
\end{equation}
Therefore the minimum value of the coupling constrained by observations is
\begin{align}
g_{a\gamma\gamma}^{\rm min}
&\simeq
2.2\times10^{-12}\,{\rm GeV^{-1}}
\left(\frac{t_{\rm obs}}{100\,{\rm h}}\right)^{-1/2}
\left(\frac{\mu_a}{1\,\mu{\rm eV}}\right)^{-5/6}
\left(\frac{B_s}{10^{14}\,{\rm G}}\right)^{-1/3} \nonumber \\
&\quad\times
\left(\frac{R}{10\,{\rm km}}\right)^{-1}
\left(\frac{d}{180\,{\rm pc}}\right)
\left(\frac{\rho_{\rm NS}}{0.4\,{\rm GeV/cm^3}}\right)^{-1/2}
\left(\frac{\Omega}{1\,{\rm s^{-1}}}\right)^{2/3}.
\end{align}
Then, assuming a minimal signal-to-noise ratio ${\rm SNR}_{\rm req}=3$ and a minimal sensitivity for the detector ${\rm SEFD}=2\,{\rm Jy}$,
for an axion mass $\mu_a = 1\,\mu{\rm eV}$ observations of the pulsar PSR J2144$-$3933 give the following limit on $g_{a\gamma}$:
\begin{equation}
g_{a\gamma\gamma}^{\rm min}
\simeq 1.8\times10^{-12}\,{\rm GeV^{-1}}.
\end{equation}

\section*{Acknowledgements}

This article is based on the work from COST Action COSMIC WISPers CA21106, supported by COST (European Cooperation in Science and Technology).  JM acknowledges support from the Science and Technology Facilities
Council (STFC) Consolidated Grant [Grant No. ST/X00077X/1] and from a United Kingdom Research and Innovation Future Leaders Fellowship [Grant No. MR/V021974/2].
The work of AL was partially supported by the research grant number 2022E2J4RK "PANTHEON: Perspectives in Astroparticle and Neutrino THEory with Old and New messengers" under the program PRIN 2022 (Mission 4, Component 1, CUP I53D23001110006) funded by the Italian Ministero dell'Universit\`a e della Ricerca (MUR) and by the European Union – Next Generation EU. AL acknowledges support from Italian MUR through the FIS 2 project FIS-2023-01577 (DD n. 23314 10-12-2024, CUP C53C24001460001), and by Istituto Nazionale di Fisica Nucleare (INFN) through the Theoretical Astroparticle Physics (TAsP) project.

\bibliographystyle{bib_style}
\bibliography{biblio}

@article{Pappas:2025zld,
    author = {Pappas, Kaliro{\"e} M. W. and others},
    title = "{High-Frequency Gravitational Wave Search with ABRACADABRA-10 cm}",
    eprint = "2505.02821",
    archivePrefix = "arXiv",
    primaryClass = "hep-ex",
    reportNumber = "FERMILAB-CONF-25-0305-T",
    month = "5",
    year = "2025"
}

@article{Domcke:2024mfu,
    author = "Domcke, Valerie and Ellis, Sebastian A. R. and Rodd, Nicholas L.",
    title = "{Magnets are Weber Bar Gravitational Wave Detectors}",
    eprint = "2408.01483",
    archivePrefix = "arXiv",
    primaryClass = "hep-ph",
    reportNumber = "CERN-TH-2024-132",
    doi = "10.1103/966v-r5fm",
    journal = "Phys. Rev. Lett.",
    volume = "134",
    number = "23",
    pages = "231401",
    year = "2025"
}

@article{Berlin:2023grv,
    author = {Berlin, Asher and Blas, Diego and Tito D'Agnolo, Raffaele and Ellis, Sebastian A. R. and Harnik, Roni and Kahn, Yonatan and Sch{\"u}tte-Engel, Jan and Wentzel, Michael},
    title = "{Electromagnetic cavities as mechanical bars for gravitational waves}",
    eprint = "2303.01518",
    archivePrefix = "arXiv",
    primaryClass = "hep-ph",
    reportNumber = "FERMILAB-PUB-22-892-SQMS-T",
    doi = "10.1103/PhysRevD.108.084058",
    journal = "Phys. Rev. D",
    volume = "108",
    number = "8",
    pages = "084058",
    year = "2023"
}

@article{Fischer:2024msc,
    author = "Fischer, Lars and others",
    title = "{First characterisation of the MAGO cavity, a superconducting RF detector for kHz{\textendash}MHz gravitational waves}",
    eprint = "2411.18346",
    archivePrefix = "arXiv",
    primaryClass = "gr-qc",
    reportNumber = "FERMILAB-PUB-24-0819-SQMS-TD, DESY-24-181",
    doi = "10.1088/1361-6382/add8da",
    journal = "Class. Quant. Grav.",
    volume = "42",
    number = "11",
    pages = "115015",
    year = "2025"
}

@article{Domcke:2022rgu,
    author = "Domcke, Valerie and Garcia-Cely, Camilo and Rodd, Nicholas L.",
    title = "{Novel Search for High-Frequency Gravitational Waves with Low-Mass Axion Haloscopes}",
    eprint = "2202.00695",
    archivePrefix = "arXiv",
    primaryClass = "hep-ph",
    reportNumber = "DESY-22-017, CERN-TH-2022-010",
    doi = "10.1103/PhysRevLett.129.041101",
    journal = "Phys. Rev. Lett.",
    volume = "129",
    number = "4",
    pages = "041101",
    year = "2022"
}

@article{Berlin:2021txa,
    author = {Berlin, Asher and Blas, Diego and Tito D'Agnolo, Raffaele and Ellis, Sebastian A. R. and Harnik, Roni and Kahn, Yonatan and Sch{\"u}tte-Engel, Jan},
    title = "{Detecting high-frequency gravitational waves with microwave cavities}",
    eprint = "2112.11465",
    archivePrefix = "arXiv",
    primaryClass = "hep-ph",
    reportNumber = "FERMILAB-PUB-21-724-SQMS-T",
    doi = "10.1103/PhysRevD.105.116011",
    journal = "Phys. Rev. D",
    volume = "105",
    number = "11",
    pages = "116011",
    year = "2022"
}

@article{Ai:2024vfa,
    author = "Ai, Wen-Yuan and Garbrecht, Bjorn and Tamarit, Carlos",
    title = "{The QCD theta-parameter in canonical quantization}",
    eprint = "2403.00747",
    archivePrefix = "arXiv",
    primaryClass = "hep-th",
    reportNumber = "KCL-PH-TH/2024-13, TUM-HEP-1499/24, MITP-24-031",
    month = "3",
    year = "2024"
}

@article{Ai:2020ptm,
    author = {Ai, Wen-Yuan and Cruz, Juan S. and Garbrecht, Bj{\"o}rn and Tamarit, Carlos},
    title = "{Consequences of the order of the limit of infinite spacetime volume and the sum over topological sectors for CP violation in the strong interactions}",
    eprint = "2001.07152",
    archivePrefix = "arXiv",
    primaryClass = "hep-th",
    reportNumber = "TUM-HEP-1249/20, CP3-20-02",
    doi = "10.1016/j.physletb.2021.136616",
    journal = "Phys. Lett. B",
    volume = "822",
    pages = "136616",
    year = "2021"
}

@article{Dine:1981rt,
  author = {Dine, Michael and Fischler, Willy and Srednicki, Mark},
  title = {A Simple Solution to the Strong CP Problem with a Harmless Axion},
  journal = {Phys. Lett. B},
  volume = {104},
  pages = {199},
  year = {1981}
}

@article{Zhitnitsky:1980tq,
  author = {Zhitnitsky, A. R.},
  title = {On Possible Suppression of the Axion Hadron Interactions},
  journal = {Sov. J. Nucl. Phys.},
  volume = {31},
  pages = {260},
  year = {1980}
}

@article{Kim:1979if,
  author = {Kim, Jihn E.},
  title = {Weak Interaction Singlet and Strong CP Invariance},
  journal = {Phys. Rev. Lett.},
  volume = {43},
  pages = {103},
  year = {1979}
}

@article{Shifman:1979if,
  author = {Shifman, M. A. and Vainshtein, A. I. and Zakharov, V. I.},
  title = {Can Confinement Ensure Natural CP Invariance of Strong Interactions?},
  journal = {Nucl. Phys. B},
  volume = {166},
  pages = {493},
  year = {1980}
}

@article{Kaplan:2019ako,
    author = "Kaplan, David E. and Rajendran, Surjeet and Riggins, Paul",
    title = "{Particle Probes with Superradiant Pulsars}",
    eprint = "1908.10440",
    archivePrefix = "arXiv",
    primaryClass = "hep-ph",
    month = "8",
    year = "2019"
}

@article{Bai:2025nzd,
    author = "Bai, Zhaoyu and Cardoso, Vitor and Chen, Yifan and Li, Yuyan and McDonald, Jamie I. and Seong, Hyeonseok",
    title = "{Stellar Superradiance and Low-Energy Absorption in Dense Nuclear Media}",
    eprint = "2512.13816",
    archivePrefix = "arXiv",
    primaryClass = "hep-ph",
    reportNumber = "DESY-25-188",
    month = "12",
    year = "2025"
}

@article{Cardoso:2017kgn,
    author = "Cardoso, Vitor and Pani, Paolo and Yu, Tien-Tien",
    title = "{Superradiance in rotating stars and pulsar-timing constraints on dark photons}",
    eprint = "1704.06151",
    archivePrefix = "arXiv",
    primaryClass = "gr-qc",
    reportNumber = "CERN-TH-2017-082",
    doi = "10.1103/PhysRevD.95.124056",
    journal = "Phys. Rev. D",
    volume = "95",
    number = "12",
    pages = "124056",
    year = "2017"
}

@article{Cardoso:2015zqa,
    author = "Cardoso, Vitor and Brito, Richard and Rosa, Joao L.",
    title = "{Superradiance in stars}",
    eprint = "1505.05509",
    archivePrefix = "arXiv",
    primaryClass = "gr-qc",
    doi = "10.1103/PhysRevD.91.124026",
    journal = "Phys. Rev. D",
    volume = "91",
    number = "12",
    pages = "124026",
    year = "2015"
}

@article{Hirata:1987hu,
      author         = "Hirata, K. S. and Kajita, T. and Koshiba, M. and Nakahata, M.
                        and Oyama, Y.",
      title          = "{Observation of a neutrino burst from the supernova SN1987A}",
      journal        = "Phys. Rev. Lett.",
      volume         = "58",
      pages          = "1490--1493",
      doi            = "10.1103/PhysRevLett.58.1490",
      year           = "1987",
      reportNumber   = "",
      SLACcitation   = "%%CITATION = PRLTA,58,1490;%%"
}

@article{Bionta:1987qt,
      author         = "Bionta, R. M. and Blewitt, G. and Bratton, C. B. and Casper, D.
                        and Ciocio, A. and Claus, R. and Cortez, B. G. and Crouch, M.
                        and Dye, S. T. and Errede, S. and Foster, G. W. and Gajewski, W.
                        and Ganezer, K. S. and Goldhaber, M. and Haines, T. J. and
                        Jones, T. W. and Kielczewska, D. and Kropp, W. R. and Learned, J.
                        G. and LoSecco, J. M. and Matthews, J. and Miller, R. and Mudan,
                        M. S. and Park, H. S. and Price, L. R. and Reines, F. and
                        Schultz, J. and Seidel, S. and Shumard, E. and Sinclair, D.
                        and Sobel, H. W. and Stone, J. L. and Sulak, L. and Svoboda,
                        R. and Thornton, G. J. and Van der Velde, J. C. and Wuest, C.
                        R.",
      title          = "{Observation of a neutrino burst in coincidence with supernova 1987A in the Large Magellanic Cloud}",
      journal        = "Phys. Rev. Lett.",
      volume         = "58",
      pages          = "1494--1496",
      doi            = "10.1103/PhysRevLett.58.1494",
      year           = "1987",
      reportNumber   = "",
      SLACcitation   = "%%CITATION = PRLTA,58,1494;%%"
}

@article{NANOGrav:2020bcs,
  author       = {Z. Arzoumanian and P. T. Baker and H. Blumer and B. B\'ecsy and A. Brazier
                  and P. R. Brook and S. Burke-Spolaor and S. Chatterjee and S. Chen
                  and J. M. Cordes and N. J. Cornish and F. Crawford and H. T. Cromartie
                  and C. M. F. Mingarelli and X. Siemens and others},
  title        = {{The NANOGrav 12.5-year Data Set: Search for an Isotropic Stochastic Gravitational-Wave Background}},
  journal      = {Astrophys. J. Lett.},
  volume       = {905},
  pages        = {L34},
  year         = {2020},
  doi          = {10.3847/2041-8213/abcddf},
  eprint       = {2009.04496},
  archivePrefix= {arXiv},
  primaryClass = {astro-ph.HE},
  note         = {Evidence for a common-spectrum process in 12.5-yr pulsar timing data}  
}

@article{Goncharov:2021oub,
  author       = {B. Goncharov and R. M. Shannon and D. J. Reardon and G. Hobbs
                  and A. Zic and M. Bailes and M. Cury\l{}o and S. Dai and M. Kerr
                  and M. E. Lower and R. N. Manchester and R. Spiewak and S. Wang and J. B. Wang
                  and L. Zhang and S. Zhang},
  title        = {{On the Evidence for a Common-spectrum Process in the Search for the Nanohertz Gravitational-Wave Background with the Parkes Pulsar Timing Array}},
  journal      = {Astrophys. J. Lett.},
  volume       = {917},
  pages        = {L20},
  year         = {2021},
  doi          = {10.3847/2041-8213/ac13f2},
  eprint       = {2107.12112},
  archivePrefix= {arXiv},
  primaryClass = {astro-ph.HE},
  note         = {PPTA DR2 common-spectrum analysis}  
}

@article{EPTA:2021crs,
  author       = {S. Chen and R. N. Caballero and Y. J. Guo and M. Kerr and A. Parthasarathy
                  and G. Shaifullah and P. T. Baker and R. Ferdman and L. Lentati and A. L. Jones
                  and J. M. Cordes and others},
  title        = {{Common-red-signal analysis with 24-yr high-precision timing of the European Pulsar Timing Array: Inferences in the stochastic gravitational-wave background search}},
  journal      = {Mon. Not. R. Astron. Soc.},
  volume       = {508},
  number       = {4},
  pages        = {4970--4993},
  year         = {2021},
  doi          = {10.1093/mnras/stab2789},
  eprint       = {2101.08477},
  archivePrefix= {arXiv},
  primaryClass = {astro-ph.IM},
  note         = {EPTA common-red signal in 24 yr data}  
}

@article{Tarafdar:2022toa,
  author       = {S. Tarafdar and B. B. P. Perera and R. N. Manchester and G. Hobbs and R. M. Shannon
                  and D. J. Reardon and N. D. R. Bhat and M. Bailes and A. Zic and others},
  title        = {{Search for an Isotropic Gravitational-Wave Background with the Parkes Pulsar Timing Array Data Release 3}},
  journal      = {Astrophys. J.},
  volume       = {936},
  pages        = {45},
  year         = {2022},
  doi          = {10.3847/1538-4357/ac7d89},
  eprint       = {2306.16215},
  archivePrefix= {arXiv},
  primaryClass = {astro-ph.HE},
  note         = {PPTA DR3 stochastic GWB search and common red process}  
}

@article{LIGOScientific:2016aoc,
    author = "Abbott, B. P. and others",
    collaboration = "LIGO Scientific, Virgo",
    title = "{Observation of Gravitational Waves from a Binary Black Hole Merger}",
    eprint = "1602.03837",
    archivePrefix = "arXiv",
    primaryClass = "gr-qc",
    reportNumber = "LIGO-P150914",
    doi = "10.1103/PhysRevLett.116.061102",
    journal = "Phys. Rev. Lett.",
    volume = "116",
    number = "6",
    pages = "061102",
    year = "2016"
}

@article{Aggarwal:2025noe,
    author = "Aggarwal, Nancy and others",
    title = "{Challenges and opportunities of gravitational-wave searches above 10 kHz}",
    eprint = "2501.11723",
    archivePrefix = "arXiv",
    primaryClass = "gr-qc",
    reportNumber = "CERN-TH-2025-014, DESY-25-007",
    doi = "10.1007/s41114-025-00060-5",
    journal = "Living Rev. Rel.",
    volume = "28",
    number = "1",
    pages = "10",
    year = "2025"
}

@article{He:2023xoh,
    author = "He, Yutong and Giri, Sambit K. and Sharma, Ramkishor and Mtchedlidze, Salome and Georgiev, Ivelin",
    title = "{Inverse Gertsenshtein effect as a probe of high-frequency gravitational waves}",
    eprint = "2312.17636",
    archivePrefix = "arXiv",
    primaryClass = "astro-ph.CO",
    reportNumber = "NORDITA-2023-066",
    doi = "10.1088/1475-7516/2024/05/051",
    journal = "JCAP",
    volume = "05",
    pages = "051",
    year = "2024"
}

@article{Ramazanov:2023nxz,
    author = "Ramazanov, Sabir and Samanta, Rome and Trenkler, Georg and Urban, Federico R.",
    title = "{Shimmering gravitons in the gamma-ray sky}",
    eprint = "2304.11222",
    archivePrefix = "arXiv",
    primaryClass = "astro-ph.HE",
    doi = "10.1088/1475-7516/2023/06/019",
    journal = "JCAP",
    volume = "06",
    pages = "019",
    year = "2023"
}

@article{Liu:2023mll,
    author = "Liu, Tao and Ren, Jing and Zhang, Chen",
    title = "{Limits on High-Frequency Gravitational Waves in Planetary Magnetospheres}",
    eprint = "2305.01832",
    archivePrefix = "arXiv",
    primaryClass = "hep-ph",
    doi = "10.1103/PhysRevLett.132.131402",
    journal = "Phys. Rev. Lett.",
    volume = "132",
    number = "13",
    pages = "131402",
    year = "2024"
}

@article{Ito:2023nkq,
    author = "Ito, Asuka and Kohri, Kazunori and Nakayama, Kazunori",
    title = "{Gravitational Wave Search through Electromagnetic Telescopes}",
    eprint = "2309.14765",
    archivePrefix = "arXiv",
    primaryClass = "gr-qc",
    reportNumber = "KEK-QUP-2023-0018, KEK-TH-2558, KEK-Cosmo-0327, TU-1205",
    doi = "10.1093/ptep/ptae004",
    journal = "PTEP",
    volume = "2024",
    number = "2",
    pages = "023E03",
    year = "2024"
}

@article{Lella:2024dus,
    author = "Lella, Alessandro and Calore, Francesca and Carenza, Pierluca and Mirizzi, Alessandro",
    title = "{Constraining gravitational-wave backgrounds from conversions into photons in the Galactic magnetic field}",
    eprint = "2406.17853",
    archivePrefix = "arXiv",
    primaryClass = "hep-ph",
    reportNumber = "LAPTH-036/24, BARI-TH/761-24",
    doi = "10.1103/PhysRevD.110.083042",
    journal = "Phys. Rev. D",
    volume = "110",
    number = "8",
    pages = "083042",
    year = "2024"
}

@article{Bowman:2018yin,
    author = "Bowman, Judd D. and Rogers, Alan E. E. and Monsalve, Raul A. and Mozdzen, Thomas J. and Mahesh, Nivedita",
    title = "{An absorption profile centred at 78 megahertz in the sky-averaged spectrum}",
    eprint = "1810.05912",
    archivePrefix = "arXiv",
    primaryClass = "astro-ph.CO",
    doi = "10.1038/nature25792",
    journal = "Nature",
    volume = "555",
    number = "7694",
    pages = "67--70",
    year = "2018"
}

@article{Fixsen_2011,
	doi = "10.1088/0004-637x/734/1/5",
	year = "2011",
	publisher = "{IOP} Publishing",
	volume = "734",
	number = "1",
	pages = "5",
	author = "D.~J. Fixsen and A. Kogut and S. Levin and M. Limon and P. Lubin and P. Mirel and M. Seiffert and J. Singal and E. Wollack and T. Villela and C.~A. Wuensche",
	title = "{{Arcade 2 measurement of the absolute sky brightness at 3-90 GHz}}",
	journal = {Astrophys. J.}
}

@article{Cillis:1996qy,
    author = "Cillis, Analia N. and Harari, Diego D.",
    title = "{Photon - graviton conversion in a primordial magnetic field and the cosmic microwave background}",
    eprint = "astro-ph/9609200",
    archivePrefix = "arXiv",
    reportNumber = "PRINT-96-170 (BUENOS-AIRES)",
    doi = "10.1103/PhysRevD.54.4757",
    journal = "Phys. Rev. D",
    volume = "54",
    pages = "4757--4759",
    year = "1996"
}

@article{Dolgov:2012be,
    author = "Dolgov, Alexander D. and Ejlli, Damian",
    title = "{Conversion of relic gravitational waves into photons in cosmological magnetic fields}",
    eprint = "1211.0500",
    archivePrefix = "arXiv",
    primaryClass = "gr-qc",
    doi = "10.1088/1475-7516/2012/12/003",
    journal = "JCAP",
    volume = "12",
    pages = "003",
    year = "2012"
}

@article{Leinson:2021ety,
    author = "Leinson, Lev B.",
    title = "{Impact of axions on the Cassiopea A neutron star cooling}",
    eprint = "2105.14745",
    archivePrefix = "arXiv",
    primaryClass = "hep-ph",
    doi = "10.1088/1475-7516/2021/09/001",
    journal = "JCAP",
    volume = "09",
    pages = "001",
    year = "2021"
}

@article{Buschmann:2019pfp,
    author = "Buschmann, Malte and Co, Raymond T. and Dessert, Christopher and Safdi, Benjamin R.",
    title = "{Axion Emission Can Explain a New Hard X-Ray Excess from Nearby Isolated Neutron Stars}",
    eprint = "1910.04164",
    archivePrefix = "arXiv",
    primaryClass = "hep-ph",
    reportNumber = "LCTP-19-26",
    doi = "10.1103/PhysRevLett.126.021102",
    journal = "Phys. Rev. Lett.",
    volume = "126",
    number = "2",
    pages = "021102",
    year = "2021"
}

@article{Fortin:2018ehg,
    author = "Fortin, Jean-Fran{\c{c}}ois and Sinha, Kuver",
    title = "{Constraining Axion-Like-Particles with Hard X-ray Emission from Magnetars}",
    eprint = "1804.01992",
    archivePrefix = "arXiv",
    primaryClass = "hep-ph",
    doi = "10.1007/JHEP06(2018)048",
    journal = "JHEP",
    volume = "06",
    pages = "048",
    year = "2018"
}

@article{Morris:1984iz,
    author = "Morris, Donald E.",
    title = "{Axion Mass Limits From Pulsar X-rays}",
    reportNumber = "LBL-18690",
    doi = "10.1103/PhysRevD.34.843",
    journal = "Phys. Rev. D",
    volume = "34",
    pages = "843",
    year = "1986"
}

@article{Leinson:2014ioa,
    author = "Leinson, L. B.",
    title = "{Axion mass limit from observations of the neutron star in Cassiopeia A}",
    eprint = "1405.6873",
    archivePrefix = "arXiv",
    primaryClass = "hep-ph",
    doi = "10.1088/1475-7516/2014/08/031",
    journal = "JCAP",
    volume = "08",
    pages = "031",
    year = "2014"
}

@article{Hamaguchi:2018oqw,
    author = "Hamaguchi, Koichi and Nagata, Natsumi and Yanagi, Keisuke and Zheng, Jiaming",
    title = "{Limit on the Axion Decay Constant from the Cooling Neutron Star in Cassiopeia A}",
    eprint = "1806.07151",
    archivePrefix = "arXiv",
    primaryClass = "hep-ph",
    reportNumber = "UT-18-13, IPMU 18-0111, IPMU-18-0111",
    doi = "10.1103/PhysRevD.98.103015",
    journal = "Phys. Rev. D",
    volume = "98",
    number = "10",
    pages = "103015",
    year = "2018"
}

@article{Chen:1994ch,
    author = "Chen, Pisin",
    title = "{Resonant photon - graviton conversion and cosmic microwave background fluctuations}",
    reportNumber = "SLAC-PUB-6494",
    doi = "10.1103/PhysRevLett.74.634",
    journal = "Phys. Rev. Lett.",
    volume = "74",
    pages = "634--637",
    year = "1995",
    note = "[Erratum: Phys.Rev.Lett. 74, 3091 (1995)]"
}

@article{Jordan2009,
    author = "S. Jordan", 
    title= " IAU Symposium", 
    volume= "259",
    journal = "Cambridge University Press, Cambridge, UK",
    year = "2009",
    pages = "369--378"
}

@ARTICLE{1992A&A...259..143P,
       author = {{Piirola}, V. and {Reiz}, A.},
        title = "{The highly magnetic (B 500 MG) white dwarf PG 1031+234 : discovery of wavelength dependent polarization and intensity variations.}",
      journal = {\aap},
         year = 1992,
        month = jun,
       volume = {259},
        pages = {143-148},
       adsurl = {https://ui.adsabs.harvard.edu/abs/1992A&A...259..143P}
}

@article{Vanlandingham_2005,
doi = {10.1086/431580},
url = {https://doi.org/10.1086/431580},
year = {2005},
month = {aug},
publisher = {},
volume = {130},
number = {2},
pages = {734},
author = {Vanlandingham, Karen M. and Schmidt, Gary D. and Eisenstein, Daniel J. and Harris, Hugh C. and Anderson, Scott F. and Hall, Patrick B. and Liebert, James and Schneider, Donald P. and Silvestri, Nicole M. and Stinson, Gregory S. and Wolfe, Michael A.},
title = {Magnetic White Dwarfs from the SDSS. II. The Second and Third Data Releases*},
journal = {The Astronomical Journal}
}

@article{Pshirkov:2009sf,
    author = "Pshirkov, M. S. and Baskaran, D.",
    title = "{Limits on High-Frequency Gravitational Wave Background from its interplay with Large Scale Magnetic Fields}",
    eprint = "0903.4160",
    archivePrefix = "arXiv",
    primaryClass = "gr-qc",
    doi = "10.1103/PhysRevD.80.042002",
    journal = "Phys. Rev. D",
    volume = "80",
    pages = "042002",
    year = "2009"
}

@article{Dandoy:2024oqg,
    author = "Dandoy, Virgile and Bert{\'o}lez-Mart{\'\i}nez, Toni and Costa, Francesco",
    title = "{High Frequency Gravitational Wave bounds from galactic neutron stars}",
    eprint = "2402.14092",
    archivePrefix = "arXiv",
    primaryClass = "gr-qc",
    doi = "10.1088/1475-7516/2024/12/023",
    journal = "JCAP",
    volume = "12",
    pages = "023",
    year = "2024"
}

@article{Hardy:2016kme,
    author = "Hardy, Edward and Lasenby, Robert",
    title = "{Stellar cooling bounds on new light particles: plasma mixing effects}",
    eprint = "1611.05852",
    archivePrefix = "arXiv",
    primaryClass = "hep-ph",
    doi = "10.1007/JHEP02(2017)033",
    journal = "JHEP",
    volume = "02",
    pages = "033",
    year = "2017"
}

@article{Gamow:1941gis,
    author = "Gamow, G. and Schoenberg, M.",
    title = "{Neutrino Theory of Stellar Collapse}",
    doi = "10.1103/PhysRev.59.539",
    journal = "Phys. Rev.",
    volume = "59",
    number = "7",
    pages = "539",
    year = "1941"
}

@article{Caputo:2024oqc,
    author = "Caputo, Andrea and Raffelt, Georg",
    title = "{Astrophysical Axion Bounds: The 2024 Edition}",
    eprint = "2401.13728",
    archivePrefix = "arXiv",
    primaryClass = "hep-ph",
    reportNumber = "MPP-2024-13, CERN-TH-2024-013",
    doi = "10.22323/1.454.0041",
    journal = "PoS",
    volume = "COSMICWISPers",
    pages = "041",
    year = "2024"
}

@article{Yakovlev:2000jp,
    author = "Yakovlev, D. G. and Kaminker, A. D. and Gnedin, Oleg Y. and Haensel, P.",
    title = "{Neutrino emission from neutron stars}",
    eprint = "astro-ph/0012122",
    archivePrefix = "arXiv",
    doi = "10.1016/S0370-1573(00)00131-9",
    journal = "Phys. Rept.",
    volume = "354",
    pages = "1",
    year = "2001"
}

@article{Yakovlev:2004iq,
    author = "Yakovlev, Dima G. and Pethick, C. J.",
    title = "{Neutron star cooling}",
    eprint = "astro-ph/0402143",
    archivePrefix = "arXiv",
    doi = "10.1146/annurev.astro.42.053102.134013",
    journal = "Ann. Rev. Astron. Astrophys.",
    volume = "42",
    pages = "169--210",
    year = "2004"
}

@article{Page:2005fq,
    author = "Page, Dany and Geppert, Ulrich and Weber, Fridolin",
    title = "{The Cooling of compact stars}",
    eprint = "astro-ph/0508056",
    archivePrefix = "arXiv",
    doi = "10.1016/j.nuclphysa.2005.09.019",
    journal = "Nucl. Phys. A",
    volume = "777",
    pages = "497--530",
    year = "2006"
}

@article{Lattimer:2004pg,
    author = "Lattimer, J. M. and Prakash, M.",
    title = "{The physics of neutron stars}",
    eprint = "astro-ph/0405262",
    archivePrefix = "arXiv",
    doi = "10.1126/science.1090720",
    journal = "Science",
    volume = "304",
    pages = "536--542",
    year = "2004"
}

@book{Haensel:2007yy,
    author = "Haensel, P. and Potekhin, A. Y. and Yakovlev, D. G.",
    title = "{Neutron stars 1: Equation of state and structure}",
    doi = "10.1007/978-0-387-47301-7",
    publisher = "Springer",
    address = "New York, USA",
    volume = "326",
    year = "2007"
}

@article{VitaliiGinzburg_1971,
doi = {10.1070/PU1971v014n02ABEH004446},
url = {https://doi.org/10.1070/PU1971v014n02ABEH004446},
year = {1971},
month = {feb},
publisher = {},
volume = {14},
number = {2},
pages = {83},
author = {Vitalii L Ginzburg},
title = {PULSARS (Theoretical Concepts)},
journal = {Soviet Physics Uspekhi},
}

@article{Page:2006ud,
    author = "Page, Dany and Reddy, Sanjay",
    title = "{Dense Matter in Compact Stars: Theoretical Developments and Observational Constraints}",
    eprint = "astro-ph/0608360",
    archivePrefix = "arXiv",
    doi = "10.1146/annurev.nucl.56.080805.140600",
    journal = "Ann. Rev. Nucl. Part. Sci.",
    volume = "56",
    pages = "327--374",
    year = "2006"
}

@article{Tetzlaff:2012rz,
    author = "Tetzlaff, Nina and Schmidt, Janos G. and Hohle, Markus M. and Neuhaeuser, Ralph",
    title = "{Neutron stars from young nearby associations the origin of RXJ1605.3+3249}",
    eprint = "1202.1388",
    archivePrefix = "arXiv",
    primaryClass = "astro-ph.GA",
    doi = "10.1071/AS11057",
    journal = "Publ. Astron. Soc. Austral.",
    volume = "29",
    pages = "98",
    year = "2012"
}

@article{Iwamoto:1984ir,
    author = "Iwamoto, N.",
    title = "{Axion Emission from Neutron Stars}",
    doi = "10.1103/PhysRevLett.53.1198",
    journal = "Phys. Rev. Lett.",
    volume = "53",
    pages = "1198--1201",
    year = "1984"
}

@article{Brinkmann:1988vi,
    author = "Brinkmann, Ralf Peter and Turner, Michael S.",
    title = "{Numerical Rates for Nucleon-Nucleon Axion Bremsstrahlung}",
    reportNumber = "FERMILAB-PUB-88-029-A",
    doi = "10.1103/PhysRevD.38.2338",
    journal = "Phys. Rev. D",
    volume = "38",
    pages = "2338",
    year = "1988"
}

@article{Iwamoto:1992jp,
    author = "Iwamoto, Naoki",
    title = "{Nucleon-nucleon bremsstrahlung of axions and pseudoscalar particles from neutron star matter}",
    reportNumber = "FPRINT-92-29",
    doi = "10.1103/PhysRevD.64.043002",
    journal = "Phys. Rev. D",
    volume = "64",
    pages = "043002",
    year = "2001"
}

@article{Raffelt:2006cw,
    author = "Raffelt, Georg G.",
    editor = "Kuster, Markus and Raffelt, Georg and Beltran, Berta",
    title = "{Astrophysical axion bounds}",
    eprint = "hep-ph/0611350",
    archivePrefix = "arXiv",
    reportNumber = "MPP-2006-172",
    doi = "10.1007/978-3-540-73518-2_3",
    journal = "Lect. Notes Phys.",
    volume = "741",
    pages = "51--71",
    year = "2008"
}

@article{Fischer:2016cyd,
    author = "Fischer, Tobias and Chakraborty, Sovan and Giannotti, Maurizio and Mirizzi, Alessandro and Payez, Alexandre and Ringwald, Andreas",
    title = "{Probing axions with the neutrino signal from the next galactic supernova}",
    eprint = "1605.08780",
    archivePrefix = "arXiv",
    primaryClass = "astro-ph.HE",
    reportNumber = "DESY-16-094",
    doi = "10.1103/PhysRevD.94.085012",
    journal = "Phys. Rev. D",
    volume = "94",
    number = "8",
    pages = "085012",
    year = "2016"
}

@article{Chang:2018rso,
    author = "Chang, Jae Hyeok and Essig, Rouven and McDermott, Samuel D.",
    title = "{Supernova 1987A Constraints on Sub-GeV Dark Sectors, Millicharged Particles, the QCD Axion, and an Axion-like Particle}",
    eprint = "1803.00993",
    archivePrefix = "arXiv",
    primaryClass = "hep-ph",
    reportNumber = "YITP-SB-18-01, FERMILAB-PUB-17-432-T",
    doi = "10.1007/JHEP09(2018)051",
    journal = "JHEP",
    volume = "09",
    pages = "051",
    year = "2018"
}

@article{Carenza:2019pxu,
    author = "Carenza, Pierluca and Fischer, Tobias and Giannotti, Maurizio and Guo, Gang and Mart{\'\i}nez-Pinedo, Gabriel and Mirizzi, Alessandro",
    title = "{Improved axion emissivity from a supernova via nucleon-nucleon bremsstrahlung}",
    eprint = "1906.11844",
    archivePrefix = "arXiv",
    primaryClass = "hep-ph",
    doi = "10.1088/1475-7516/2019/10/016",
    journal = "JCAP",
    volume = "10",
    number = "10",
    pages = "016",
    year = "2019",
    note = "[Erratum: JCAP 05, E01 (2020)]"
}

@article{Carenza:2020cis,
    author = "Carenza, Pierluca and Fore, Bryce and Giannotti, Maurizio and Mirizzi, Alessandro and Reddy, Sanjay",
    title = "{Enhanced Supernova Axion Emission and its Implications}",
    eprint = "2010.02943",
    archivePrefix = "arXiv",
    primaryClass = "hep-ph",
    reportNumber = "INT-PUB-20-039",
    doi = "10.1103/PhysRevLett.126.071102",
    journal = "Phys. Rev. Lett.",
    volume = "126",
    number = "7",
    pages = "071102",
    year = "2021"
}

@article{Suzuki:2021ium,
    author = "Suzuki, Hiromasa and Bamba, Aya and Shibata, Shinpei",
    title = "{Quantitative Age Estimation of Supernova Remnants and Associated Pulsars}",
    eprint = "2104.10052",
    archivePrefix = "arXiv",
    primaryClass = "astro-ph.HE",
    doi = "10.3847/1538-4357/abfb02",
    journal = "Astrophys. J.",
    volume = "914",
    number = "2",
    pages = "103",
    year = "2021"
}

@article{Potekhin:2020ttj,
    author = "Potekhin, A. Y. and Zyuzin, D. A. and Yakovlev, D. G. and Beznogov, M. V. and Shibanov, Yu. A.",
    title = "{Thermal luminosities of cooling neutron stars}",
    eprint = "2006.15004",
    archivePrefix = "arXiv",
    primaryClass = "astro-ph.HE",
    doi = "10.1093/mnras/staa1871",
    journal = "Mon. Not. Roy. Astron. Soc.",
    volume = "496",
    number = "4",
    pages = "5052--5071",
    year = "2020"
}

@article{Pires:2019qsk,
    author = "Pires, A. M. and Schwope, A. D. and Haberl, F. and Zavlin, V. E. and Motch, C. and Zane, S.",
    title = "{A deep XMM-Newton look on the thermally emitting isolated neutron star RX J1605.3+3249}",
    eprint = "1901.08533",
    archivePrefix = "arXiv",
    primaryClass = "astro-ph.HE",
    doi = "10.1051/0004-6361/201834801",
    journal = "Astron. Astrophys.",
    volume = "623",
    pages = "A73",
    year = "2019"
}

@article{Potekhin:2015qsa,
    author = "Potekhin, A. Y. and Pons, J. A. and Page, Dany",
    title = "{Neutron stars - cooling and transport}",
    eprint = "1507.06186",
    archivePrefix = "arXiv",
    primaryClass = "astro-ph.HE",
    doi = "10.1007/s11214-015-0180-9",
    journal = "Space Sci. Rev.",
    volume = "191",
    number = "1-4",
    pages = "239--291",
    year = "2015"
}

@article{Potekhin:2010aii,
    author = "Potekhin, Aleksandr Y.",
    title = "{The physics of neutron stars}",
    eprint = "1102.5735",
    archivePrefix = "arXiv",
    primaryClass = "astro-ph.SR",
    doi = "10.3367/UFNe.0180.201012c.1279",
    journal = "Phys. Usp.",
    volume = "53",
    pages = "1235--1256",
    year = "2010"
}

@article{Ito:2023fcr,
    author = "Ito, Asuka and Kohri, Kazunori and Nakayama, Kazunori",
    title = "{Probing high frequency gravitational waves with pulsars}",
    eprint = "2305.13984",
    archivePrefix = "arXiv",
    primaryClass = "gr-qc",
    reportNumber = "KEK-QUP-2023-0011, KEK-TH-2529, KEK-Cosmo-0314, TU-1192",
    doi = "10.1103/PhysRevD.109.063026",
    journal = "Phys. Rev. D",
    volume = "109",
    number = "6",
    pages = "063026",
    year = "2024"
}

@article{McDonald:2024nxj,
    author = "McDonald, Jamie I. and Ellis, Sebastian A. R.",
    title = "{Resonant conversion of gravitational waves in neutron star magnetospheres}",
    eprint = "2406.18634",
    archivePrefix = "arXiv",
    primaryClass = "hep-ph",
    doi = "10.1103/PhysRevD.110.103003",
    journal = "Phys. Rev. D",
    volume = "110",
    number = "10",
    pages = "103003",
    year = "2024"
}

@article{Domcke:2020yzq,
    author = "Domcke, Valerie and Garcia-Cely, Camilo",
    title = "{Potential of radio telescopes as high-frequency gravitational wave detectors}",
    eprint = "2006.01161",
    archivePrefix = "arXiv",
    primaryClass = "astro-ph.CO",
    reportNumber = "DESY-20-097, CERN-TH-2020-082",
    doi = "10.1103/PhysRevLett.126.021104",
    journal = "Phys. Rev. Lett.",
    volume = "126",
    number = "2",
    pages = "021104",
    year = "2021"
}

@article{Dessert:2022yqq,
    author = "Dessert, Christopher and Dunsky, David and Safdi, Benjamin R.",
    title = "{Upper limit on the axion-photon coupling from magnetic white dwarf polarization}",
    eprint = "2203.04319",
    archivePrefix = "arXiv",
    primaryClass = "hep-ph",
    doi = "10.1103/PhysRevD.105.103034",
    journal = "Phys. Rev. D",
    volume = "105",
    number = "10",
    pages = "103034",
    year = "2022"
}

@article{McDonald:2023ohd,
    author = "McDonald, J. I. and Garbrecht, B. and Millington, P.",
    title = "{Axion-photon conversion in 3D media and astrophysical plasmas}",
    eprint = "2307.11812",
    archivePrefix = "arXiv",
    primaryClass = "hep-ph",
    doi = "10.1088/1475-7516/2023/12/031",
    journal = "JCAP",
    volume = "12",
    pages = "031",
    year = "2023"
}

@article{Dessert:2019sgw,
    author = "Dessert, Christopher and Long, Andrew J. and Safdi, Benjamin R.",
    title = "{X-ray Signatures of Axion Conversion in Magnetic White Dwarf Stars}",
    eprint = "1903.05088",
    archivePrefix = "arXiv",
    primaryClass = "hep-ph",
    reportNumber = "LCTP-19-05",
    doi = "10.1103/PhysRevLett.123.061104",
    journal = "Phys. Rev. Lett.",
    volume = "123",
    number = "6",
    pages = "061104",
    year = "2019"
}

@article{Fiorillo:2025zzx,
    author = "Fiorillo, Damiano F. G. and Lella, Alessandro and O'Hare, Ciaran A. J. and Vitagliano, Edoardo",
    title = "{Leading Bounds on Micrometer to Picometer Fifth Forces from Neutron Star Cooling}",
    eprint = "2506.19906",
    archivePrefix = "arXiv",
    primaryClass = "hep-ph",
    reportNumber = "BARI-TH/776-25",
    doi = "10.1103/tlqz-713s",
    journal = "Phys. Rev. Lett.",
    volume = "135",
    number = "21",
    pages = "211003",
    year = "2025"
}

@article{Lella:2023bfb,
    author = "Lella, Alessandro and Carenza, Pierluca and Co', Giampaolo and Lucente, Giuseppe and Giannotti, Maurizio and Mirizzi, Alessandro and Rauscher, Thomas",
    title = "{Getting the most on supernova axions}",
    eprint = "2306.01048",
    archivePrefix = "arXiv",
    primaryClass = "hep-ph",
    doi = "10.1103/PhysRevD.109.023001",
    journal = "Phys. Rev. D",
    volume = "109",
    number = "2",
    pages = "023001",
    year = "2024"
}

@article{Lella:2022uwi,
    author = "Lella, Alessandro and Carenza, Pierluca and Lucente, Giuseppe and Giannotti, Maurizio and Mirizzi, Alessandro",
    title = "{Protoneutron stars as cosmic factories for massive axionlike particles}",
    eprint = "2211.13760",
    archivePrefix = "arXiv",
    primaryClass = "hep-ph",
    doi = "10.1103/PhysRevD.107.103017",
    journal = "Phys. Rev. D",
    volume = "107",
    number = "10",
    pages = "103017",
    year = "2023"
}

@article{Fiorillo:2025gnd,
    author = "Fiorillo, Damiano F. G. and Gil Muyor, {\'A}ngel and Janka, Hans-Thomas and Raffelt, Georg G. and Vitagliano, Edoardo",
    title = "{Axion-photon conversion in transient compact stars: Systematics, constraints, and opportunities}",
    eprint = "2509.13322",
    archivePrefix = "arXiv",
    primaryClass = "hep-ph",
    doi = "10.1088/1475-7516/2026/03/053",
    journal = "JCAP",
    volume = "03",
    pages = "053",
    year = "2026"
}

@article{Springmann:2024ret,
    author = "Springmann, Konstantin and Stadlbauer, Michael and Stelzl, Stefan and Weiler, Andreas",
    title = "{Universal bound on QCD axions from supernovae}",
    eprint = "2410.19902",
    archivePrefix = "arXiv",
    primaryClass = "hep-ph",
    reportNumber = "TUM-HEP-1531/24",
    doi = "10.1103/18t2-1w3b",
    journal = "Phys. Rev. D",
    volume = "112",
    number = "7",
    pages = "075009",
    year = "2025"
}

@article{Buschmann:2021juv,
    author = "Buschmann, Malte and Dessert, Christopher and Foster, Joshua W. and Long, Andrew J. and Safdi, Benjamin R.",
    title = "{Upper Limit on the QCD Axion Mass from Isolated Neutron Star Cooling}",
    eprint = "2111.09892",
    archivePrefix = "arXiv",
    primaryClass = "hep-ph",
    doi = "10.1103/PhysRevLett.128.091102",
    journal = "Phys. Rev. Lett.",
    volume = "128",
    number = "9",
    pages = "091102",
    year = "2022"
}

@article{Gaia:2021gsq,
    author = "Brown, A. G. A. and others",
    collaboration = "Gaia",
    title = "{Gaia Early Data Release 3}",
    eprint = "2012.01533",
    archivePrefix = "arXiv",
    primaryClass = "astro-ph.GA",
    doi = "10.1051/0004-6361/202039657",
    journal = "Astron. Astrophys.",
    volume = "649",
    pages = "A1",
    year = "2021",
    note = "[Erratum: Astron.Astrophys. 650, C3 (2021)]"
}

@article{1995MNRAS.277..971B,
	Adsurl = {http://adsabs.harvard.edu/abs/1995MNRAS.277..971B},
	Author = {{Barstow}, M.~A. and {Jordan}, S. and {O'Donoghue}, D. and {Burleigh}, M.~R. and {Napiwotzki}, R. and {Harrop-Allin}, M.~K.},
	Doi = {10.1093/mnras/277.3.971},
	Journal = {MNRAS},
	Month = dec,
	Pages = {971-985},
	Title = {{RE J0317-853: the hottest known highly magnetic DA white dwarf}},
	Volume = 277,
	Year = 1995}

@article{2010A&A...524A..36K,
	Adsurl = {https://ui.adsabs.harvard.edu/\#abs/2010A&A...524A..36K},
	Archiveprefix = {arXiv},
	Author = {{K{\"u}lebi}, B. and {Jordan}, S. and {Nelan}, E. and {Bastian}, U. and {Altmann}, M.},
	Doi = {10.1051/0004-6361/201015237},
	Eid = {A36},
	Eprint = {1007.4978},
	Journal = {Astron. Astrophys.},
	Month = Dec,
	Pages = {A36},
	Primaryclass = {astro-ph.SR},
	Title = {{Constraints on the origin of the massive, hot, and rapidly rotating magnetic white dwarf RE J 0317-853 from an HST parallax measurement}},
	Volume = {524},
	Year = 2010}

@article{doi:10.1093/pasj/65.4.73,
	Author = {Harayama, Atsushi and Terada, Yukikatsu and Ishida, Manabu and Hayashi, Takayuki and Bamba, Aya and Tashiro, Makoto S.},
	Journal = {Publications of the Astronomical Society of Japan},
	Number = {4},
	Pages = {73},
	Title = {Search for Non-Thermal Emissions from an Isolated Magnetic White Dwarf, EUVE J0317 855, with Suzaku},
	Volume = {65},
	Year = {2013}}

@article{McDonald:2024uuh,
    author = "McDonald, J. I. and Millington, P.",
    title = "{Axion-photon mixing in 3D: classical equations and geometric optics}",
    eprint = "2407.11192",
    archivePrefix = "arXiv",
    primaryClass = "hep-ph",
    doi = "10.1088/1475-7516/2024/09/072",
    journal = "JCAP",
    volume = "09",
    pages = "072",
    year = "2024"
}

@article{Burleigh:1998pqa,
	Archiveprefix = {arXiv},
	Author = {Burleigh, M. R. and Jordan, S. and Schweizer, W.},
	Doi = {10.1086/311794},
	Eprint = {astro-ph/9810109},
	Journal = {Astrophys. J.},
	Pages = {L37},
	Primaryclass = {astro-ph},
	Slaccitation = {%%CITATION = ASTRO-PH/9810109;%%},
	Title = {{Phase-resolved far-ultraviolet hst spectroscopy of the peculiar magnetic white dwarf re j0317-853}},
	Volume = {510},
	Year = {1999}}

@article{Lai:2006af,
    author = "Lai, Dong and Heyl, Jeremy",
    title = "{Probing Axions with Radiation from Magnetic Stars}",
    eprint = "astro-ph/0609775",
    archivePrefix = "arXiv",
    doi = "10.1103/PhysRevD.74.123003",
    journal = "Phys. Rev. D",
    volume = "74",
    pages = "123003",
    year = "2006"
}

@article{Gill:2011yp,
    author = "Gill, Ramandeep and Heyl, Jeremy S.",
    title = "{Constraining the photon-axion coupling constant with magnetic white dwarfs}",
    eprint = "1105.2083",
    archivePrefix = "arXiv",
    primaryClass = "astro-ph.HE",
    doi = "10.1103/PhysRevD.84.085001",
    journal = "Phys. Rev. D",
    volume = "84",
    pages = "085001",
    year = "2011"
}

@article{Gines:2024ekm,
    author = "Gin{\'e}s, Estanis Utrilla and Noordhuis, Dion and Weniger, Christoph and Witte, Samuel J.",
    title = "{Numerical analysis of resonant axion-photon mixing}",
    eprint = "2405.08865",
    archivePrefix = "arXiv",
    primaryClass = "hep-ph",
    doi = "10.1103/PhysRevD.110.083007",
    journal = "Phys. Rev. D",
    volume = "110",
    number = "8",
    pages = "083007",
    year = "2024"
}

@article{Battye:2023oac,
    author = "Battye, R. A. and Keith, M. J. and McDonald, J. I. and Srinivasan, S. and Stappers, B. W. and Weltevrede, P.",
    title = "{Searching for time-dependent axion dark matter signals in pulsars}",
    eprint = "2303.11792",
    archivePrefix = "arXiv",
    primaryClass = "astro-ph.CO",
    doi = "10.1103/PhysRevD.108.063001",
    journal = "Phys. Rev. D",
    volume = "108",
    number = "6",
    pages = "063001",
    year = "2023"
}

@article{McDonald:2023shx,
    author = "McDonald, J. I. and Witte, S. J.",
    title = "{Generalized ray tracing for axions in astrophysical plasmas}",
    eprint = "2309.08655",
    archivePrefix = "arXiv",
    primaryClass = "hep-ph",
    doi = "10.1103/PhysRevD.108.103021",
    journal = "Phys. Rev. D",
    volume = "108",
    number = "10",
    pages = "103021",
    year = "2023"
}

@article{Chadha-Day:2022inf,
    author = {Chadha-Day, Francesca and Garbrecht, Bj\"orn and McDonald, Jamie},
    title = "{Superradiance in stars: non-equilibrium approach to damping of fields in stellar media}",
    eprint = "2207.07662",
    archivePrefix = "arXiv",
    primaryClass = "hep-ph",
    reportNumber = "IPPP/22/46",
    doi = "10.1088/1475-7516/2022/12/008",
    journal = "JCAP",
    volume = "12",
    pages = "008",
    year = "2022"
}

@article{Battye:2021xvt,
    author = "Battye, R. A. and Garbrecht, B. and McDonald, J. I. and Srinivasan, S.",
    title = "{Radio line properties of axion dark matter conversion in neutron stars}",
    eprint = "2104.08290",
    archivePrefix = "arXiv",
    primaryClass = "hep-ph",
    doi = "10.1007/JHEP09(2021)105",
    journal = "JHEP",
    volume = "09",
    pages = "105",
    year = "2021"
}

@article{Noordhuis:2022ljw,
    author = "Noordhuis, Dion and Prabhu, Anirudh and Witte, Samuel J. and Chen, Alexander Y. and Cruz, F\'abio and Weniger, Christoph",
    title = "{Novel Constraints on Axions Produced in Pulsar Polar Cap Cascades}",
    eprint = "2209.09917",
    archivePrefix = "arXiv",
    primaryClass = "hep-ph",
    month = "9",
    year = "2022"
}

@article{Dessert:2021bkv,
    author = "Dessert, Christopher and Long, Andrew J. and Safdi, Benjamin R.",
    title = "{No evidence for axions from Chandra observation of magnetic white dwarf}",
    eprint = "2104.12772",
    archivePrefix = "arXiv",
    primaryClass = "hep-ph",
    month = "4",
    year = "2021"
}

@article{Huang:2018lxq,
    author = "Huang, Fa Peng and Kadota, Kenji and Sekiguchi, Toyokazu and Tashiro, Hiroyuki",
    title = "{Radio telescope search for the resonant conversion of cold dark matter axions from the magnetized astrophysical sources}",
    eprint = "1803.08230",
    archivePrefix = "arXiv",
    primaryClass = "hep-ph",
    reportNumber = "CTPU-PTC-18-08, RESCEU-7-18",
    doi = "10.1103/PhysRevD.97.123001",
    journal = "Phys. Rev. D",
    volume = "97",
    number = "12",
    pages = "123001",
    year = "2018"
}

@ARTICLE{Battye2022,
       author = {{Battye}, R.~A. and {Darling}, J. and {McDonald}, J.~I. and {Srinivasan}, S.},
        title = "{Towards robust constraints on axion dark matter using PSR J1745-2900}",
      journal = {\prd},
         year = 2022,
        month = jan,
       volume = {105},
       number = {2},
          eid = {L021305},
        pages = {L021305},
          doi = {10.1103/PhysRevD.105.L021305},
archivePrefix = {arXiv},
       eprint = {2107.01225},
 primaryClass = {astro-ph.CO},
       adsurl = {https://ui.adsabs.harvard.edu/abs/2022PhRvD.105b1305B}
}

@article{Witte:2021arp,
     author = {{Witte}, Samuel J. and {Noordhuis}, Dion and {Edwards}, Thomas D.~P. and {Weniger}, Christoph},
        title = "{Axion-photon conversion in neutron star magnetospheres: The role of the plasma in the Goldreich-Julian model}",
      journal = {\prd},
         year = 2021,
        month = nov,
       volume = {104},
       number = {10},
          eid = {103030},
        pages = {103030},
          doi = {10.1103/PhysRevD.104.103030},
archivePrefix = {arXiv},
       eprint = {2104.07670},
 primaryClass = {hep-ph},
       adsurl = {https://ui.adsabs.harvard.edu/abs/2021PhRvD.104j3030W}
}

@article{Arvanitaki:2009fg,
    author = "Arvanitaki, Asimina and Dimopoulos, Savas and Dubovsky, Sergei and Kaloper, Nemanja and March-Russell, John",
    title = "{String Axiverse}",
    eprint = "0905.4720",
    archivePrefix = "arXiv",
    primaryClass = "hep-th",
    doi = "10.1103/PhysRevD.81.123530",
    journal = "Phys. Rev. D",
    volume = "81",
    pages = "123530",
    year = "2010"
}

@article{Prabhu:2020yif,
    author = "Prabhu, Anirudh and Rapidis, Nicholas M.",
    title = "{Resonant Conversion of Dark Matter Oscillons in Pulsar Magnetospheres}",
    eprint = "2005.03700",
    archivePrefix = "arXiv",
    primaryClass = "astro-ph.CO",
    doi = "10.1088/1475-7516/2020/10/054",
    journal = "JCAP",
    volume = "10",
    pages = "054",
    year = "2020"
}

@ARTICLE{FosterSETI2022,
       author = {{Foster}, Joshua W. and {Witte}, Samuel J. and {Lawson}, Matthew and {Linden}, Tim and {Gajjar}, Vishal and {Weniger}, Christoph and {Safdi}, Benjamin R.},
        title = "{Extraterrestrial Axion Search with the Breakthrough Listen Galactic Center Survey}",
      journal = {arXiv e-prints},
         year = 2022,
        month = feb,
          eid = {arXiv:2202.08274},
        pages = {arXiv:2202.08274},
archivePrefix = {arXiv},
       eprint = {2202.08274},
 primaryClass = {astro-ph.CO},
       adsurl = {https://ui.adsabs.harvard.edu/abs/2022arXiv220208274F}
}

@article{Foster:2020pgt,
    author = "Foster, Joshua W. and Kahn, Yonatan and Macias, Oscar and Sun, Zhiquan and Eatough, Ralph P. and Kondratiev, Vladislav I. and Peters, Wendy M. and Weniger, Christoph and Safdi, Benjamin R.",
    title = "{Green Bank and Effelsberg Radio Telescope Searches for Axion Dark Matter Conversion in Neutron Star Magnetospheres}",
    eprint = "2004.00011",
    archivePrefix = "arXiv",
    primaryClass = "astro-ph.CO",
    reportNumber = "LCTP-20-04",
    doi = "10.1103/PhysRevLett.125.171301",
    journal = "Phys. Rev. Lett.",
    volume = "125",
    number = "17",
    pages = "171301",
    year = "2020"
}

@article{Darling:2020uyo,
    author = "Darling, Jeremy",
    title = "{New Limits on Axionic Dark Matter from the Magnetar PSR J1745-2900}",
    eprint = "2008.11188",
    archivePrefix = "arXiv",
    primaryClass = "astro-ph.CO",
    doi = "10.3847/2041-8213/abb23f",
    journal = "Astrophys. J. Lett.",
    volume = "900",
    number = "2",
    pages = "L28",
    year = "2020"
}

@article{Leroy:2019ghm,
    author = {Leroy, Mika\"el and Chianese, Marco and Edwards, Thomas D. P. and Weniger, Christoph},
    title = "{Radio Signal of Axion-Photon Conversion in Neutron Stars: A Ray Tracing Analysis}",
    eprint = "1912.08815",
    archivePrefix = "arXiv",
    primaryClass = "hep-ph",
    doi = "10.1103/PhysRevD.101.123003",
    journal = "Phys. Rev. D",
    volume = "101",
    number = "12",
    pages = "123003",
    year = "2020"
}

@article{Arza:2026rsl,
    author = "Arza, A. and others",
    title = "{The COSMIC WISPers White Paper: The physics case for Weakly Interacting Slim Particles}",
    eprint = "2603.03433",
    archivePrefix = "arXiv",
    primaryClass = "hep-ph",
    reportNumber = "BARI-TH/784-26, CERN-TH-2026-016, IPPP/26/13, IFT-UAM/CSIC-26-13, KCL-PH-TH/2026-04, KEK-Cosmo-0411, KEK-TH-2804, LAPTH-008/26, MPP-2026-21, RESCEU-5/26, SLAC-PUB-260219, ST/T006994/1, ST/Y004531/1",
    month = "3",
    year = "2026"
}

@article{Igoshev:2021ewx,
    author = "Igoshev, Andrei P. and Popov, Sergei B. and Hollerbach, Rainer",
    title = "{Evolution of Neutron Star Magnetic Fields}",
    eprint = "2109.05584",
    archivePrefix = "arXiv",
    primaryClass = "astro-ph.HE",
    doi = "10.3390/universe7090351",
    journal = "Universe",
    volume = "7",
    number = "9",
    pages = "351",
    year = "2021"
}

@article{Battye:2019aco,
    author = "Battye, Richard A. and Garbrecht, Bjoern and McDonald, Jamie I. and Pace, Francesco and Srinivasan, Sankarshana",
    title = "{Dark matter axion detection in the radio/mm-waveband}",
    eprint = "1910.11907",
    archivePrefix = "arXiv",
    primaryClass = "astro-ph.CO",
    doi = "10.1103/PhysRevD.102.023504",
    journal = "Phys. Rev. D",
    volume = "102",
    number = "2",
    pages = "023504",
    year = "2020"
}

@article{Safdi:2018oeu,
      author         = "Safdi, Benjamin R. and Sun, Zhiquan and Chen, Alexander
                        Y.",
      title          = "{Detecting Axion Dark Matter with Radio Lines from
                        Neutron Star Populations}",
      journal        = "Phys. Rev.",
      volume         = "D99",
      year           = "2019",
      number         = "12",
      pages          = "123021",
      doi            = "10.1103/PhysRevD.99.123021",
      eprint         = "1811.01020",
      archivePrefix  = "arXiv",
      primaryClass   = "astro-ph.CO",
      reportNumber   = "LCTP-18-22",
      SLACcitation   = "%%CITATION = ARXIV:1811.01020;%%"
}

@article{Pshirkov:2007st,
      author         = "Pshirkov, M. S. and Popov, S. B.",
      title          = "{Conversion of Dark matter axions to photons in
                        magnetospheres of neutron stars}",
      journal        = "J. Exp. Theor. Phys.",
      volume         = "108",
      year           = "2009",
      pages          = "384-388",
      doi            = "10.1134/S1063776109030030",
      eprint         = "0711.1264",
      archivePrefix  = "arXiv",
      primaryClass   = "astro-ph",
      SLACcitation   = "%%CITATION = ARXIV:0711.1264;%%"
}

@article{Raffelt:1987im,
      author         = "Raffelt, Georg and Stodolsky, Leo",
      title          = "{Mixing of the Photon with Low Mass Particles}",
      journal        = "Phys. Rev.",
      volume         = "D37",
      year           = "1988",
      pages          = "1237",
      doi            = "10.1103/PhysRevD.37.1237",
      reportNumber   = "MPI-PAE/PTh-54/87",
      SLACcitation   = "%%CITATION = PHRVA,D37,1237;%%"
}

@PREAMBLE{
 "\providecommand{\noopsort}[1]{}" 
 # "\providecommand{\singleletter}[1]{#1}%" 
}

@ARTICLE{darling2020apj,
       author = {{Darling}, Jeremy},
        title = "{New Limits on Axionic Dark Matter from the Magnetar PSR J1745-2900}",
      journal = {\apjl},
         year = 2020,
        month = sep,
       volume = {900},
       number = {2},
          eid = {L28},
        pages = {L28},
          doi = {10.3847/2041-8213/abb23f},
archivePrefix = {arXiv},
       eprint = {2008.11188},
 primaryClass = {astro-ph.CO},
       adsurl = {https://ui.adsabs.harvard.edu/abs/2020ApJ...900L..28D}
}

@ARTICLE{foster2020,
       author = {{Foster}, Joshua W. and {Kahn}, Yonatan and {Macias}, Oscar and {Sun}, Zhiquan and {Eatough}, Ralph P. and {Kondratiev}, Vladislav I. and {Peters}, Wendy M. and {Weniger}, Christoph and {Safdi}, Benjamin R.},
        title = "{Green Bank and Effelsberg Radio Telescope Searches for Axion Dark Matter Conversion in Neutron Star Magnetospheres}",
      journal = {\prl},
         year = 2020,
        month = oct,
       volume = {125},
       number = {17},
          eid = {171301},
        pages = {171301},
          doi = {10.1103/PhysRevLett.125.171301},
archivePrefix = {arXiv},
       eprint = {2004.00011},
 primaryClass = {astro-ph.CO},
       adsurl = {https://ui.adsabs.harvard.edu/abs/2020PhRvL.125q1301F}
}

@article{Goldreich:1969sb,
    author = "Goldreich, Peter and Julian, William H.",
    title = "{Pulsar electrodynamics}",
    doi = "10.1086/150119",
    journal = "Astrophys. J.",
    volume = "157",
    pages = "869",
    year = "1969"
}

@ARTICLE{Peccei1977,

       author = {{Peccei}, R.~D. and {Quinn}, Helen R.},
       title  = "{CP Conservation in the Presence of Pseudoparticles}",
      journal = {\prl},
         year = 1977,
        month = jun,
       volume = {38},
       number = {25},
        pages = {1440-1443},
          doi = {10.1103/PhysRevLett.38.1440},
       adsurl = {https://ui.adsabs.harvard.edu/abs/1977PhRvL..38.1440P}
}

@article{Crewther:1979pi,
    author = "Crewther, R. J. and Di Vecchia, P. and Veneziano, G. and Witten, Edward",
    title = "{Chiral Estimate of the Electric Dipole Moment of the Neutron in Quantum Chromodynamics}",
    reportNumber = "CERN-TH-2735",
    doi = "10.1016/0370-2693(79)90128-X",
    journal = "Phys. Lett. B",
    volume = "88",
    pages = "123",
    year = "1979",
    note = "[Erratum: Phys.Lett.B 91, 487 (1980)]"
}

@article{Sannino:2026wgx,
    author = "Sannino, Francesco",
    title = "{Strong CP and the QCD Axion: Lecture Notes via Effective Field Theory}",
    eprint = "2601.19735",
    archivePrefix = "arXiv",
    primaryClass = "hep-ph",
    month = "1",
    year = "2026"
}

@article{Arvanitaki:2010sy,
    author = "Arvanitaki, Asimina and Dubovsky, Sergei",
    title = "{Exploring the String Axiverse with Precision Black Hole Physics}",
    eprint = "1004.3558",
    archivePrefix = "arXiv",
    primaryClass = "hep-th",
    doi = "10.1103/PhysRevD.83.044026",
    journal = "Phys. Rev. D",
    volume = "83",
    pages = "044026",
    year = "2011"
}

@article{Witte:2024drg,
    author = "Witte, Samuel J. and Mummery, Andrew",
    title = "{Stepping up superradiance constraints on axions}",
    eprint = "2412.03655",
    archivePrefix = "arXiv",
    primaryClass = "hep-ph",
    doi = "10.1103/PhysRevD.111.083044",
    journal = "Phys. Rev. D",
    volume = "111",
    number = "8",
    pages = "083044",
    year = "2025"
}

@article{Baryakhtar:2020gao,
    author = "Baryakhtar, Masha and Galanis, Marios and Lasenby, Robert and Simon, Olivier",
    title = "{Black hole superradiance of self-interacting scalar fields}",
    eprint = "2011.11646",
    archivePrefix = "arXiv",
    primaryClass = "hep-ph",
    doi = "10.1103/PhysRevD.103.095019",
    journal = "Phys. Rev. D",
    volume = "103",
    number = "9",
    pages = "095019",
    year = "2021"
}

@article{Leaver:1985ax,
    author = "Leaver, E. W.",
    title = "{An Analytic representation for the quasi normal modes of Kerr black holes}",
    doi = "10.1098/rspa.1985.0119",
    journal = "Proc. Roy. Soc. Lond. A",
    volume = "402",
    pages = "285--298",
    year = "1985"
}

@article{Zouros:1979iw,
    author = "Zouros, T. J. M. and Eardley, D. M.",
    title = "{INSTABILITIES OF MASSIVE SCALAR PERTURBATIONS OF A ROTATING BLACK HOLE}",
    doi = "10.1016/0003-4916(79)90237-9",
    journal = "Annals Phys.",
    volume = "118",
    pages = "139--155",
    year = "1979"
}

@ARTICLE{Zeldovich:1971,
       author = {{Zel'dovich}, Ya. B.},
        title = "{Generation of Waves by a Rotating Body}",
      journal = {Pis’ma Zh. Eksp. Teor. Fiz.},
         year = 1971,
       volume = {14},
        pages = {280},
        note  = "[\href{http://www.jetpletters.ru/ps/1604/article_24607.pdf}{JETP Lett. 14 (1971) 180}]",
       adsurl = {https://ui.adsabs.harvard.edu/abs/1971JETPL..14..180Z}
}

@ARTICLE{Zeldovich:1972,
       author = {{Zel'dovich}, Ya. B.},
        title = "{Amplification of Cylindrical Electromagnetic Waves Reflected from a Rotating Body}",
      journal = {Zh. Eksp. Teor. Fiz.},
         year = 1972,
       volume = {62},
        pages = {2076},
        note  = "[\href{http://jetp.ras.ru/cgi-bin/dn/e_035_06_1085.pdf}{Sov. Phys. JETP 35 (1972) 1085}]",
       adsurl = {https://ui.adsabs.harvard.edu/abs/1971JETPL..14..180Z}
}

@article{Endlich:2016jgc,
    author = "Endlich, Solomon and Penco, Riccardo",
    title = "{A Modern Approach to Superradiance}",
    eprint = "1609.06723",
    archivePrefix = "arXiv",
    primaryClass = "hep-th",
    doi = "10.1007/JHEP05(2017)052",
    journal = "JHEP",
    volume = "05",
    pages = "052",
    year = "2017"
}

@article{Detweiler:1980uk,
    author = "Detweiler, Steven L.",
    title = "{KLEIN-GORDON EQUATION AND ROTATING BLACK HOLES}",
    doi = "10.1103/PhysRevD.22.2323",
    journal = "Phys. Rev. D",
    volume = "22",
    pages = "2323--2326",
    year = "1980"
}

@article{Althaus:2010pi,
    author = "Althaus, Leandro G. and Corsico, Alejandro H. and Isern, Jordi and a-Berro, Enrique Garci",
    title = "{Evolutionary and pulsational properties of white dwarf stars}",
    eprint = "1007.2659",
    archivePrefix = "arXiv",
    primaryClass = "astro-ph.SR",
    doi = "10.1007/s00159-010-0033-1",
    journal = "Astron. Astrophys. Rev.",
    volume = "18",
    pages = "471--566",
    year = "2010"
}

@article{Winget:2003xf,
    author = "Winget, D. E. and Sullivan, D. J. and Metcalfe, T. S. and Kawaler, S. D. and Montgomery, M. H.",
    title = "{A strong test of electro-weak theory using pulsating db white dwarf stars as plasmon neutrino detectors}",
    eprint = "astro-ph/0312303",
    archivePrefix = "arXiv",
    doi = "10.1086/382591",
    journal = "Astrophys. J. Lett.",
    volume = "602",
    pages = "L109--L112",
    year = "2004"
}

@article{Isern:1997na,
    author = "Isern, J. and Mochkovitch, R. and Garcia-Berro, E. and Hernanz, Margarita",
    title = "{The physics of crystallizing white dwarfs}",
    eprint = "astro-ph/9703028",
    archivePrefix = "arXiv",
    doi = "10.1086/304425",
    journal = "Astrophys. J.",
    volume = "485",
    pages = "308",
    year = "1997"
}

@ARTICLE{1994ApJ...434..641S,
       author = {{Segretain}, L. and {Chabrier}, G. and {Hernanz}, M. and {Garcia-Berro}, E. and {Isern}, J. and {Mochkovitch}, R.},
        title = "{Cooling Theory of Crystallized White Dwarfs}",
      journal = {\apj},
         year = 1994,
        month = oct,
       volume = {434},
        pages = {641},
          doi = {10.1086/174766},
       adsurl = {https://ui.adsabs.harvard.edu/abs/1994ApJ...434..641S}
}

@article{Domcke:2025qlw,
    author = "Domcke, Valerie and Garcia-Cely, Camilo and Lee, Sung Mook",
    title = "{Gravitational wave scattering on magnetic fields}",
    eprint = "2507.16609",
    archivePrefix = "arXiv",
    primaryClass = "gr-qc",
    reportNumber = "CERN-TH-2025-142",
    doi = "10.1088/1475-7516/2025/11/016",
    journal = "JCAP",
    volume = "11",
    pages = "016",
    year = "2025"
}

@ARTICLE{2010CoPP...50...82P,
       author = {{Potekhin}, A.~Y. and {Chabrier}, G.},
        title = "{Thermodynamic Functions of Dense Plasmas: Analytic Approximations for Astrophysical Applications}",
      journal = {Contributions to Plasma Physics},
         year = 2010,
        month = jan,
       volume = {50},
       number = {1},
        pages = {82-87},
          doi = {10.1002/ctpp.201010017},
archivePrefix = {arXiv},
       eprint = {1001.0690},
 primaryClass = {physics.plasm-ph},
       adsurl = {https://ui.adsabs.harvard.edu/abs/2010CoPP...50...82P}
}

@ARTICLE{1989ApJ...347..934D,
       author = {{D'Antona}, Francesca and {Mazzitelli}, Italo},
        title = "{The Fastest Evolving White Dwarfs}",
      journal = {\apj},
         year = 1989,
        month = dec,
       volume = {347},
        pages = {934},
          doi = {10.1086/168185},
       adsurl = {https://ui.adsabs.harvard.edu/abs/1989ApJ...347..934D}
}

@ARTICLE{2021ApJ...913...72J,
       author = {{Jermyn}, Adam S. and {Schwab}, Josiah and {Bauer}, Evan and {Timmes}, F.~X. and {Potekhin}, Alexander Y.},
        title = "{Skye: A Differentiable Equation of State}",
      journal = {\apj},
         year = 2021,
        month = may,
       volume = {913},
       number = {1},
          eid = {72},
        pages = {72},
          doi = {10.3847/1538-4357/abf48e},
archivePrefix = {arXiv},
       eprint = {2104.00691},
 primaryClass = {astro-ph.SR},
       adsurl = {https://ui.adsabs.harvard.edu/abs/2021ApJ...913...72J}
}

@article{Ferrario:2015oda,
    author = "Ferrario, Lilia and de Martino, Domitilla and Gaensicke, Boris",
    title = "{Magnetic White Dwarfs}",
    eprint = "1504.08072",
    archivePrefix = "arXiv",
    primaryClass = "astro-ph.SR",
    doi = "10.1007/s11214-015-0152-0",
    journal = "Space Sci. Rev.",
    volume = "191",
    number = "1-4",
    pages = "111--169",
    year = "2015"
}

@ARTICLE{1989MNRAS.237...39H,
       author = {{Hameury}, J.~M. and {King}, A.~R. and {Lasota}, J.~P.},
        title = "{Consequences of mass transfert fluctuations in close binaries.}",
      journal = {\mnras},
         year = 1989,
        month = mar,
       volume = {237},
        pages = {39-47},
          doi = {10.1093/mnras/237.1.39},
       adsurl = {https://ui.adsabs.harvard.edu/abs/1989MNRAS.237...39H}
}

@article{10.1111/j.1365-8711.1998.t01-1-01913.x,
    author = {Li, Jianke and Wickramasinghe, Dayal T.},
    title = {Magnetic braking in magnetic binary stars},
    journal = {Monthly Notices of the Royal Astronomical Society},
    volume = {300},
    number = {3},
    pages = {718-732},
    year = {1998},
    month = {11},
    issn = {0035-8711},
    doi = {10.1111/j.1365-8711.1998.t01-1-01913.x},
    url = {https://doi.org/10.1111/j.1365-8711.1998.t01-1-01913.x},
    eprint = {https://academic.oup.com/mnras/article-pdf/300/3/718/3019616/300-3-718.pdf},
}

@article{Bagnulo_2022,
doi = {10.3847/2041-8213/ac84d3},
url = {https://doi.org/10.3847/2041-8213/ac84d3},
year = {2022},
month = {aug},
publisher = {The American Astronomical Society},
volume = {935},
number = {1},
pages = {L12},
author = {Bagnulo, Stefano and Landstreet, John D.},
title = {Multiple Channels for the Onset of Magnetism in Isolated White Dwarfs},
journal = {The Astrophysical Journal Letters}
}

@article{10.1111/j.1365-2966.2004.08603.x,
    author = {Wickramasinghe, D. T. and Ferrario, Lilia},
    title = {The origin of the magnetic fields in white dwarfs},
    journal = {Monthly Notices of the Royal Astronomical Society},
    volume = {356},
    number = {4},
    pages = {1576-1582},
    year = {2005},
    month = {02},
    issn = {0035-8711},
    doi = {10.1111/j.1365-2966.2004.08603.x},
    url = {https://doi.org/10.1111/j.1365-2966.2004.08603.x},
    eprint = {https://academic.oup.com/mnras/article-pdf/356/4/1576/3316378/356-4-1576.pdf},
}

@article{Isern:2019nrg,
    author = "Isern, Jordi",
    editor = "Barstow, Martin A. and Kleinman, Scot J. and Provencal, Judith L. and Ferrario, Lilia",
    title = "{White Dwarfs as Advanced Physics Laboratories. The Axion case}",
    eprint = "2002.08069",
    archivePrefix = "arXiv",
    primaryClass = "astro-ph.SR",
    doi = "10.1017/S1743921320000873",
    journal = "IAU Symp.",
    volume = "357",
    pages = "138--153",
    year = "2019"
}

@article{Isern:2022vdx,
    author = "Isern, Jordi and Torres, Santiago and Rebassa-Mansergas, Alberto",
    title = "{White Dwarfs as Physics Laboratories: Lights and Shadows}",
    eprint = "2202.02052",
    archivePrefix = "arXiv",
    primaryClass = "astro-ph.HE",
    doi = "10.3389/fspas.2022.815517",
    journal = "Front. Astron. Space Sci.",
    volume = "9",
    pages = "815517",
    year = "2022"
}

@article{Carenza:2024ehj,
    author = "Carenza, Pierluca and Giannotti, Maurizio and Isern, Jordi and Mirizzi, Alessandro and Straniero, Oscar",
    title = "{Axion astrophysics}",
    eprint = "2411.02492",
    archivePrefix = "arXiv",
    primaryClass = "hep-ph",
    reportNumber = "BARI-TH/66-24",
    doi = "10.1016/j.physrep.2025.02.002",
    journal = "Phys. Rept.",
    volume = "1117",
    pages = "1--102",
    year = "2025"
}

@book{Raffelt:1996wa,
  author    = {Georg G. Raffelt},
  title     = {Stars as Laboratories for Fundamental Physics: The Astrophysics of Neutrinos, Axions, and Other Weakly Interacting Particles},
  publisher = {University of Chicago Press},
  address   = {Chicago},
  year      = {1996},
  isbn      = {0-226-70272-3},
  note      = {Theoretical Astrophysics; 686~pp., 188 line drawings, 34 tables}
}

@article{Bekenstein:1998nt,
    author = "Bekenstein, Jacob D. and Schiffer, Marcelo",
    title = "{The Many faces of superradiance}",
    eprint = "gr-qc/9803033",
    archivePrefix = "arXiv",
    doi = "10.1103/PhysRevD.58.064014",
    journal = "Phys. Rev. D",
    volume = "58",
    pages = "064014",
    year = "1998"
}

@article{Sedrakian:2018kdm,
    author = "Sedrakian, Armen",
    title = "{Axion cooling of neutron stars. II. Beyond hadronic axions}",
    eprint = "1810.00190",
    archivePrefix = "arXiv",
    primaryClass = "astro-ph.HE",
    doi = "10.1103/PhysRevD.99.043011",
    journal = "Phys. Rev. D",
    volume = "99",
    number = "4",
    pages = "043011",
    year = "2019"
}

@article{Sedrakian:2015krq,
    author = "Sedrakian, Armen",
    title = "{Axion cooling of neutron stars}",
    eprint = "1512.07828",
    archivePrefix = "arXiv",
    primaryClass = "astro-ph.HE",
    doi = "10.1103/PhysRevD.93.065044",
    journal = "Phys. Rev. D",
    volume = "93",
    number = "6",
    pages = "065044",
    year = "2016"
}

@article{OHare:2020wah,
    author = "O'Hare, Ciaran A. J. and Vitagliano, Edoardo",
    title = "{Cornering the axion with $CP$-violating interactions}",
    eprint = "2010.03889",
    archivePrefix = "arXiv",
    primaryClass = "hep-ph",
    reportNumber = "CPPC-2020-16",
    doi = "10.1103/PhysRevD.102.115026",
    journal = "Phys. Rev. D",
    volume = "102",
    number = "11",
    pages = "115026",
    year = "2020"
}

@article{Raffelt:2012sp,
    author = "Raffelt, Georg",
    title = "{Limits on a CP-violating scalar axion-nucleon interaction}",
    eprint = "1205.1776",
    archivePrefix = "arXiv",
    primaryClass = "hep-ph",
    reportNumber = "MPP-2012-74",
    doi = "10.1103/PhysRevD.86.015001",
    journal = "Phys. Rev. D",
    volume = "86",
    pages = "015001",
    year = "2012"
}

@article{Brito:2015oca,
    author = "Brito, Richard and Cardoso, Vitor and Pani, Paolo",
    title = "{Superradiance}: {New Frontiers in Black Hole
Physics}",
    eprint = "1501.06570",
    archivePrefix = "arXiv",
    primaryClass = "gr-qc",
    doi = "10.1007/978-3-319-19000-6",
    journal = "Lect. Notes Phys.",
    volume = "906",
    pages = "pp.1--237",
    year = "2015"
}

@ARTICLE{Woltjer1964,
       author = {{Woltjer}, L.},
        title = "{X-Rays and Type I Supernova Remnants.}",
      journal = {\apj},
         year = 1964,
        month = oct,
       volume = {140},
        pages = {1309-1313},
          doi = {10.1086/148028},
       adsurl = {https://ui.adsabs.harvard.edu/abs/1964ApJ...140.1309W}
}

@article{Pacini:1967epn,
    author = "Pacini, F.",
    title = "{Energy Emission from a Neutron Star}",
    doi = "10.1038/216567a0",
    journal = "Nature",
    volume = "216",
    number = "5115",
    pages = "567--568",
    year = "1967"
}

@article{Gold:1968zf,
    author = "Gold, T.",
    title = "{Rotating neutron stars as the origin of the pulsating radio sources}",
    doi = "10.1038/218731a0",
    journal = "Nature",
    volume = "218",
    pages = "731--732",
    year = "1968"
}

@article{Hulse:1974eb,
    author = "Hulse, R. A. and Taylor, J. H.",
    title = "{Discovery of a pulsar in a binary system}",
    doi = "10.1086/181708",
    journal = "Astrophys. J. Lett.",
    volume = "195",
    pages = "L51--L53",
    year = "1975"
}

@ARTICLE{HulseTaylorGWs,
       author = {{Taylor}, J.~H. and {Weisberg}, J.~M.},
        title = "{A new test of general relativity - Gravitational radiation and the binary pulsar PSR 1913+16}",
      journal = {\apj},
         year = 1982,
        month = feb,
       volume = {253},
        pages = {908-920},
          doi = {10.1086/159690},
       adsurl = {https://ui.adsabs.harvard.edu/abs/1982ApJ...253..908T}
}

@article{Kerr:1963ud,
    author = "Kerr, Roy P.",
    title = "{Gravitational field of a spinning mass as an example of algebraically special metrics}",
    doi = "10.1103/PhysRevLett.11.237",
    journal = "Phys. Rev. Lett.",
    volume = "11",
    pages = "237--238",
    year = "1963"
}

@article{Baade:1934wuu,
    author = "Baade, W. and Zwicky, F.",
    title = "{Remarks on Super-Novae and Cosmic Rays}",
    doi = "10.1103/PhysRev.46.76.2",
    journal = "Phys. Rev.",
    volume = "46",
    number = "1",
    pages = "76",
    year = "1934"
}

@article{Schwarzschild:1916uq,
    author = "Schwarzschild, Karl",
    title = "{On the gravitational field of a mass point according to Einstein's theory}",
    eprint = "physics/9905030",
    archivePrefix = "arXiv",
    journal = "Sitzungsber. Preuss. Akad. Wiss. Berlin (Math. Phys. )",
    volume = "1916",
    pages = "189--196",
    year = "1916"
}

@article{Hewish:1968bj,
    author = "Hewish, A. and Bell, S. J. and Pilkington, J. D. H and Scott, P. F. and Collins, R. A.",
    title = "{Observation of a rapidly pulsating radio source}",
    doi = "10.1038/217709a0",
    journal = "Nature",
    volume = "217",
    pages = "709--713",
    year = "1968"
}

@article{Michell,
author = {Michell, John },
title = {VII. On the means of discovering the distance, magnitude, \&amp;c. of the fixed stars, in consequence of the diminution of the velocity of their light, in case such a diminution should be found to take place in any of them, and such other data should be procured from observations, as would be farther necessary for that purpose. By the Rev. John Michell, B.D. F.R.S. In a letter to Henry Cavendish, Esq. F.R.S. and A.S},
journal = {Philosophical Transactions of the Royal Society of London},
volume = {74},
number = {},
pages = {35-57},
year = {1784},
doi = {10.1098/rstl.1784.0008},

URL = {https://royalsocietypublishing.org/doi/abs/10.1098/rstl.1784.0008},
eprint = {https://royalsocietypublishing.org/doi/pdf/10.1098/rstl.1784.0008}

}

@ARTICLE{1965AJ.....70..754G,
       author = {{Gardner}, F.~F. and {Milne}, D.~K.},
        title = "{The supernova of A.D. 1006}",
      journal = {\aj},
         year = 1965,
        month = nov,
       volume = {70},
        pages = {754},
          doi = {10.1086/109813},
       adsurl = {https://ui.adsabs.harvard.edu/abs/1965AJ.....70..754G}
}

@article{Hook:2018iia,
    author = "Hook, Anson and Kahn, Yonatan and Safdi, Benjamin R. and Sun, Zhiquan",
    title = "{Radio Signals from Axion Dark Matter Conversion in Neutron Star  Magnetospheres}",
    eprint = "1804.03145",
    archivePrefix = "arXiv",
    primaryClass = "hep-ph",
    reportNumber = "LCTP-18-09, PUPT-2558",
    doi = "10.1103/PhysRevLett.121.241102",
    journal = "Phys. Rev. Lett.",
    volume = "121",
    number = "24",
    pages = "241102",
    year = "2018"
}

@article{Cukanovaite_2023,
   title={Local stellar formation history from the 40 pc white dwarf sample},
   volume={522},
   ISSN={1365-2966},
   url={http://dx.doi.org/10.1093/mnras/stad1020},
   DOI={10.1093/mnras/stad1020},
   number={2},
   journal={Monthly Notices of the Royal Astronomical Society},
   publisher={Oxford University Press (OUP)},
   author={Cukanovaite, E and Tremblay, P-E and Toonen, S and Temmink, K D and Manser, Christopher J and O’Brien, M W and McCleery, J},
   year={2023},
   month=apr, pages={1643–1661} }

@article{Fleury:2025ahw,
    author = "Fleury, Leesa and Obertas, Alysa and Richer, Harvey and Heyl, Jeremy",
    title = "{Axion Constraints from White Dwarf Cooling in 47 Tucanae}",
    eprint = "2511.21676",
    archivePrefix = "arXiv",
    primaryClass = "astro-ph.SR",
    month = "11",
    year = "2025"
}

@ARTICLE{1983Natur.303..781W,
       author = {{Winget}, D.~E. and {Hansen}, C.~J. and {van Horn}, H.~M.},
        title = "{Do pulsating PG1159-035 stars put constraints on stellar evolution?}",
      journal = {\nat},
         year = 1983,
        month = jun,
       volume = {303},
       number = {5920},
        pages = {781-782},
          doi = {10.1038/303781a0},
       adsurl = {https://ui.adsabs.harvard.edu/abs/1983Natur.303..781W}
}

@article{Isern:1992gia,
    author = "Isern, J. and Hernanz, M. and Garcia-Berro, E.",
    title = "{Axion cooling of white dwarfs}",
    doi = "10.1086/186416",
    journal = "Astrophys. J. Lett.",
    volume = "392",
    pages = "L23",
    year = "1992"
}

@ARTICLE{1968ApJ...153..151L,
       author = {{Landolt}, Arlo U.},
        title = "{A New Short-Period Blue Variable}",
      journal = {\apj},
         year = 1968,
        month = jul,
       volume = {153},
        pages = {151},
          doi = {10.1086/149645},
       adsurl = {https://ui.adsabs.harvard.edu/abs/1968ApJ...153..151L}
}

@ARTICLE{2022MNRAS.511.1574R,
       author = {{Romero}, Alejandra D. and {Kepler}, S.~O. and {Hermes}, J.~J. and {Amaral}, Larissa Antunes and {Uzundag}, Murat and {Bogn{\'a}r}, Zs{\'o}fia and {Bell}, Keaton J. and {VanWyngarden}, Madison and {Baran}, Andy and {Pelisoli}, Ingrid and {Oliveira}, Gabriela da Rosa and {Koester}, Detlev and {Klippel}, T.~S. and {Fraga}, Luciano and {Bradley}, Paul A. and {Vu{\v{c}}kovi{\'c}}, Maja and {Heintz}, Tyler M. and {Reding}, Joshua S. and {Kaiser}, B.~C. and {Charpinet}, St{\'e}phane},
        title = "{Discovery of 74 new bright ZZ Ceti stars in the first three years of TESS}",
      journal = {\mnras},
         year = 2022,
        month = apr,
       volume = {511},
       number = {2},
        pages = {1574-1590},
          doi = {10.1093/mnras/stac093},
archivePrefix = {arXiv},
       eprint = {2201.04158},
 primaryClass = {astro-ph.SR},
       adsurl = {https://ui.adsabs.harvard.edu/abs/2022MNRAS.511.1574R}
}

@INPROCEEDINGS{1991ASIC..336..153F,
       author = {{Fontaine}, G. and {Brassard}, P. and {Wesemael}, F. and {Kepler}, S.~O. and {Wood}, M.~A.},
        title = "{On the interpretation of the dP/dt measurement in G117-B15A}",
    booktitle = {White Dwarfs},
         year = 1991,
       editor = {{Vauclair}, Gerard and {Sion}, Edward},
       series = {NATO Advanced Study Institute (ASI) Series C},
       volume = {336},
        month = jan,
        pages = {153},
       adsurl = {https://ui.adsabs.harvard.edu/abs/1991ASIC..336..153F}
}

@article{Corsico:2019nmr,
    author = "C\'orsico, Alejandro H. and Althaus, Leandro G. and Miller Bertolami, Marcelo M. and Kepler, S. O.",
    title = "{Pulsating white dwarfs: new insights}",
    eprint = "1907.00115",
    archivePrefix = "arXiv",
    primaryClass = "astro-ph.SR",
    doi = "10.1007/s00159-019-0118-4",
    journal = "Astron. Astrophys. Rev.",
    volume = "27",
    number = "1",
    pages = "7",
    year = "2019"
}

@article{Nakagawa:1987pga,
    author = "Nakagawa, Masayuki and Kohyama, Yasuharu and Itoh, Naoki",
    title = "{Axion Bremsstrahlung in Dense Stars}",
    doi = "10.1086/165724",
    journal = "Astrophys. J.",
    volume = "322",
    pages = "291",
    year = "1987"
}

@article{Nakagawa:1988rhp,
    author = "Nakagawa, Masayuki and Adachi, Tomoo and Kohyama, Yasuharu and Itoh, Naoki",
    title = "{Axion bremsstrahlung in dense stars. II - Phonon contributions}",
    doi = "10.1086/166085",
    journal = "Astrophys. J.",
    volume = "326",
    pages = "241",
    year = "1988"
}

@article{Carenza:2021osu,
    author = "Carenza, Pierluca and Lucente, Giuseppe",
    title = "{Revisiting axion-electron bremsstrahlung emission rates in astrophysical environments}",
    eprint = "2104.09524",
    archivePrefix = "arXiv",
    primaryClass = "hep-ph",
    doi = "10.1103/PhysRevD.103.123024",
    journal = "Phys. Rev. D",
    volume = "103",
    number = "12",
    pages = "123024",
    year = "2021"
}

@ARTICLE{1983ApJ...275..858I,
       author = {{Itoh}, N. and {Kohyama}, Y.},
        title = "{Neutrino-pair bremsstrahlung in dense stars. I. Liquid metal case.}",
      journal = {\apj},
         year = 1983,
        month = dec,
       volume = {275},
        pages = {858-866},
          doi = {10.1086/161579},
       adsurl = {https://ui.adsabs.harvard.edu/abs/1983ApJ...275..858I}
}

@ARTICLE{1991ApJ...378L..45K,
       author = {{Kepler}, S.~O. and {Winget}, D.~E. and {Nather}, R.~E. and {Bradley}, P.~A. and {Grauer}, A.~D. and {Fontaine}, G. and {Bergeron}, P. and {Vauclair}, G. and {Claver}, C.~F. and {Marar}, T.~M.~K. and {Seetha}, S. and {Ashoka}, B.~N. and {Mazeh}, T. and {Leibowitz}, E. and {Dolez}, N. and {Chevreton}, M. and {Barstow}, M.~A. and {Clemens}, J.~C. and {Kleinman}, S.~J. and {Sansom}, A.~E. and {Tweedy}, R.~W. and {Kanaan}, A. and {Hine}, B.~P. and {Provencal}, J.~L. and {Wesemael}, F. and {Wood}, M.~A. and {Brassard}, P. and {Solheim}, J. -E. and {Emanuelsen}, P. -I.},
        title = "{A Detection of the Evolutionary Time Scale of the DA White Dwarf G117-B15A with the Whole Earth Telescope}",
      journal = {\apjl},
         year = 1991,
        month = sep,
       volume = {378},
        pages = {L45},
          doi = {10.1086/186138},
       adsurl = {https://ui.adsabs.harvard.edu/abs/1991ApJ...378L..45K}
}

@article{2000ApJ...534L.185K,
    author = "Kepler, S. O. and Mukadam, Anjum and Winget, D. E. and Nather, R. E. and Metcalfe, T. S. and Reed, M. D. and Kawaler, S. D. and Bradley, Paul A.",
    title = "{Evolutionary timescale of the dav g117-b15a: the most stable optical clock known}",
    eprint = "astro-ph/0003478",
    archivePrefix = "arXiv",
    doi = "10.1086/312664",
    journal = "Astrophys. J. Lett.",
    volume = "534",
    pages = "L185",
    year = "2000"
}

@article{2005ApJ...634.1311K,
    author = "Kepler, S. O. and Costa, J. E. S. and Castanheira, B. G. and Winget, D. E. and Mullally, Fergal and Nather, R. E. and Kilic, Mukremin and von Hippel, Ted and Mukadam, Anjum S. and Sullivan, Denis J.",
    title = "{Measuring the evolution of the most stable optical clock g 117-b15a}",
    eprint = "astro-ph/0507487",
    archivePrefix = "arXiv",
    doi = "10.1086/497002",
    journal = "Astrophys. J.",
    volume = "634",
    pages = "1311--1318",
    year = "2005"
}

@INPROCEEDINGS{2012ASPC..462..322K,
       author = {{Kepler}, S.~O.},
        title = "{White Dwarf Stars: Pulsations and Magnetism}",
    booktitle = {Progress in Solar/Stellar Physics with Helio- and Asteroseismology},
         year = 2012,
       editor = {{Shibahashi}, H. and {Takata}, M. and {Lynas-Gray}, A.~E.},
       series = {Astronomical Society of the Pacific Conference Series},
       volume = {462},
        month = sep,
        pages = {322},
       adsurl = {https://ui.adsabs.harvard.edu/abs/2012ASPC..462..322K}
}

@ARTICLE{2021ApJ...906....7K,
       author = {{Kepler}, S.~O. and {Winget}, D.~E. and {Vanderbosch}, Zachary P. and {Castanheira}, Barbara Garcia and {Hermes}, J.~J. and {Bell}, Keaton J. and {Mullally}, Fergal and {Romero}, Alejandra D. and {Montgomery}, M.~H. and {DeGennaro}, Steven and {Winget}, Karen I. and {Chandler}, Dean and {Jeffery}, Elizabeth J. and {Fritzen}, Jamile K. and {Williams}, Kurtis A. and {Chote}, Paul and {Zola}, Staszek},
        title = "{The Pulsating White Dwarf G117-B15A: Still the Most Stable Optical Clock Known}",
      journal = {\apj},
         year = 2021,
        month = jan,
       volume = {906},
       number = {1},
          eid = {7},
        pages = {7},
          doi = {10.3847/1538-4357/abc626},
archivePrefix = {arXiv},
       eprint = {2010.16062},
 primaryClass = {astro-ph.SR},
       adsurl = {https://ui.adsabs.harvard.edu/abs/2021ApJ...906....7K}
}

@article{2001NewA....6..197C,
    author = "Corsico, Alejandro H. and Benvenuto, Omar G. and Althaus, Leandro G. and Isern, Jordi and Garcia-Berro, Enrique",
    title = "{The Potential of the variable DA white dwarf G117 - B15A as a tool for fundamental physics}",
    eprint = "astro-ph/0104103",
    archivePrefix = "arXiv",
    doi = "10.1016/S1384-1076(01)00055-0",
    journal = "New Astron.",
    volume = "6",
    pages = "197--213",
    year = "2001"
}

@article{2012MNRAS.424.2792C,
    author = "Corsico, Alejandro H. and Althaus, Leandro G. and Bertolami, Marcelo M. Miller and Romero, Alejandra D. and Garcia-Berro, Enrique and Isern, Jordi and Kepler, S. O.",
    title = "{The rate of cooling of the pulsating white dwarf star G117$-$B15A: a new asteroseismological inference of the axion mass}",
    eprint = "1205.6180",
    archivePrefix = "arXiv",
    primaryClass = "astro-ph.SR",
    doi = "10.1111/j.1365-2966.2012.21401.x",
    journal = "Mon. Not. Roy. Astron. Soc.",
    volume = "424",
    pages = "2792",
    year = "2012"
}

@article{2012MNRAS.420.1462R,
    author = "Romero, Alejandra D. and Corsico, Alejandro H. and Althaus, Leandro G. and Kepler, S. O. and Castanheira, Barbara G. and Bertolami, Marcelo M. Miller",
    title = "{Toward ensemble asteroseismology of ZZ Ceti stars with fully evolutionary models}",
    eprint = "1109.6682",
    archivePrefix = "arXiv",
    primaryClass = "astro-ph.SR",
    doi = "10.1111/j.1365-2966.2011.20134.x",
    journal = "Mon. Not. Roy. Astron. Soc.",
    volume = "420",
    pages = "1462",
    year = "2012"
}

@article{2008ApJ...675.1512B,
    author = "Bischoff-Kim, A. and Montgomery, M. H. and Winget, D. E.",
    title = "{Strong limits on the DFSZ axion mass with G117-B15A}",
    eprint = "0711.2041",
    archivePrefix = "arXiv",
    primaryClass = "astro-ph",
    doi = "10.1086/526398",
    journal = "Astrophys. J.",
    volume = "675",
    pages = "1512",
    year = "2008"
}

@INPROCEEDINGS{2018phos.confE..28B,
       author = {{Bischoff-Kim}, Agnes},
        title = "{Non-luminous sources of cooling in pulsating white dwarfs}",
    booktitle = {PHysics of Oscillating STars. Proceedings from the PHOST (PHysics of Oscillating STars) symposium hosted by the Oceanographic Observatory in Banyuls-sur-mer (France) from 2-7 September 2018. This conference honours the life work of Professor Hiromoto Shibahashi},
         year = 2018,
        month = sep,
          eid = {28},
        pages = {28},
          doi = {10.5281/zenodo.1715917},
       adsurl = {https://ui.adsabs.harvard.edu/abs/2018phos.confE..28B}
}

@article{2012JCAP...12..010C,
    author = "Corsico, A. H. and Althaus, L. G. and Romero, A. D. and Mukadam, A. S. and Garcia-Berro, E. and Isern, J. and Kepler, S. O. and Corti, M. A.",
    title = "{An independent limit on the axion mass from the variable white dwarf star R548}",
    eprint = "1211.3389",
    archivePrefix = "arXiv",
    primaryClass = "astro-ph.SR",
    doi = "10.1088/1475-7516/2012/12/010",
    journal = "JCAP",
    volume = "12",
    pages = "010",
    year = "2012"
}

@article{2016JCAP...07..036C,
    author = "C\'orsico, Alejandro H. and Romero, Alejandra D. and Althaus, Leandro G. and Garc\'\i{}a-Berro, Enrique and Isern, Jordi and Kepler, S. O. and Miller Bertolami, Marcelo M. and Sullivan, Denis J. and Chote, Paul",
    title = "{An asteroseismic constraint on the mass of the axion from the period drift of the pulsating DA white dwarf star L19-2}",
    eprint = "1605.06458",
    archivePrefix = "arXiv",
    primaryClass = "astro-ph.SR",
    doi = "10.1088/1475-7516/2016/07/036",
    journal = "JCAP",
    volume = "07",
    pages = "036",
    year = "2016"
}

@ARTICLE{2013ApJ...771...17M,
       author = {{Mukadam}, Anjum S. and {Bischoff-Kim}, Agnes and {Fraser}, Oliver and {C{\'o}rsico}, A.~H. and {Montgomery}, M.~H. and {Kepler}, S.~O. and {Romero}, A.~D. and {Winget}, D.~E. and {Hermes}, J.~J. and {Riecken}, T.~S. and {Kronberg}, M.~E. and {Winget}, K.~I. and {Falcon}, Ross E. and {Chandler}, D.~W. and {Kuehne}, J.~W. and {Sullivan}, D.~J. and {Reaves}, D. and {von Hippel}, T. and {Mullally}, F. and {Shipman}, H. and {Thompson}, S.~E. and {Silvestri}, N.~M. and {Hynes}, R.~I.},
        title = "{Measuring the Evolutionary Rate of Cooling of ZZ Ceti}",
      journal = {\apj},
         year = 2013,
        month = jul,
       volume = {771},
       number = {1},
          eid = {17},
        pages = {17},
          doi = {10.1088/0004-637X/771/1/17},
       adsurl = {https://ui.adsabs.harvard.edu/abs/2013ApJ...771...17M}
}

@INPROCEEDINGS{2015ASPC..493..199S,
       author = {{Sullivan}, D.~J. and {Chote}, P.},
        title = "{The Frequency Stability of the Pulsating White Dwarf L19-2}",
    booktitle = {19th European Workshop on White Dwarfs},
         year = 2015,
       editor = {{Dufour}, P. and {Bergeron}, P. and {Fontaine}, G.},
       series = {Astronomical Society of the Pacific Conference Series},
       volume = {493},
        month = jun,
        pages = {199},
       adsurl = {https://ui.adsabs.harvard.edu/abs/2015ASPC..493..199S}
}

@article{2016JCAP...08..062B,
    author = "Battich, Tiara and C\'orsico, Alejandro Hugo and Althaus, Leandro Gabriel and Miller Bertolami, Marcelo Miguel",
    title = "{First axion bounds from a pulsating helium-rich white dwarf star}",
    eprint = "1605.07668",
    archivePrefix = "arXiv",
    primaryClass = "astro-ph.SR",
    doi = "10.1088/1475-7516/2016/08/062",
    journal = "JCAP",
    volume = "08",
    pages = "062",
    year = "2016"
}

@ARTICLE{1988ApJ...332..891L,
   author = {{Liebert}, J. and {Dahn}, C.~C. and {Monet}, D.~G.},
    title = "{The luminosity function of white dwarfs}",
  journal = {\apj},
     year = 1988,
    month = sep,
   volume = 332,
    pages = {891-909},
      doi = {10.1086/166699},
   adsurl = {http://cdsads.u-strasbg.fr/abs/1988ApJ...332..891L}
}

@ARTICLE{1992MNRAS.255..521E,
   author = {{Evans}, D.~W.},
    title = "{The APM Proper Motion Project. I - High proper motion stars}",
  journal = {\mnras},
     year = 1992,
    month = apr,
   volume = 255,
    pages = {521-538},
      doi = {10.1093/mnras/255.3.521},
   adsurl = {http://adsabs.harvard.edu/abs/1992MNRAS.255..521E}
}

@ARTICLE{1996Natur.382..692O,
   author = {{Oswalt}, T.~D. and {Smith}, J.~A. and {Wood}, M.~A. and {Hintzen}, P.
	},
    title = "{A lower limit of 9.5 Gyr on the age of the Galactic disk from the oldest white dwarf stars}",
  journal = {\nat},
     year = 1996,
    month = aug,
   volume = 382,
    pages = {692-694},
      doi = {10.1038/382692a0},
   adsurl = {http://adsabs.harvard.edu/abs/1996Natur.382..692O}
}

@ARTICLE{1998ApJ...497..294L,
   author = {{Leggett}, S.~K. and {Ruiz}, M.~T. and {Bergeron}, P.},
    title = "{The Cool White Dwarf Luminosity Function and the Age of the Galactic Disk}",
  journal = {\apj},
     year = 1998,
    month = apr,
   volume = 497,
    pages = {294-302},
      doi = {10.1086/305463},
   adsurl = {http://adsabs.harvard.edu/abs/1998ApJ...497..294L}
}

@article{1999MNRAS.306..736K,
    author = "Knox, R. A. and Hawkins, M. R. S. and Hambly, N. C",
    title = "{A survey for cool white dwarfs and the age of the galactic disc}",
    eprint = "astro-ph/9903345",
    archivePrefix = "arXiv",
    doi = "10.1046/j.1365-8711.1999.02625.x",
    journal = "Mon. Not. Roy. Astron. Soc.",
    volume = "306",
    pages = "736",
    year = "1999"
}

@ARTICLE{1987ApJ...315L..77W,
       author = {{Winget}, D.~E. and {Hansen}, C.~J. and {Liebert}, James and {van Horn}, H.~M. and {Fontaine}, G. and {Nather}, R.~E. and {Kepler}, S.~O. and {Lamb}, D.~Q.},
        title = "{An Independent Method for Determining the Age of the Universe}",
      journal = {\apjl},
         year = 1987,
        month = apr,
       volume = {315},
        pages = {L77},
          doi = {10.1086/184864},
       adsurl = {https://ui.adsabs.harvard.edu/abs/1987ApJ...315L..77W}
}

@ARTICLE{2011MNRAS.417...93R,
   author = {{Rowell}, N. and {Hambly}, N.~C.},
    title = "{White dwarfs in the SuperCOSMOS Sky Survey: the thin disc, thick disc and spheroid luminosity functions}",
  journal = {\mnras},
     year = 2011,
    month = oct,
   volume = 417,
    pages = {93-113},
      doi = {10.1111/j.1365-2966.2011.18976.x},
   adsurl = {http://cdsads.u-strasbg.fr/abs/2011MNRAS.417...93R}
}

@article{2014JCAP...10..069M,
    author = "Miller Bertolami, Marcelo M. and Melendez, Brenda E. and Althaus, Leandro G. and Isern, Jordi",
    title = "{Revisiting the axion bounds from the Galactic white dwarf luminosity function}",
    eprint = "1406.7712",
    archivePrefix = "arXiv",
    primaryClass = "hep-ph",
    doi = "10.1088/1475-7516/2014/10/069",
    journal = "JCAP",
    volume = "10",
    pages = "069",
    year = "2014"
}

@ARTICLE{2021A&A...649A...6G,
       author = {{Gaia Collaboration} and {Smart}, R.~L. and others},
        title = "{Gaia Early Data Release 3. The Gaia Catalogue of Nearby Stars}",
      journal = {\aap},
         year = 2021,
        month = may,
       volume = {649},
          eid = {A6},
        pages = {A6},
          doi = {10.1051/0004-6361/202039498},
archivePrefix = {arXiv},
       eprint = {2012.02061},
 primaryClass = {astro-ph.SR},
       adsurl = {https://ui.adsabs.harvard.edu/abs/2021A&A...649A...6G}
}

@ARTICLE{2009A&A...508..339K,
       author = {{Krzesinski}, J. and {Kleinman}, S.~J. and {Nitta}, A. and {H{\"u}gelmeyer}, S. and {Dreizler}, S. and {Liebert}, J. and {Harris}, H.},
        title = "{A hot white dwarf luminosity function from the Sloan Digital Sky Survey}",
      journal = {\aap},
         year = 2009,
        month = dec,
       volume = {508},
       number = {1},
        pages = {339-344},
          doi = {10.1051/0004-6361/200912094},
       adsurl = {https://ui.adsabs.harvard.edu/abs/2009A&A...508..339K}
}

@article{2010A&ARv..18..471A,
    author = "Althaus, Leandro G. and Corsico, Alejandro H. and Isern, Jordi and a-Berro, Enrique Garci",
    title = "{Evolutionary and pulsational properties of white dwarf stars}",
    eprint = "1007.2659",
    archivePrefix = "arXiv",
    primaryClass = "astro-ph.SR",
    doi = "10.1007/s00159-010-0033-1",
    journal = "Astron. Astrophys. Rev.",
    volume = "18",
    pages = "471--566",
    year = "2010"
}

@ARTICLE{2019A&ARv..27....7C,
       author = {{C{\'o}rsico}, Alejandro H. and {Althaus}, Leandro G. and {Miller Bertolami}, Marcelo M. and {Kepler}, S.~O.},
        title = "{Pulsating white dwarfs: new insights}",
      journal = {\aapr},
         year = 2019,
        month = sep,
       volume = {27},
       number = {1},
          eid = {7},
        pages = {7},
          doi = {10.1007/s00159-019-0118-4},
archivePrefix = {arXiv},
       eprint = {1907.00115},
 primaryClass = {astro-ph.SR},
       adsurl = {https://ui.adsabs.harvard.edu/abs/2019A&ARv..27....7C}
}

@ARTICLE{2020FrASS...7...47C,
       author = {{C{\'o}rsico}, Alejandro H.},
        title = "{White-dwarf asteroseismology with the Kepler space telescope}",
      journal = {Frontiers in Astronomy and Space Sciences},
         year = 2020,
        month = aug,
       volume = {7},
          eid = {47},
        pages = {47},
          doi = {10.3389/fspas.2020.00047},
archivePrefix = {arXiv},
       eprint = {2006.04955},
 primaryClass = {astro-ph.SR},
       adsurl = {https://ui.adsabs.harvard.edu/abs/2020FrASS...7...47C}
}

@ARTICLE{2021RvMP...93a5001A,
       author = {{Aerts}, C.},
        title = "{Probing the interior physics of stars through asteroseismology}",
      journal = {Reviews of Modern Physics},
         year = 2021,
        month = jan,
       volume = {93},
       number = {1},
          eid = {015001},
        pages = {015001},
          doi = {10.1103/RevModPhys.93.015001},
archivePrefix = {arXiv},
       eprint = {1912.12300},
 primaryClass = {astro-ph.SR},
       adsurl = {https://ui.adsabs.harvard.edu/abs/2021RvMP...93a5001A}
}

@article{2006AJ....131..571H,
    author = "Harris, Hugh C. and others",
    title = "{The white dwarf luminosity function from sdss imaging data}",
    eprint = "astro-ph/0510820",
    archivePrefix = "arXiv",
    doi = "10.1086/497966",
    journal = "Astron. J.",
    volume = "131",
    pages = "571--581",
    year = "2006"
}

@article{Isern:2018uce,
    author = "Isern, Jordi and Garcia-Berro, Enrique and Torres, Santiago and Cojocaru, Roxana and Catalan, Silvia",
    title = "{Axions and the luminosity function of white dwarfs: the thin and thick discs, and the halo}",
    eprint = "1805.00135",
    archivePrefix = "arXiv",
    primaryClass = "astro-ph.SR",
    doi = "10.1093/mnras/sty1162",
    journal = "Mon. Not. Roy. Astron. Soc.",
    volume = "478",
    number = "2",
    pages = "2569--2575",
    year = "2018"
}

@ARTICLE{2017AJ....153...10M,
       author = {{Munn}, Jeffrey A. and {Harris}, Hugh C. and {von Hippel}, Ted and {Kilic}, Mukremin and {Liebert}, James W. and {Williams}, Kurtis A. and {DeGennaro}, Steven and {Jeffery}, Elizabeth and {Dame}, Kyra and {Gianninas}, A. and {Brown}, Warren R.},
        title = "{A Deep Proper Motion Catalog Within the Sloan Digital Sky Survey Footprint. II. The White Dwarf Luminosity Function}",
      journal = {\aj},
         year = 2017,
        month = jan,
       volume = {153},
       number = {1},
          eid = {10},
        pages = {10},
          doi = {10.3847/1538-3881/153/1/10},
archivePrefix = {arXiv},
       eprint = {1611.06275},
 primaryClass = {astro-ph.SR},
       adsurl = {https://ui.adsabs.harvard.edu/abs/2017AJ....153...10M}
}

@ARTICLE{2017ApJ...837..162K,
       author = {{Kilic}, Mukremin and {Munn}, Jeffrey A. and {Harris}, Hugh C. and {von Hippel}, Ted and {Liebert}, James W. and {Williams}, Kurtis A. and {Jeffery}, Elizabeth and {DeGennaro}, Steven},
        title = "{The Ages of the Thin Disk, Thick Disk, and the Halo from Nearby White Dwarfs}",
      journal = {\apj},
         year = 2017,
        month = mar,
       volume = {837},
       number = {2},
          eid = {162},
        pages = {162},
          doi = {10.3847/1538-4357/aa62a5},
archivePrefix = {arXiv},
       eprint = {1702.06984},
 primaryClass = {astro-ph.SR},
       adsurl = {https://ui.adsabs.harvard.edu/abs/2017ApJ...837..162K}
}

@ARTICLE{2019Natur.565..202T,
       author = {{Tremblay}, Pier-Emmanuel and {Fontaine}, Gilles and {Gentile Fusillo}, Nicola Pietro and {Dunlap}, Bart H. and {G{\"a}nsicke}, Boris T. and {Hollands}, Mark A. and {Hermes}, J.~J. and {Marsh}, Thomas R. and {Cukanovaite}, Elena and {Cunningham}, Tim},
        title = "{Core crystallization and pile-up in the cooling sequence of evolving white dwarfs}",
      journal = {\nat},
         year = 2019,
        month = jan,
       volume = {565},
       number = {7738},
        pages = {202-205},
          doi = {10.1038/s41586-018-0791-x},
archivePrefix = {arXiv},
       eprint = {1908.00370},
 primaryClass = {astro-ph.SR},
       adsurl = {https://ui.adsabs.harvard.edu/abs/2019Natur.565..202T}
}

@article{Straniero:2020iyi,
    author = "Straniero, O. and Pallanca, C. and Dalessandro, E. and Dominguez, I. and Ferraro, F. R. and Giannotti, M. and Mirizzi, A. and Piersanti, L.",
    title = "{The RGB tip of galactic globular clusters and the revision of the axion-electron coupling bound}",
    eprint = "2010.03833",
    archivePrefix = "arXiv",
    primaryClass = "astro-ph.SR",
    doi = "10.1051/0004-6361/202038775",
    journal = "Astron. Astrophys.",
    volume = "644",
    pages = "A166",
    year = "2020"
}

@article{Capozzi:2020cbu,
    author = "Capozzi, Francesco and Raffelt, Georg",
    title = "{Axion and neutrino bounds improved with new calibrations of the tip of the red-giant branch using geometric distance determinations}",
    eprint = "2007.03694",
    archivePrefix = "arXiv",
    primaryClass = "astro-ph.SR",
    reportNumber = "MPP-2020-106",
    doi = "10.1103/PhysRevD.102.083007",
    journal = "Phys. Rev. D",
    volume = "102",
    number = "8",
    pages = "083007",
    year = "2020"
}

@article{Bottaro:2023gep,
    author = "Bottaro, Salvatore and Caputo, Andrea and Raffelt, Georg and Vitagliano, Edoardo",
    title = "{Stellar limits on scalars from electron-nucleus bremsstrahlung}",
    eprint = "2303.00778",
    archivePrefix = "arXiv",
    primaryClass = "hep-ph",
    reportNumber = "CERN-TH-2023-035",
    doi = "10.1088/1475-7516/2023/07/071",
    journal = "JCAP",
    volume = "07",
    pages = "071",
    year = "2023"
}

@article{Yamamoto:2023zlu,
    author = "Yamamoto, Yasuhiro and Yoshioka, Koichi",
    title = "{Stellar cooling limits on light scalar boson revisited}",
    eprint = "2303.03123",
    archivePrefix = "arXiv",
    primaryClass = "hep-ph",
    reportNumber = "KUNS-2955",
    doi = "10.1016/j.physletb.2023.138027",
    journal = "Phys. Lett. B",
    volume = "843",
    pages = "138027",
    year = "2023"
}

@article{Weidemann:1968uw,
    author = "Weidemann, V.",
    title = "{White dwarfs}",
    doi = "10.1146/annurev.aa.06.090168.002031",
    journal = "Ann. Rev. Astron. Astrophys.",
    volume = "6",
    pages = "351--372",
    year = "1968"
}

\end{document}